\documentclass{aa}  
\usepackage{float}
\usepackage{array}
\usepackage[switch]{lineno}
\usepackage{graphicx,booktabs}
\usepackage{multirow}
\usepackage{color}
\usepackage{txfonts}
\usepackage{numprint}
\usepackage{textcomp,gensymb}
\usepackage{subcaption}
\usepackage{comment}
\usepackage{tabularx}
\usepackage{siunitx}
\usepackage{booktabs}
\usepackage{amsmath}
\usepackage{hyperref}
\usepackage[dvipsnames]{xcolor}
\usepackage{parskip}
\usepackage{rotating}
\usepackage{pdflscape} 
\usepackage{lscape}
\newcommand{\farc}{\mbox{\ensuremath{^{\prime\prime}}}}

\newcommand{\polint}{PI}     
\newcommand{\polfrac}{PF}    
\newcommand{\polang}{PA}  

\newcommand{\DeltaAng}{\mathcal{S}} 

\newcommand{\alma}{ALMA} 
\newcommand{\polfracdust}{\polfrac_{\rm dust}} 
\newcommand{\polfracsynch}{\polfrac_{\rm syn}} 
\newcommand{\NH}{N_{H}} 
\newcommand{\pearson}{\rho} 

\newcommand{\rev}[1]{\textcolor{black}{#1}}
\newcommand{\newrev}[1]{\textcolor{black}{#1}}

\usepackage{afterpage}

\begin{document} 

   \title{The Dust Polarisation and Magnetic Field Structure in the Centre of NGC253 with ALMA}

   \subtitle{}

   \author{Davide~Belfiori
          \inst{\ref{UniBO}}$^{,}$\inst{\ref{IRA}}
          \and
          Rosita~Paladino\inst{\ref{IRA}}
          \and
          Annie~Hughes\inst{\ref{IRAP}}
          \and
          Jean-Philippe~Bernard\inst{\ref{IRAP}}
          \and 
        Dana~Alina\inst{\ref{Astana}}
        \and
        Ivana Be\v{s}li\'c \inst{\ref{LERMA}}
          \and
          Enrique Lopez Rodriguez\inst{\ref{KIPAC}\ref{USC}}
          \and
          Mark D. Gorski\inst{\ref{CIERA}}
          \and
          Serena A. Cronin\inst{\ref{Maryland}}
          \and 
          Alberto D. Bolatto\inst{\ref{Maryland}}
          }

   \institute{
   DIFA, Universit\`{a} di Bologna, via Gobetti 93/2, I-40129 Bologna, Italy \label{UniBO}
   \and
   INAF, Istituto di Radioastronomia, Via Gobetti 101, I-40129 Bologna, Italy \label{IRA}
   \and
   IRAP, Universit\'e de Toulouse, CNRS, CNES, UPS, (Toulouse), France\label{IRAP}
    \and
    Department of Physics, School of Sciences and Humanities, Nazarbayev University, Astana 010000, Kazakhstan\label{Astana}
    \and
    LERMA, Observatoire de Paris, PSL Research University, CNRS, Sorbonne Universités, 75014 Paris, France\label{LERMA}
    \and
    Kavli Institute for Particle Astrophysics \& Cosmology (KIPAC), Stanford University, Stanford, CA 94305, USA\label{KIPAC}
    \and
    Department of Physics \& Astronomy, University of South Carolina, Columbia, SC 29208, USA\label{USC}
    \and
    Center for Interdisciplinary Exploration and Research in Astrophysics (CIERA) Northwestern University, Evanston, IL 60208, USA\label{CIERA}
    \and
    Department of Astronomy, University of Maryland, College Park, MD 20742, USA\label{Maryland}
             }
    


  \abstract
   { Magnetic fields have an impact on galaxy evolution at multiple scales. They are particularly important for starburst galaxies, where they play a crucial role in shaping the interstellar medium (ISM), influencing star formation processes and interacting with galactic outflows.}
   {The primary aim of this study is to obtain a parsec-scale map of dust polarisation and B-field structure within the central starburst region of NGC253. This includes examining the relationship between the morphology of B-fields, galactic outflows and the spatial distribution of super star clusters (SSC), to understand their combined effects on the galaxy's star formation and ISM.}
   {We used ALMA full polarisation data in Bands 4 ($\sim$ 145 GHz) and 7 ($\sim$ 345 GHz) with resolution of $\sim 25$ and $\sim 5$ pc scale, respectively. The Stokes I, Q and U maps of the two bands have been used to compute the polarised intensity ($\polint$), polarisation fraction ($\polfrac$), B-field orientation on the plane of the sky and dispersion angle function ($\DeltaAng$) maps. We computed the pixel-by-pixel uncertainties of these maps taking into account the covariance between the Stokes parameters I, Q and U. The uncertainty allows us to detect values of $\polfrac$ as low as $\sim 0.1 \: \%$ with a S/N greater than 3.
   Through a SED fitting analysis including archival data, we investigated the main emitting components that contribute to the total and polarised emission in several areas of the starburst region. }
   {
   According to our SED-fitting analysis, the observed Band 4 emission is a combination of dust, synchrotron \rev{and} free-free, while Band 7 traces only dust. 
   \rev{The} $\polfrac$ of the synchrotron component measures $\sim \: 2\%$, while that of the dust component is $\sim 0.3\%$.
   The B-fields orientation maps in both bands at common resolution show that the same B-fields structure is traced by dust and synchrotron emission at scales $\sim$ 25 pc.  
   The B-field morphology suggests a coupling with the multiphase outflow, while the distribution of $\polfrac$ in Band 7 showed to be correlated with the presence of super starclusters (SSC). We observed a significant anti-correlation between polarisation fraction and column density in both Bands 4 and 7. A negative correlation between $\polfrac$ and $\DeltaAng$ was observed in Band 4 but was nearly absent in Band 7 at native resolution, suggesting that the tangling of B-field geometry along the plane of the sky is the main cause of depolarisation at $\sim 25$ pc scales, while other factors play a role at $\sim 5$ pc scales. 
   }
   {}

   \maketitle
%

\section{Introduction} 
\label{intro}

Decades of observational and theoretical research have confirmed the ubiquitous existence of magnetic fields (B-fields) in external galaxies \citep{Lopez-Rodriguez2022b, Shukurov_Subramanian_2021, Beck2019, Krause2014, Basu2013, Moss2012}. 
From a theoretical point of view, these fields significantly influence various astrophysical processes and thus affect the evolution of galaxies at multiple scales. Their effects include driving mass inflows of gas into the centres of galaxies \citep{Kim2012}, and controlling the collapse of molecular clouds, which is critical for star formation \citep{Hennebelle2019}.
Observations have also established a close correlation between B-fields and the properties of the interstellar medium (ISM) in their host galaxies. Studies have indicated links between the strength of B-fields and the star formation rate (SFR) in galaxies \citep{Beck2019}, the pitch angle of B-fields and the molecular gas in spiral arms \citep{VanEck2015}, the strength of large-scale ordered B-fields and the rotational velocity of galaxies \citep{Tabatabaei2013}, as well as the strength of B-fields and both gas density and SFR density \citep{Chyzy2017}.


Spiral galaxies have large-scale ordered magnetic fields, with an average strength of about $5 \pm 2\,\mu G$, while the total magnetic field strength \rev{(which includes both the ordered and perturbed component)} is around $17 \pm 14\,\mu G$ \citep{Beck2019}.
In face-on spiral galaxies, the magnetic field generally shows a kiloparsec-scale spiral pattern, with polarised emissions mainly aligned with interarm regions and peaking at the inner edges of spiral arms. This is due to a higher level of turbulence within the arms compared to interarm regions \citep{Fletcher2011, Beck2015}. In edge-on spiral galaxies, the magnetic field exhibits a dual morphology: one component parallel to the disk's midplane and another forming an X-shaped structure extending several kiloparsecs above and below the disk \citep{Krause2020}. The exact origin of the X-shaped magnetic fields in these halos remains unknown.

B-fields are especially important for starburst galaxies. The strongest B-fields ($50$ to $300$ $\mu G$) have been observed in the core of the starburst sources \citep{Lopez-Rodriguez2023, Lopez-Rodriguez2021, Thompson2006}. The energy densities derived from synchrotron radiation suggest that magnetic energy can exceed thermal and kinetic energies, implying that magnetic fields are essential to the dynamics and evolution of these starburst sources \citep{Heesen2011}. FIR polarimetric observations indicated an equipartition between turbulent magnetic and turbulent kinetic energy in the central $\sim \: 1$ kpc galactic outflows \citep{Lopez-Rodriguez2023, Lopez-Rodriguez2021}. 
The large scale B-field structures in starburst systems have been observed to be aligned with the outflow activity \citep{Heesen2011, Lopez-Rodriguez2021}. This may have an effect on the galaxy's subsequent evolution, since the outflow can drag material from the galaxy disk out into the circumgalactic medium (CGM) \citep{Lopez-Rodriguez2021, Lopez-Rodriguez2023}, thereby enriching it \citep{MartinAlvarez2021}. Analysis of simulations and observations of starburst systems have further suggested that B-fields play an important role in the post-starburst phase, and are potentially responsible for the quenching of star formation activity by preventing the collapse of giant molecular clouds \citep{Pattle2023, Tabatabaei2013}.

Most of our knowledge of B-fields in external galaxies comes from radio polarimetric observations in the $3-20$ cm wavelength range \rev{\citep{Stein2019, Beck2015}}. At these wavelengths, the emission is dominated by synchrotron radiation arising mostly in the warm and diffuse phase of the ISM of galaxies \citep{Beck2013book}. Observations of polarised light at far infrared (FIR) and sub-millimeter wavelengths offer an alternative probe of the magnetic field structure in colder and denser environments. In the presence of magnetic fields, the major axis of the elongated dust grains tends to align perpendicular to the B-field. The physical process leading to alignment remains debated, proposed mechanisms include paramagnetic alignment \citep{Davis1951} and radiative torques  \citep[e.g.][]{Hoang2014}. The thermal emission from dust in the FIR and millimeter wavelength range is thus linearly polarised, with the E-vector pointing perpendicular to the field. The measured polarisation angle (PA) rotated by 90$^{\circ}$ traces the B-field orientation, projected in the plane of the sky.

Efforts to map the dust polarisation in external galaxies have significantly increased over the past decade, largely thanks to the improved resolution and sensitivity provided by the High-Angular Wideband Camera Plus (HAWC+; \citealp{Harper2018, Dowell2010}) onboard the Stratospheric Observatory For Infrared Astronomy (SOFIA). With HAWC+, the Survey on extragALactic magnetiSm with SOFIA (SALSA) Legacy Program \citep[SALSA;][]{Lopez-Rodriguez2022} mapped the polarised dust emission from fourteen nearby galaxies ($d<20$\,Mpc, including NGC253) at 0.1$-$1\,kpc scales in the 53 to 214 $\mu$m wavelength range. The first results from  SALSA indicate typical polarisation fractions of $3$\% in galaxy disks, with starburst galaxies tending to exhibit lower polarisation fractions than normal spiral galaxies. The B-field orientations inferred from the SALSA observations tend to be less ordered than the results obtained from polarised synchrotron emission at GHz frequencies \citep{Surgent2023}, suggesting that the magnetic field structure in the cold gas traced by dust polarisation may be more dominated by morphological complexity due to star formation activity and kinematic turbulence than the warmer, diffuse ISM \citep{Borlaff2023}. For the three starburst galaxies in the SALSA sample, the dust polarisation measurements indicated extraplanar B-field structures perpendicular to the galaxy disk, physically associated with outflowing material in those systems \citep{Lopez-Rodriguez2021, Lopez-Rodriguez2023}.

In this paper, we analyse polarisation observations made with ALMA Bands~4 ($\sim$ 145 GHz) and~7 ($\sim$ 345 GHz) across the central $\sim200$\,pc of the nearby starburst galaxy NGC253. The primary aims of this study are to obtain a parsec-scale map of the dust polarisation and B-field structure within the dense star-forming gas, and identify whether there is evidence for a physical connection between the observed magnetic field structure and dust polarisation properties and the starburst activity and/or outflow. 
The paper is organized as follows: we present our target and describe the ALMA observations in Section~\ref{ss:data}. In this Section, we describe the polarisation calibration and imaging procedures, including the method that we used to characterise the noise and hence estimate the uncertainties on the polarisation quantities that we infer.  
We present our ALMA maps of polarisation fraction and polarisation angle in Section~\ref{polimages}, including a comparison of the polarisation properties at matched resolution. We also quantify the dispersion in the polarisation angles at the two bands. In order to interpret the polarisation results at the two ALMA bands, we performed a region-based total intensity SED fitting analysis to estimate the relative contributions of dust, synchrotron and free-free emission at Band~4; this analysis is presented in Section~\ref{analysis}. In Section~\ref{discussion}, we discuss the observed magnetic field structure in relation to other ISM and star formation (SF) components of NGC253, the relative contributions of dust and synchrotron emission, and compare our results for the polarisation fraction and polarisation angle dispersion to Planck results for our Galaxy. We summarize our main conclusions in Section~\ref{conclusion}. A table listing the archival data that we used for our analysis is presented, with the corresponding project codes, in Appendix~\ref{appA}.

\section{Data}
\label{ss:data}

\subsection{Target: NGC253 }

\begin{figure}[tbp]
  \begin{center}
  \begin{subfigure}{0.8\linewidth}
    \includegraphics[width=\linewidth, height=0.80\linewidth]{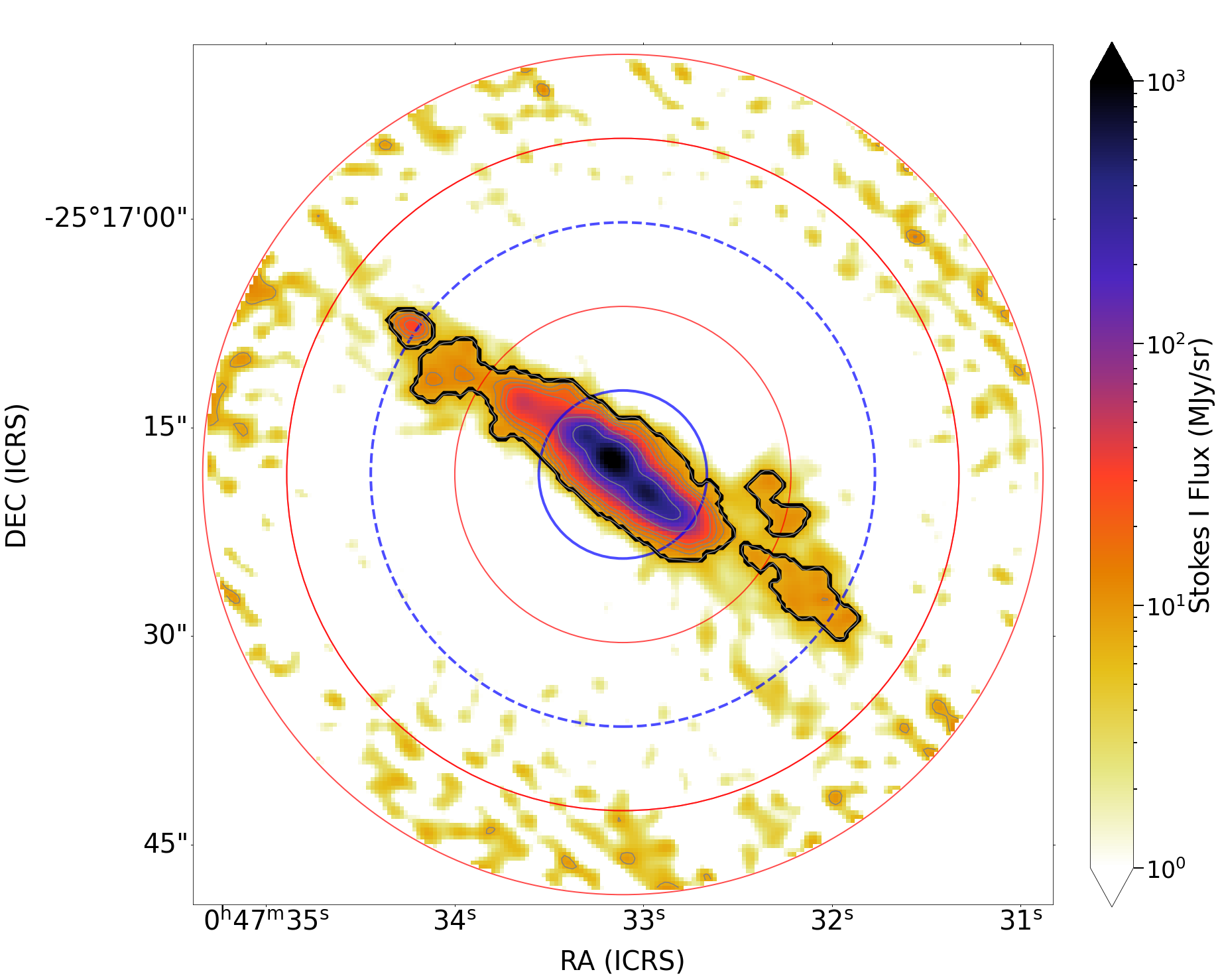}
  \end{subfigure}

  \vspace{1em} 

  \begin{subfigure}{0.8\linewidth}
    \includegraphics[width=\linewidth, height=0.80\linewidth]{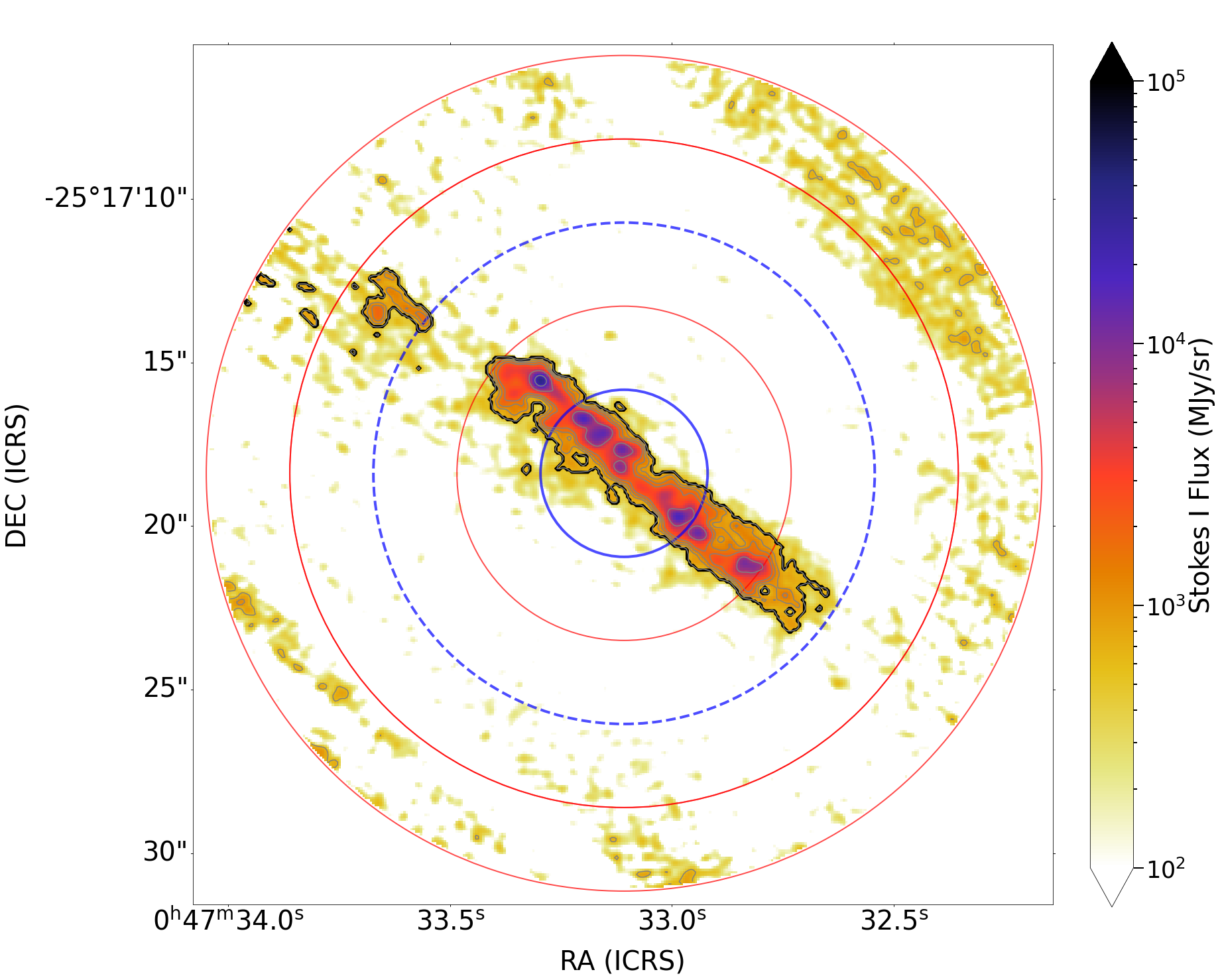}
  \end{subfigure}

  \end{center}
  
\caption{ {\bf Total intensity images.} Band~4 ($\sim$ 145 GHz) is shown on the top panel, Band~7 ($\sim$ 345 GHz) on the bottom panel.
The thick black contour corresponds to $3 \times \sigma_I$, while the grey thin contours correspond to $[5, 7, 10, 30, 100] \times \sigma_I$.
The concentric circles overlaid on the images are the annular regions used for the noise analysis (see Section~\ref{Noise}). \rev{The thick and dotted blue circles correspond to 1/3 of the HPBW and the full HPBW of the image, respectively}. The red circles have as radii multiples of $n * 1/3$ of the HPBW, where n ranges from 2 to 5. }
\label{fig:StokesI_analysis}
\end{figure}

NGC253 is a nearby, highly inclined, \rev{tilted}, starburst galaxy \citep[\rev{distance} $D=3.5$ Mpc, {inclination} $i \sim 78^{\circ}$, \rev{position angle of the long axis $\theta \sim 52^{\circ}$};][]{Rekola2005, Pence1980, Lucero2015}. As the nearest system hosting an exceptional multiphase galactic wind, it has been extensively studied in many wavebands, including X-ray \citep{Strickland2000}, optical \citep{Watson1996}, infrared \citep{Sugai2003}, sub-/millimeter \citep{Bolatto2013}, and GHz radio, including polarisation \citep{Turner1985, Heesen2009, Heesen2011}. The central starburst activity in NGC253 is fuelled by gas accretion along the bar \citep{Paglione2004}, resulting in a SFR of $\approx 2 \: M_{\odot} \: \text{yr}^{-1}$ \rev{\citep{Bendo2015, Leroy2015, Ott2005}} across the central few hundred parsecs. This intense star formation drives a massive outflow \citep[][]{Walter2017}. \rev{In this region signs of shock activities have also been found \citep{Humire2025}.}
The mass outflow rate has been estimated to be $14-39 \: M_{\odot} \: \text{yr}^{-1}$ in the cool molecular phase \citep{Krieger2019} and approximately $2.8 \: M_{\odot} \: \text{yr}^{-1}$ for the northwest side and around $3.2 \: M_{\odot} \: \text{yr}^{-1}$ for the southeast side in the warm ionized phase \citep{Lopez2023}. 
Recently, the molecular gas content of NGC253's outflow and the nascent superstar clusters (SSCs) that account for a large fraction of the star formation activity in the nuclear starburst have been studied in detail with ALMA  \citep[e.g.][]{Bolatto2013, Leroy2015, Zschaechner2018, Leroy2018, Krieger2019, Levy2021, Mills2021, Levy2022}. These studies have shown that the outflowing molecular gas is located mostly around the edge of the wind. The most prominent outflow structure in ALMA molecular emission line maps is the South-West streamer, from which $\sim 350 \: GHz$ dust continuum emission has also been detected. Estimates of the molecular mass outflow rate range between $\sim 3-9 M_{\odot} \,\rm{yr}^{-1}$ \citep{Bolatto2013} and $\sim 25-50 M_{\odot} \,\rm{yr}^{-1}$ \citep{Zschaechner2018}, i.e. a mass-loading factor of $\sim5$. The SSCs in the centre of NGC253 are young $\sim0.01-3$\,Myr, with typical stellar masses of $10^{4} - 10^{6}$\,M$_{\odot}$ and FWHM sizes of $0.4 - 0.7$\,pc. As projected on the sky, the SSCs appear to be aligned along a thin linear structure with length $\sim150$ pc. Modelling of the SSCs' morphology and kinematics suggests that they are distributed on an elliptical ring with a semimajor (semiminor) axis of 110 (60) pc, which we observe edge-on \citep{Levy2022}.

At far-infrared to radio wavelengths, NGC253's integrated spectral energy distribution (SED) is well-described by a combination of synchrotron, free-free and thermal dust emission, with a very steep synchrotron spectrum at high radio frequencies \citep{Peel2011}. Thermal dust emission seems to dominate the global SED at the frequencies of ALMA Band~4 (145~GHz) and Band~7 (343.5~GHz). As one of the nearest galaxies with an extremely bright dust continuum 
\citep[7.4~Jy at 353~GHz, ][]{PlanckVII2011}, NGC253 is an ideal target for extragalactic dust polarisation studies. SOFIA/HAWC+ observations by SALSA at 89 and 154\,$\mu$m indicate that on $\sim$100\,pc spatial scales, the thermal dust emission is $\sim1$\% polarised in the central region, with higher polarisation fractions further out in the disk. The orientation of the B-field inferred from the SALSA measurements suggests that the magnetic field traced by dust \citep{Lopez-Rodriguez2024} is preferentially aligned with the galactic disk, except in the central region where the field also shows perpendicular structure \citep{Lopez-Rodriguez2023, Lopez-Rodriguez2022}.

\subsection{ALMA Band 4 and 7 Observations} \label{observations}

We obtained full polarisation observations of the central region of NGC253 with ALMA in Band~4 \citep{Band4_rec} and Band~7 \citep{Band7_rec}. The observations consisted of a single-pointing with a primary beam FWHM of $\sim$42 and $\sim$18$\arcsec$, respectively (see Figure \ref{fig:StokesI_analysis}).
The four cross-correlation visibilities (XX, XY, YX, YY) were recorded using the default frequency setup for continuum polarisation observations, TDM mode, with four 2\,GHz spectral windows, each divided into 64 channels.
The central frequencies of the two bands are reported in Table \ref{tab:obs}.
The observations were conducted with the C43-1 for Band 4 and C43-4 for Band~7, providing a nominal angular resolution of 1.7~$\arcsec$ and 0.3~$\arcsec$, and nominal maximum recoverable scale of $\sim$15~$\arcsec$ and 3.8~$\arcsec$, respectively. 

The observations were executed using the session observing scheme, which is standard for polarisation projects with ALMA. Sessions consist of continuous execution of the same schedule until the required parallactic angle coverage for the polarisation calibrator has been achieved. 
A parallactic angle coverage of more than 60~degrees is needed to ensure a proper calibration of the instrumental polarisation (see following section).
In general, this coverage is achieved with two to three executions of the schedule. Table~\ref{tab:obs} reports the details of the different sessions observed and the relevant properties derived in each session for the polarisation calibrator J0006-0623 that was used for both bands. This calibrator was observed at regular intervals in the standard phase calibrator - science target loop. The phase calibrator used for both bands was the quasar J0038-2459, while flux and bandpass calibrations were performed on quasar J2258-2758, or on J0006-0623.  In Band~7, observations of an additional quasar J0046-2691 were included, providing an independent check on the calibration solutions.

\begin{table}[tbp]

    \begin{tabular}{c|c|c|c|c|c}
    \toprule
    
    Band & $\nu$ & Date & EVPA & $\polfrac$ & $\polang$ \\
         & GHz & &deg &\% &deg\\
    \midrule
    
    4 & 145 & 22.01.2019 & 70 & 3.78  & 18\\ 
    &&17.11.2019& 120 & 3.95 & 71.6 \\
  
    \midrule
    7&343.5& 28.04.2019 &100 & 2.73& 36.9\\
    && 30.04.2019 &100 & 2.93 & 36.5 \\
    \bottomrule

    \end{tabular}
    \caption{ {\bf Properties of the observing sessions.} \rev{Column 1 refers to the first Local Oscillator frequencies of the two bands}. Column 4 reports the parallactic angle range (EVPA) covered during each session. Columns 5 and 6 report the polarisation fraction ($\polfrac$) and the polarisation angle ($\polang$) of the polarisation calibrator J0006-0623. The results are in agreement with the monitoring results see: \url{http://www.alma.cl/~skameno/AMAPOLA}.}
    \label{tab:obs}
\end{table}

\subsection{Calibration}\label{calib}

The data were calibrated using the CASA software package 
\citep[version 5.6.1, ][]{casa2022} and the standard calibration scripts provided by the observatory. The current calibration scheme for polarisation projects is summarized in  \cite{nagai2016} and in the ALMA polarisation casaguide \footnote{\url{https://casaguides.nrao.edu/index.php?title=3C286_polarisation}}.
The total intensity calibration, deriving frequency and time dependent gains, is performed for each observing execution, while the polarisation calibration is done per session (combining consecutive executions). This calibration assumes the polarisation calibrator has no circular polarisation (Stokes V = 0) and it derives the linear polarisation properties of the calibrator as well as the instrumental polarisation from the data, exploiting the parallactic angle variation. 
Table~\ref{tab:obs} reports the calibrator's linear polarisation ratio and polarisation angle (EVPA; $\polang$) determined from the different sessions. We observed a $90^{\circ}$ wrap of the EVPA in our Band~4 observations. This is due to an intrinsic variation of the calibrator's properties at all frequencies, as reported in the monitoring survey (see \url{http://www.alma.cl/~skameno/AMAPOLA}).

The instrumental polarisation gains that we derived ($Dterms<4$\% for each antenna) were applied to all sources. To confirm the quality of the calibration, an image of the polarisation calibrator for each session was made. The Stokes I flux is normalized by the calibration procedure, so we measure a peak of $\sim$ 1~Jy/beam in both bands. The Stokes V images show no signs of residual instrumental polarisation. They have an RMS of 0.015~mJy/beam and~0.030 mJy/beam in Band~4 and Band~7, respectively.

\subsection{Full-Stokes imaging}\label{imag}

The calibrated sessions for each band were combined and imaged using the CASA task {\it tclean}.
Continuum full Stokes (I, Q, U, V) images of the target and the calibrators were obtained in multi-frequency mode (mfs), using the {\it clarkstokes} deconvolver algorithm in which a Clark clean operates separately on each Stokes plane. 
A Briggs weighting scheme \citep{Briggs1995} with robust = 0.2 and 0.5 was used for imaging of Band~4 and~7 respectively. The resulting images in Band~4 have a synthesized beam of 1.54 $\times$ 1.33 $\arcsec$ (26.1 $\times $ 22.1 pc), with a beam position angle bpa=81.2~deg, and in Band~7 a beam of 0.32 $\times$ 0.28$\arcsec$ (5.4 $\times 4.8$\,pc), with a bpa=-88.5~deg.
The RMS measured in Stokes I, Q and U, are reported in Table~\ref{tab:imag}; 
these values are consistent with expectations from the radiometer formula. Stokes I images, shown in Figure~\ref{fig:StokesI_analysis}, are dynamic range limited in both Bands 4 and 7 with a signal-to-noise ratio (SNR) of 500 and 180, respectively.

In order to compare the results at the two different frequencies, we convolved the data to a common resolution of 1.54 $\times$ 1.33$\arcsec$ (i.e. the Band~4 resolution). Additionally, the spectral energy distribution analysis presented in Section~\ref{analysis} required us to homogenize the data at all frequencies to the resolution of the archival VLA L-band data ($2.2 \times 2.2\arcsec$, $\sim37$\,pc). Table~\ref{tab:imag} reports the noise characteristics of the images at the different resolutions that we use in our analysis.

\begin{table}[tbp]

    \begin{tabular}{c|c|c|c|c|c|c}
    \toprule
    
    Band & bmaj & bmin & bpa &  $\sigma_{I}$ & $\sigma_{Q}$&  $\sigma_{U}$\\
         & [\farc] & [\farc] & [deg] & \multicolumn{3}{c}{[MJy/sr]}\\
    \midrule
    
    4  & 1.54 & 1.33 & 81.2 &  2.98 & 0.37 & 0.36 \\
      & 2.2 & 2.2 & 0 & 2.51 & 0.15 & 0.15 \\
   
    \midrule
    7 & 0.32 & 0.28 & -88.5 & 227.67 & 23.24 & 24.34 \\
      & 1.54 & 1.33 & 81.2 & 149.05 & 3.18 & 4.80 \\
      & 2.2 & 2.2 & 0 & 121.67 & 2.16 & 3.38 \\
    \bottomrule
    
    \end{tabular}
    \caption{{\bf{Properties of final images.}} RMS of the Band~4 and Band~7 Stokes I, Q and U images measured at the different resolutions used in this analysis. For each band, the properties of the images at the native resolution are reported in the first row.}    
    \label{tab:imag}
\end{table}

\subsection{Noise characterization} \label{Noise}

From the Stokes Q and U images, we constructed images of the linearly polarised intensity $PI=\sqrt{Q^2+U^2}$, the polarisation fraction $PF=PI/I$, and the polarisation angle $PA=1/2 \arctan{(U/Q)}$.

In polarisation studies, strong assumptions are often made 
about the noise properties of the measurements, such as 
neglecting the correlation between the total and polarised intensities 
and between the noise in Stokes Q and U images.
Nevertheless the covariance between the Stokes parameters can have a complex effect on the non-linear quantities derived, such as PF, PI and PA, especially at low SNR \citep[e.g.,][]{montier2015a}.
For this reason, we decided to compute the uncertainties taking into account the full covariance matrix of the Stokes I, Q and U components. This is defined as:

\begin{equation}
\Sigma = 
\begin{bmatrix}
 \sigma_{II} & \sigma_{IQ} & \sigma_{IU} \\
 \sigma_{IQ}  & \sigma_{QQ} & \sigma_{QU} \\
 \sigma_{IU}  & \sigma_{QU} &  \sigma_{UU}\\
\end{bmatrix} ,
\label{eqn:corrmat}
\end{equation}

\noindent where $\sigma_{XY}$ is the covariance between the variables X and Y. We assume that circular polarisation (i.e., Stokes V) can be neglected. We compute the uncertainties on $PF$, $PI$ and $PA$ taking into account the full covariance matrix. In this case, the noise in our derived quantities may be defined as follows: 

\begin{multline}
\sigma_{\polfrac}^2 = \frac{1}{\polfrac^2 I^4} (Q^2 \sigma_{QQ} + U^2 \sigma_{UU} + \polfrac^4 I^2 \sigma_{II} + \\  2Q U \sigma_{QU} - 2IQ \polfrac^2 \sigma_{IQ} - 2 I U \polfrac^2 \sigma_{IU} )
\end{multline}

\begin{equation}
\sigma_{\polint}^2 = \frac{1}{\polint^2} ( Q^2 \sigma_{QQ} + U^2 \sigma_{UU} + 2QU\sigma_{QU} ) \: \: \: [(MJy/sr)^2]
\end{equation}

\begin{equation}
\sigma_{\polang}^2 = \frac{1}{4}\frac{Q^2 \sigma_{UU} + U^2 \sigma_{QQ} - 2QU\sigma_{QU}}{(Q^2 + U^2)^2} \: \: \: [rad^2]
\end{equation}

\noindent
See \cite{montier2015a} and appendix B1 of \cite{planckcollaborationXIX} for a full discussion on the impact of noise on estimates of $PF$, $PI$ and $PA$.

\begin{table}[tbp]
\centering
\scalebox{0.7}{
\begin{tabular}{|c|c|c|c|c|c|c|}
\hline
\textbf{Ring} & \textbf{$\sigma_{II}$} & \textbf{$\sigma_{IQ}$} & \textbf{$\sigma_{IU}$} & \textbf{$\sigma_{QQ}$} & \textbf{$\sigma_{UU}$} & \textbf{$\sigma_{QU}$} \\ \hline
\multicolumn{7}{|c|}{ Band 4 [(MJy/sr)$^{2}$]} \\ 
\hline
0 (inner 1/3) & 6.70 & -0.06 & -0.03 & 0.04 & 0.04 & 6.9E-03 \\ \hline
1 & 7.55 & 0.09 & -0.13 & 0.03 & 0.03 & -3.0E-03 \\ \hline
2 & 6.66 & 0.03 & -0.04 & 0.04 & 0.04 & 2.1E-03 \\ \hline
3 & 4.80 & 4.14E-03 & 0.02 & 0.08 & 0.08 & 2.5E-03 \\ \hline
4 & 10.31 & 0.03 & -0.08 & 0.21 & 0.20 & 9.6E-03 \\ \hline
\multicolumn{7}{|c|}{ Band 7 [(MJy/sr)$^{\rev{2}}$]} \\ 
\hline
0 (inner 1/3) & 3.96E+04 & -257.28 & -14.19 & 98.58 & 106.45 & 15.46 \\ \hline
1 & 3.24E+04 & 56.80 & -28.09 & 105.77 & 130.86 & -1.82 \\ \hline
2 & 2.79E+04 & 121.77 & 50.77 & 168.56 & 202.37 & 7.78 \\ \hline
3 & 3.42E+04 & -9.98 & -24.79 & 351.89 & 408.84 & 0.13 \\ \hline
4 & 7.13E+04 & 4.54 & -156.03 & 833.56 & 939.74 & 4.57 \\ \hline

\end{tabular}
}
\caption{ {\bf Elements of the noise covariance matrix.} These values were computed at native resolution as measured in the ALMA maps of NGC253. Each row corresponds to a different ring shown on Figure \ref{fig:StokesI_analysis}.}
\label{tab:cov_matrix}
\end{table}

Since the ALMA data are provided with no per-pixel estimate of the noise correlation matrix, we estimate the noise correlation matrix elements of Equation~\ref{eqn:corrmat} by computing covariances in signal-free regions of the I,Q,U maps.
We divided the images in five annuli of increasing radius (see Figure \ref{fig:StokesI_analysis}).
The central region is a circle with a radius equal to one-third of the Half Power Beam Width (HPBW) of the antenna primary beam. The subsequent concentric rings are determined by circles with radii $r_i=n \times 1/3 \times HPBW$, with $n$ ranging from 2 to 5. In each region, we computed the covariance matrix excluding the pixels with total intensity greater than 0.16~mJy/beam and 0.47~mJy/beam, corresponding to three times the RMS of the total intensity Band~4 and~7 images, respectively. The covariance matrix elements thus computed for each ring are listed in Table~\ref{tab:cov_matrix}. The derived values are then employed to estimate the uncertainties in $\polfrac$, $\polint$ and $\polang$ for all pixels of that ring. These values are also used to debias the measured $PF$, following the modified asymptotic estimator formula proposed in \cite{Plaszczynski2014}. 
\rev{The variances of the covariance matrix elements $\sigma_{QQ}$ and $\sigma_{UU}$ are roughly constant in the first three rings, as it is shown on Table \ref{tab:cov_matrix}. This would suggest the absence of large systematics outside the 1/3 of the HPBW.}

\rev{In order to facilitate a more direct comparison of the data in the two Bands, we converted the $\polint$ and total intensity estimates and relative uncertainties to MJy/sr. The scale factors of the conversion from Jy/beam to \newrev{MJy/sr} are $18310$ and $431673$ for Band 4 and 7, respectively.}

\begin{figure*}[tbp]
    \centering
    \begin{subfigure}{0.48\linewidth} 
        \centering
        \includegraphics[width=\linewidth, height=0.85\linewidth]{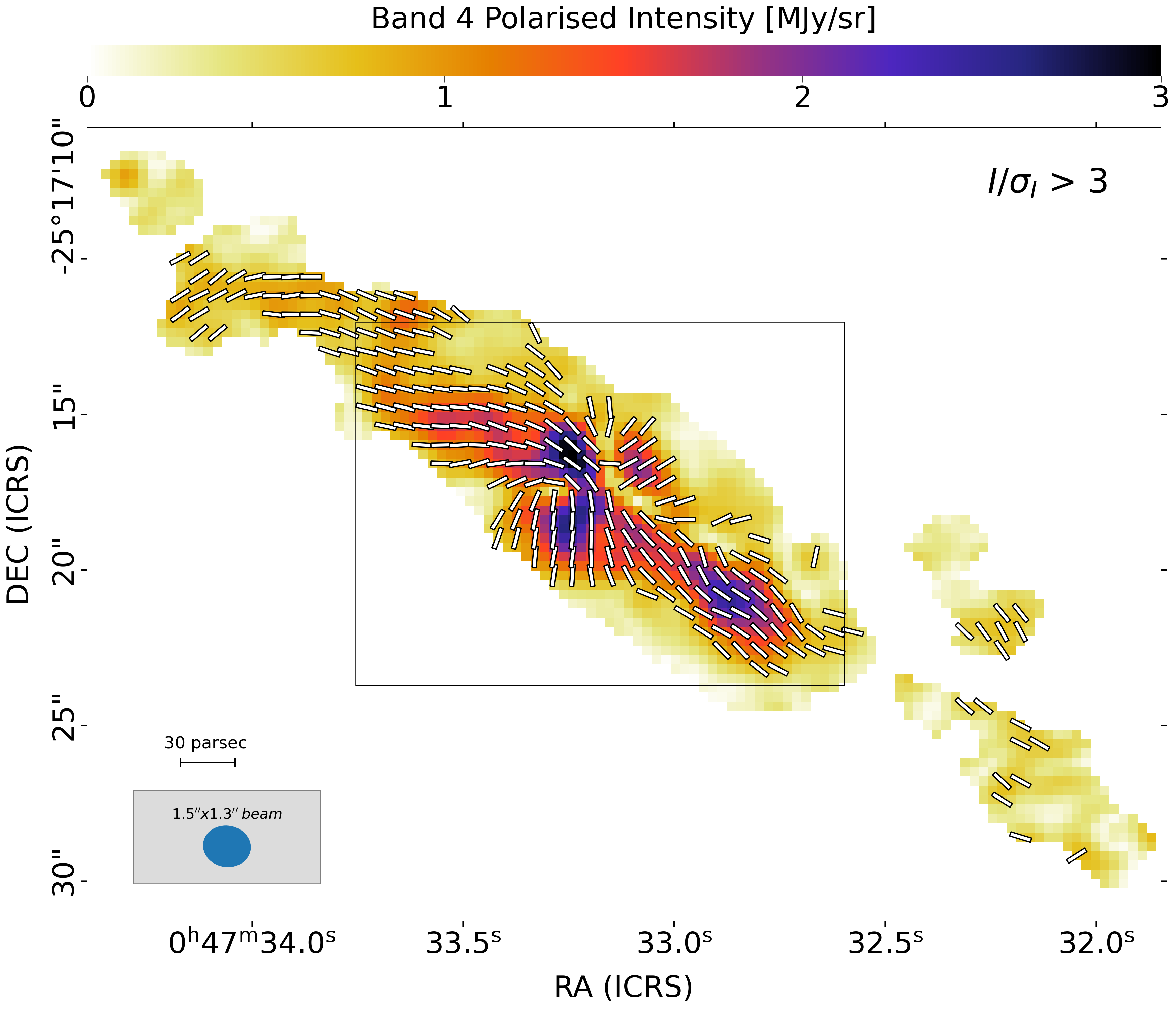} 
    \end{subfigure}\hfil
    \begin{subfigure}{0.48\linewidth} 
        \centering
        \includegraphics[width=\linewidth, height=0.85\linewidth]{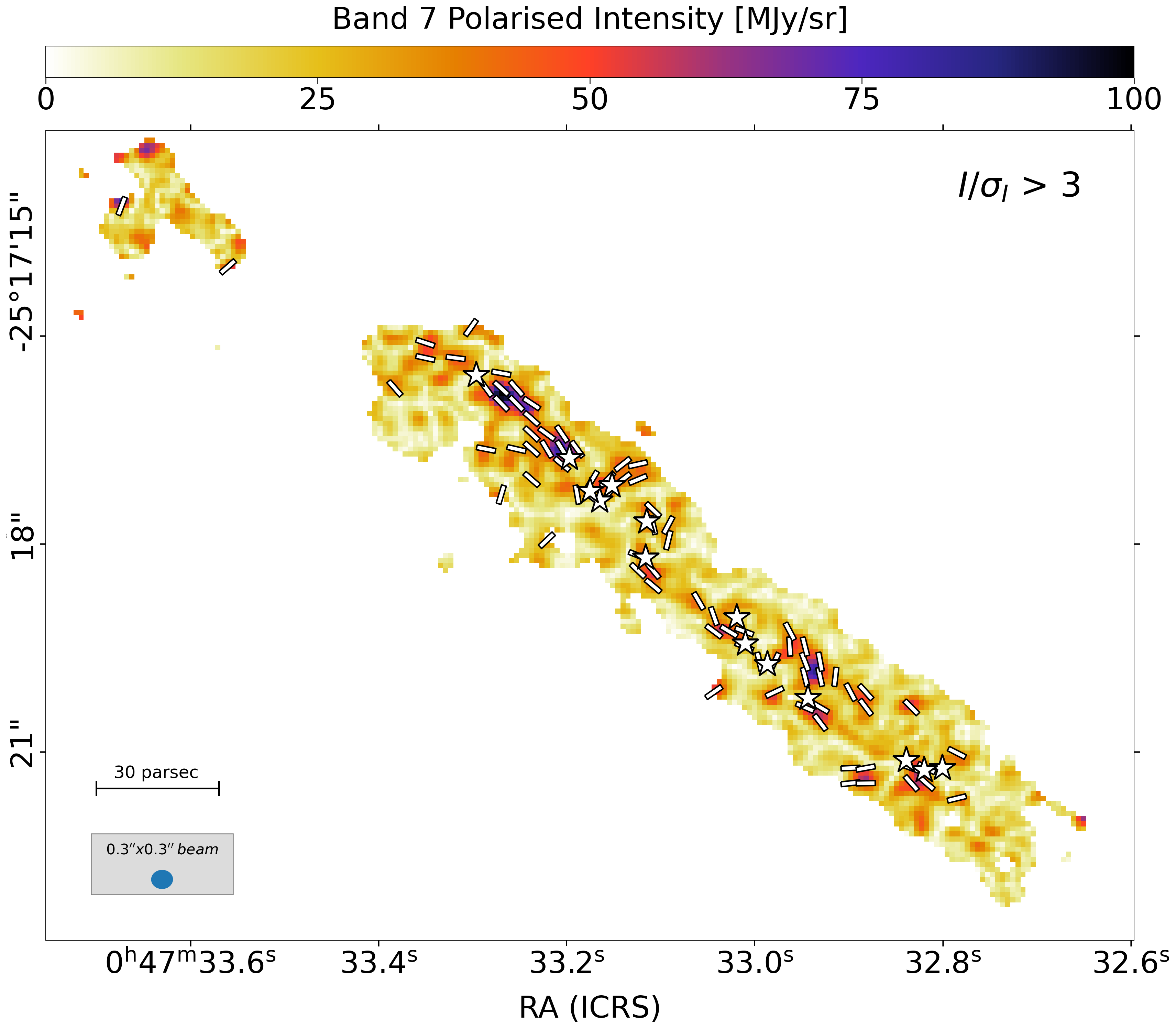} 
    \end{subfigure}

\medskip

    \begin{subfigure}{0.48\linewidth} 
        \centering
        \includegraphics[width=\linewidth, height=0.85\linewidth]{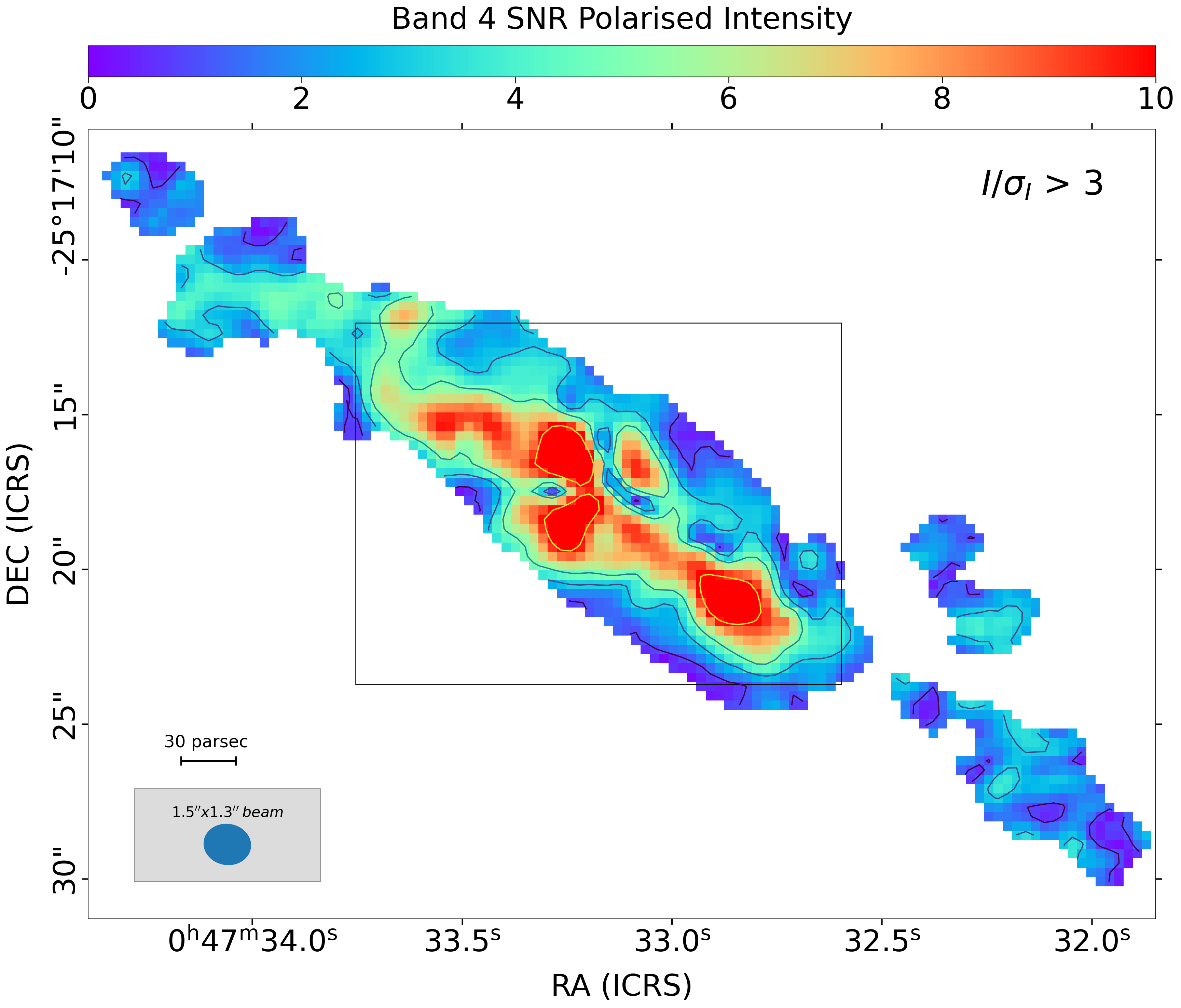} 
    \end{subfigure}\hfil
    \begin{subfigure}{0.48\linewidth} 
        \centering
        \includegraphics[width=\linewidth, height=0.85\linewidth]{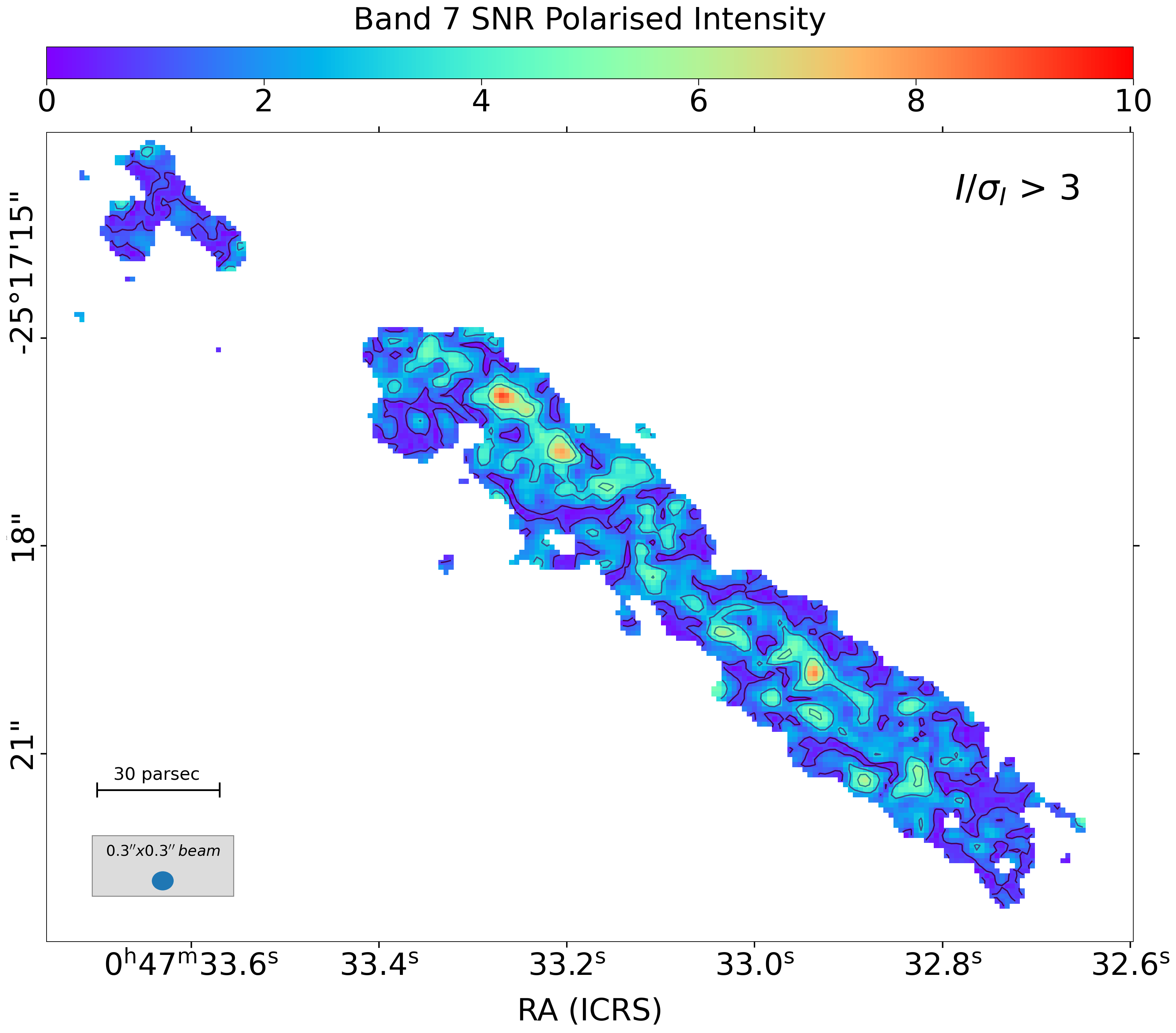} 
    \end{subfigure}
    
\medskip

    \begin{subfigure}{0.37\linewidth} 
        \centering
        \includegraphics[width=\linewidth, height=0.73\linewidth]{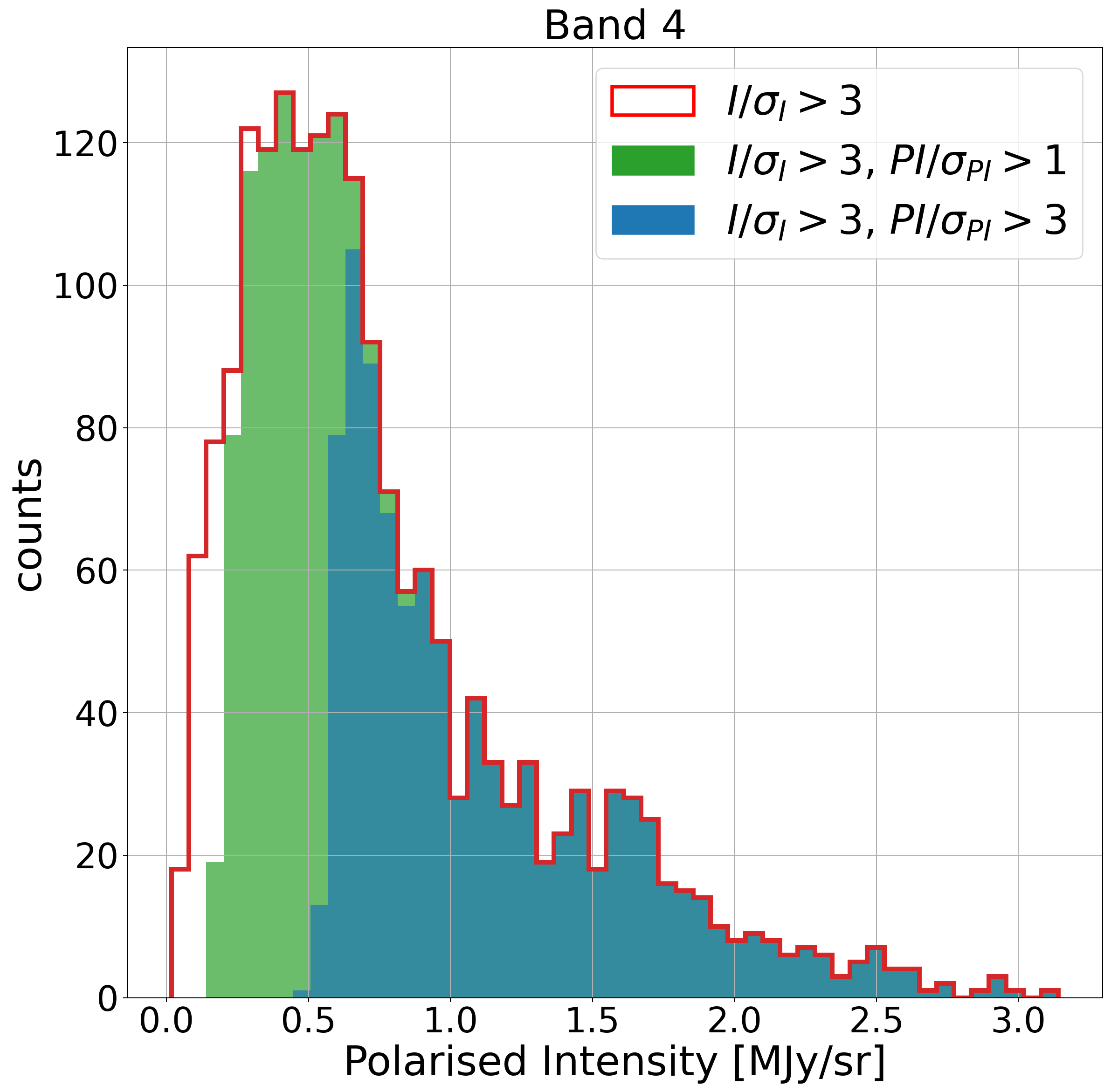} 
    \end{subfigure}\hfil
    \begin{subfigure}{0.37\linewidth} 
        \centering
        \includegraphics[width=\linewidth, height=0.73\linewidth]{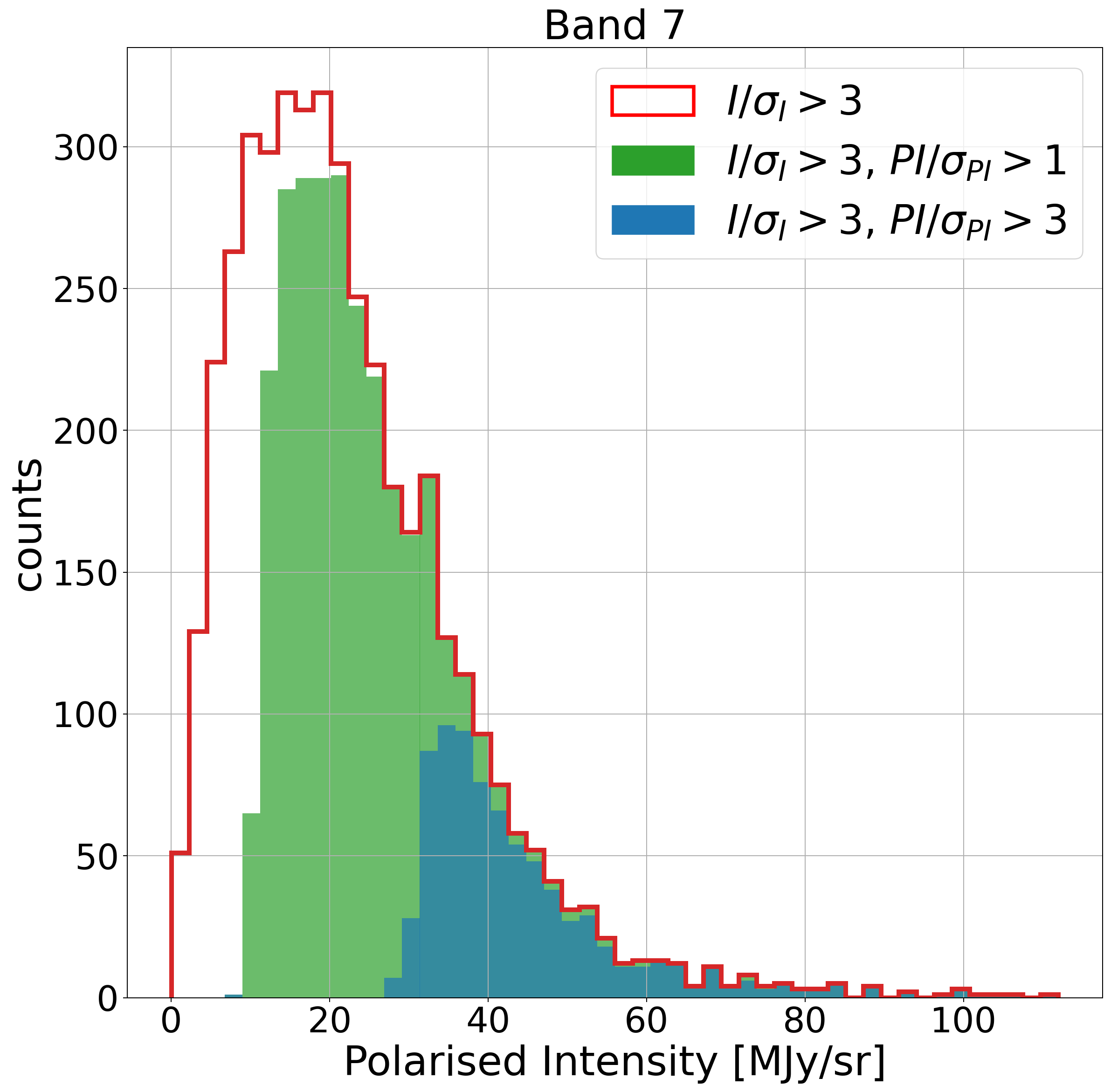} 
    \end{subfigure}

\caption{{\bf{Polarised Intensity in the centre of NGC253.}} {\it{Upper panels:}} Maps of the linearly polarised intensity. \rev{The vectors indicate the magnetic field orientation and are displayed only at pixels where $\polint/\sigma_{\polint} > 3$}.
{\it{Centre panels:}} S/N maps of the polarised intensity with contours corresponding to $[1, 3, 5, 10] \times \polint/\sigma_{\polint}$. White stars in the lower left panel indicate the positions of the SSCs catalogued by \citealp{Leroy2018}. The synthesized beam of the data is indicated in the bottom left corner, and the pixel selection threshold is shown in the top right corner. Both maps have been obtained selecting pixels with total intensity above $3\sigma$.
{\it{Bottom panels:}} Histograms of polarised intensity pixel values in the map. The empty histogram shows the distribution of pixels with total intensity above $3\sigma$, while the solid green and blue histograms are the distribution of pixels with polarised intensity above $1\sigma$ and $3\sigma$, respectively. 
The Band~4 and Band~7 results are shown in the top row and the bottom row, respectively.} 
\label{fig:largeP_analysis}
\end{figure*}

\begin{figure*}[tbp]
    \centering
    \begin{subfigure}{0.48\linewidth} 
        \centering
        \includegraphics[width=\linewidth, height=0.87\linewidth]{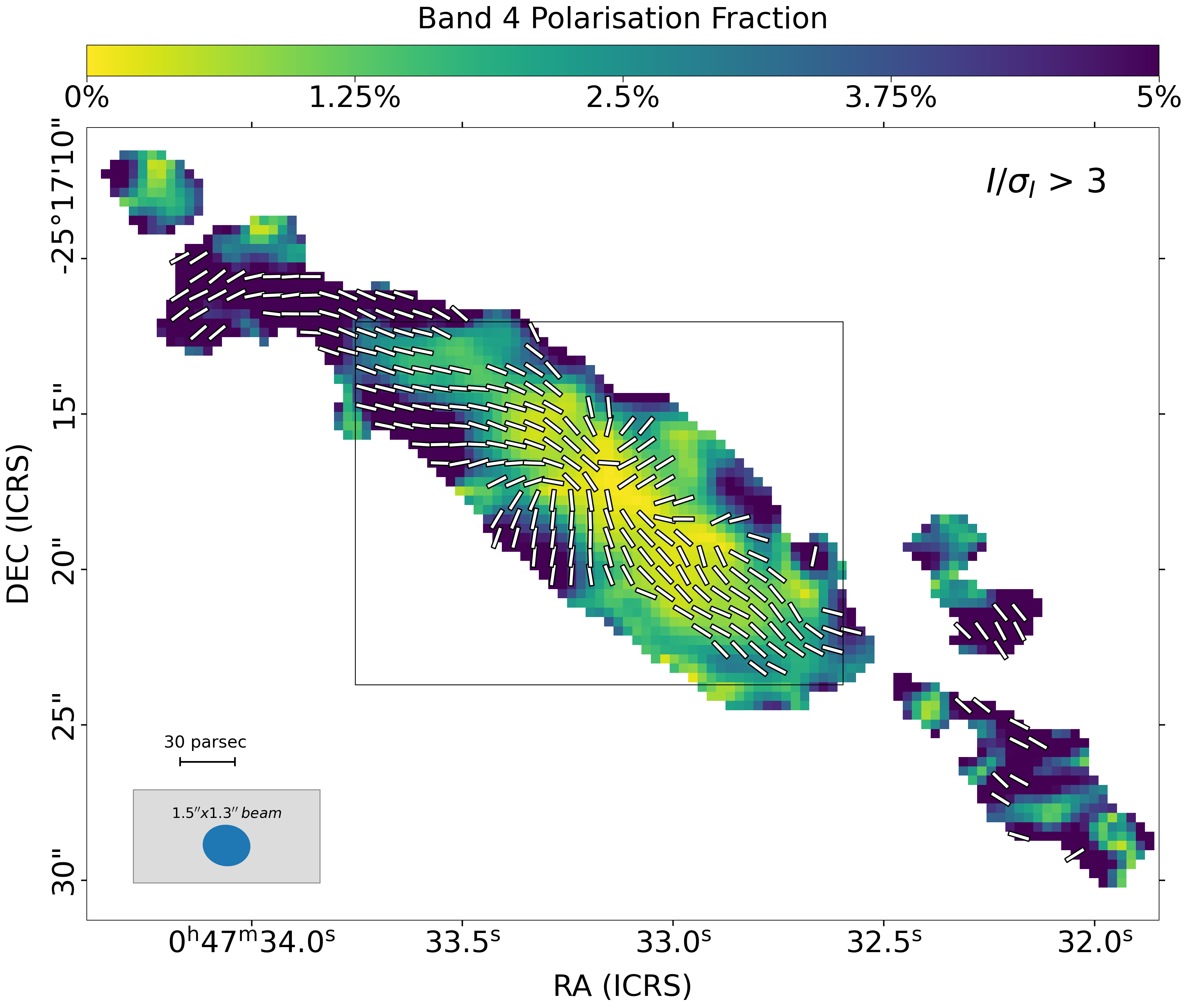} 
    \end{subfigure}\hfil
    \begin{subfigure}{0.48\linewidth} 
        \centering
        \includegraphics[width=\linewidth, height=0.87\linewidth]{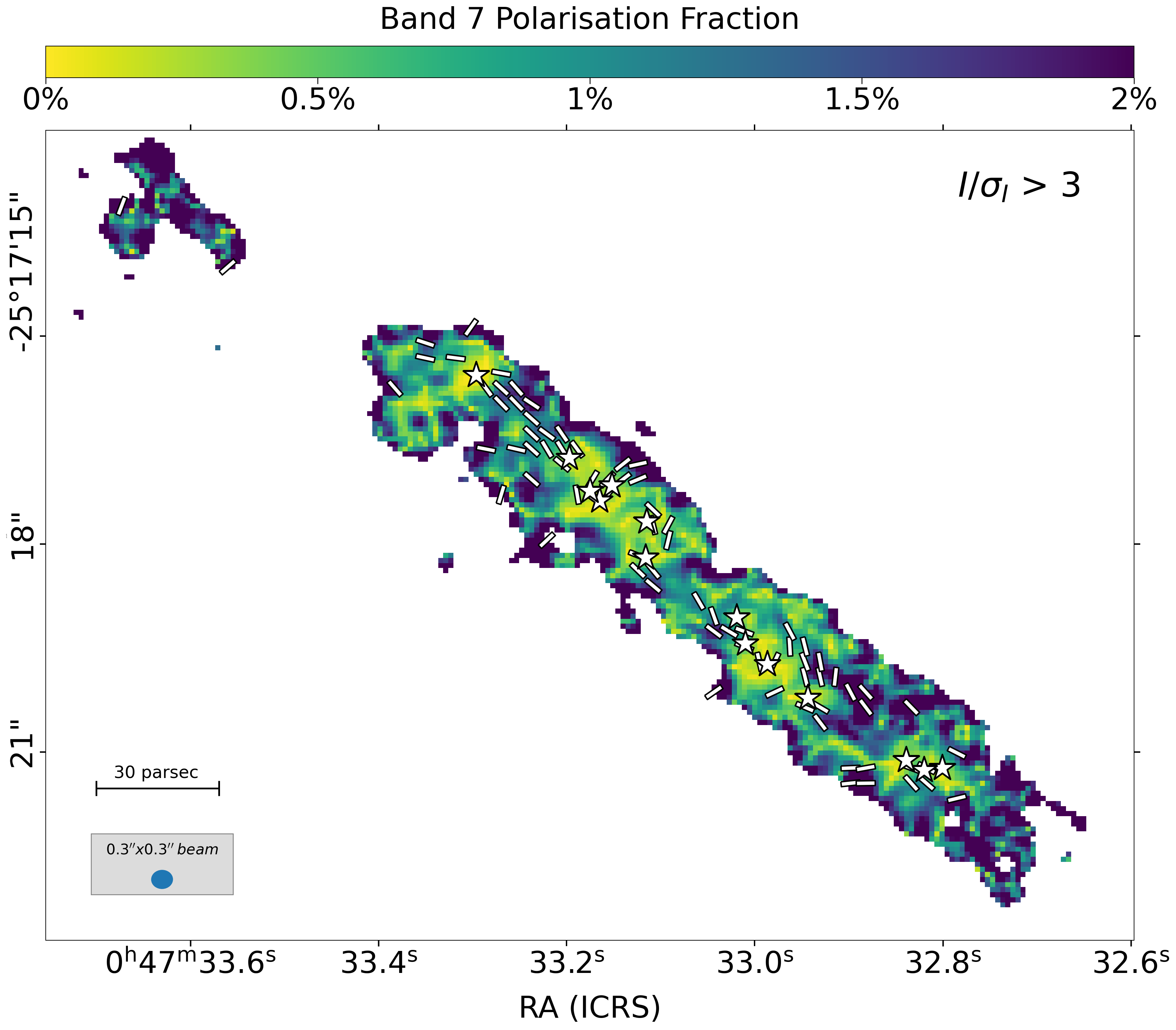} 
    \end{subfigure}

\medskip

    \begin{subfigure}{0.48\linewidth} 
        \centering
        \includegraphics[width=\linewidth, height=0.87\linewidth]{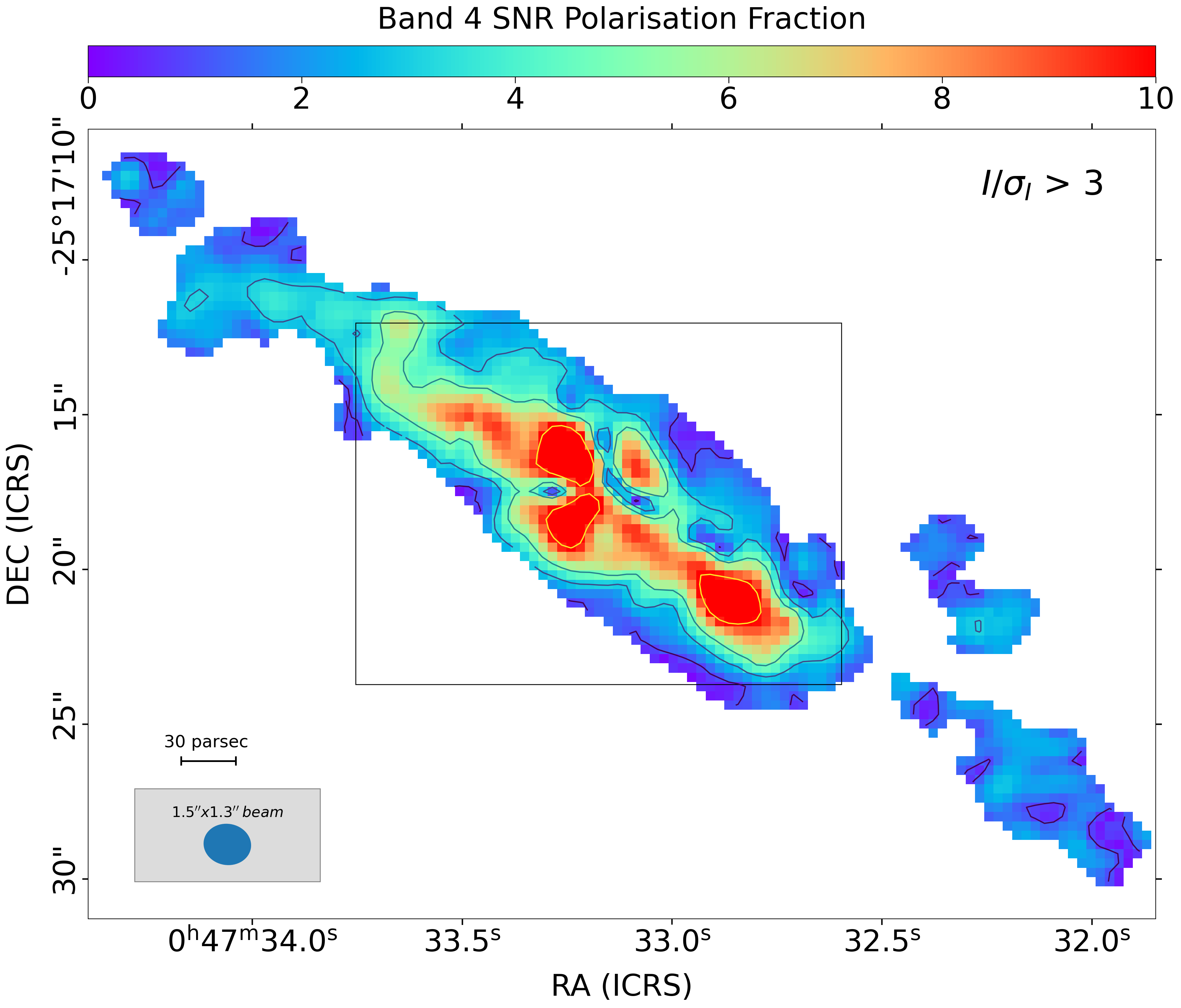} 
    \end{subfigure}\hfil
    \begin{subfigure}{0.48\linewidth} 
        \centering
        \includegraphics[width=\linewidth, height=0.87\linewidth]{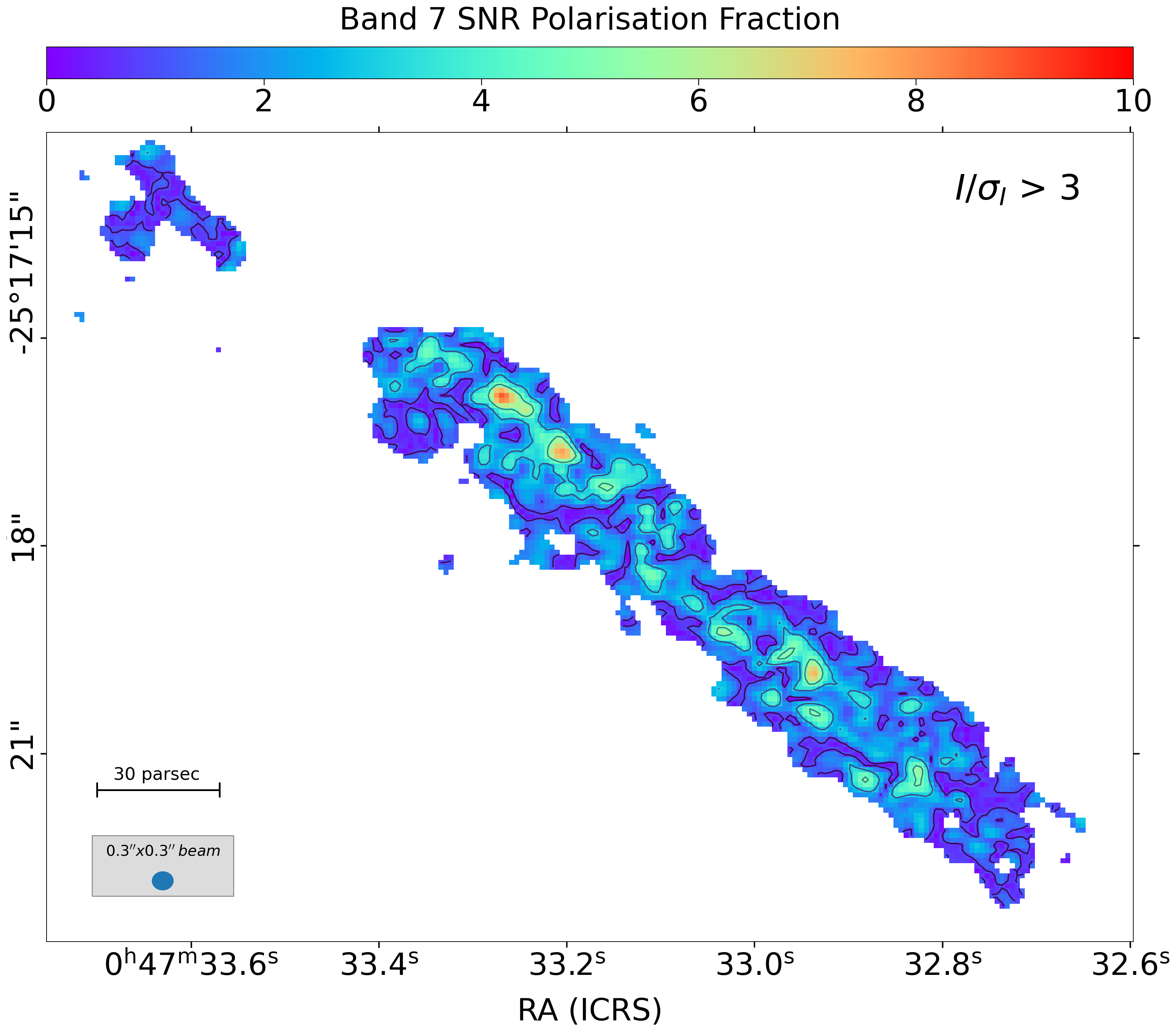} 
    \end{subfigure}
    
\medskip

    \begin{subfigure}{0.37\linewidth} 
        \centering
        \includegraphics[width=\linewidth, height=0.75\linewidth]{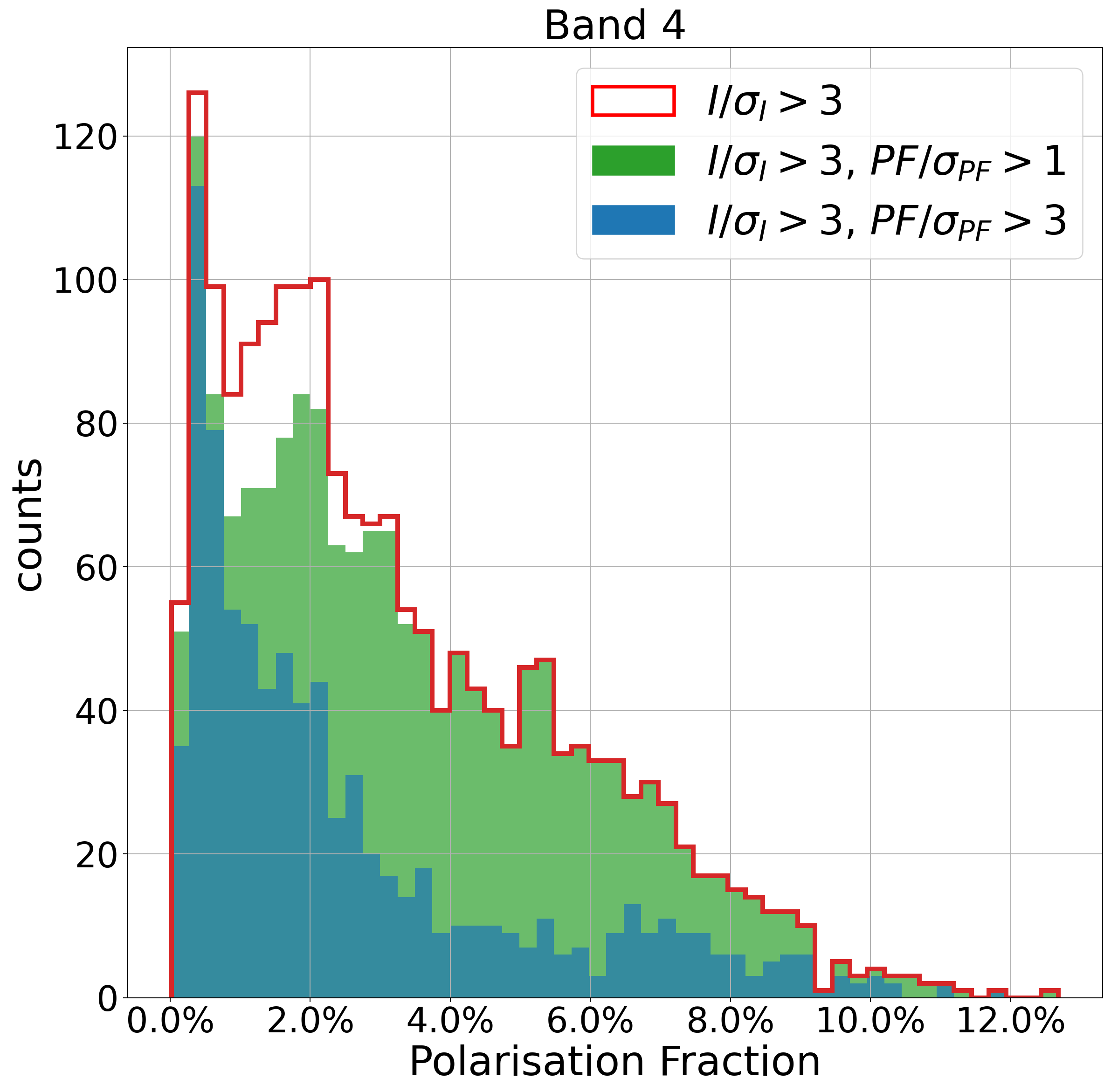} 
    \end{subfigure}\hfil
    \begin{subfigure}{0.37\linewidth} 
        \centering
        \includegraphics[width=\linewidth, height=0.75\linewidth]{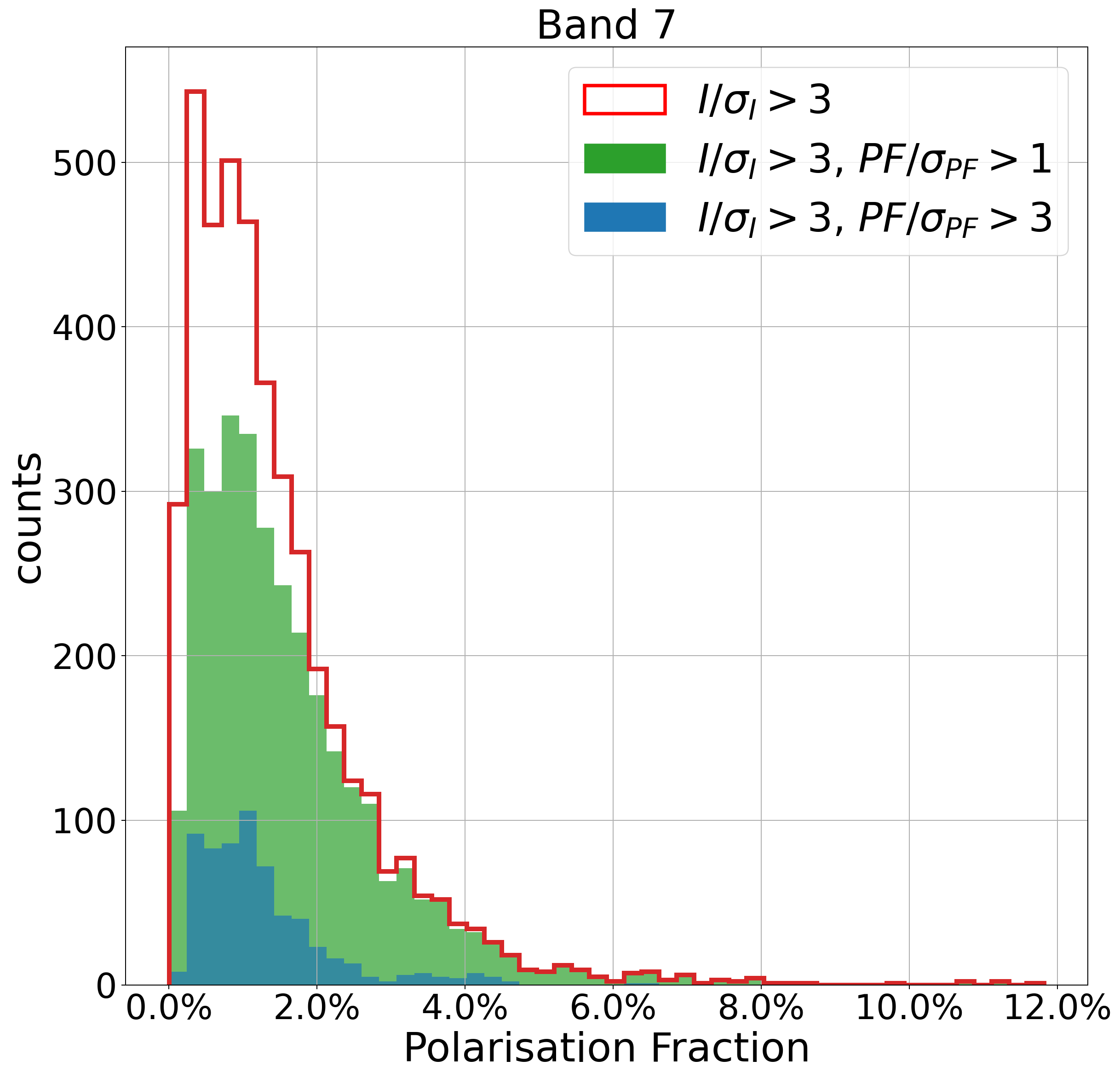} 
    \end{subfigure}
    
\caption{{\bf{Polarisation fraction in the centre of NGC253.}} {\it{Upper panels:}} Maps of the polarisation fraction. \rev{The vectors indicate the magnetic field orientation and are displayed only at pixels where $\polint/\sigma_{\polint} > 3$}. {\it{Centre panels:}} SNR maps of polarisation fraction with contours corresponding to $[1, 3, 5, 10] \times \polfrac/\sigma_{\polfrac}$.
White stars in the lower left panel indicate the positions of the SSCs catalogued by \citealp{Leroy2018}.
Both maps have been obtained selecting pixels with total intensity above 3 sigma.
{\it{Bottom panels:}} Histograms of the polarisation fraction pixel values in each map. 
The Band~4 and Band~7 results are shown in the left column and the right column, respectively. The plot annotations and pixel selection criteria are the same as in Figure~\ref{fig:largeP_analysis}.}
\label{fig:smallp_analysis}
\end{figure*}

\begin{figure*}[tbp]
    \centering
    \begin{subfigure}{0.48\linewidth} 
        \centering
        \includegraphics[width=\linewidth, height=0.87\linewidth]{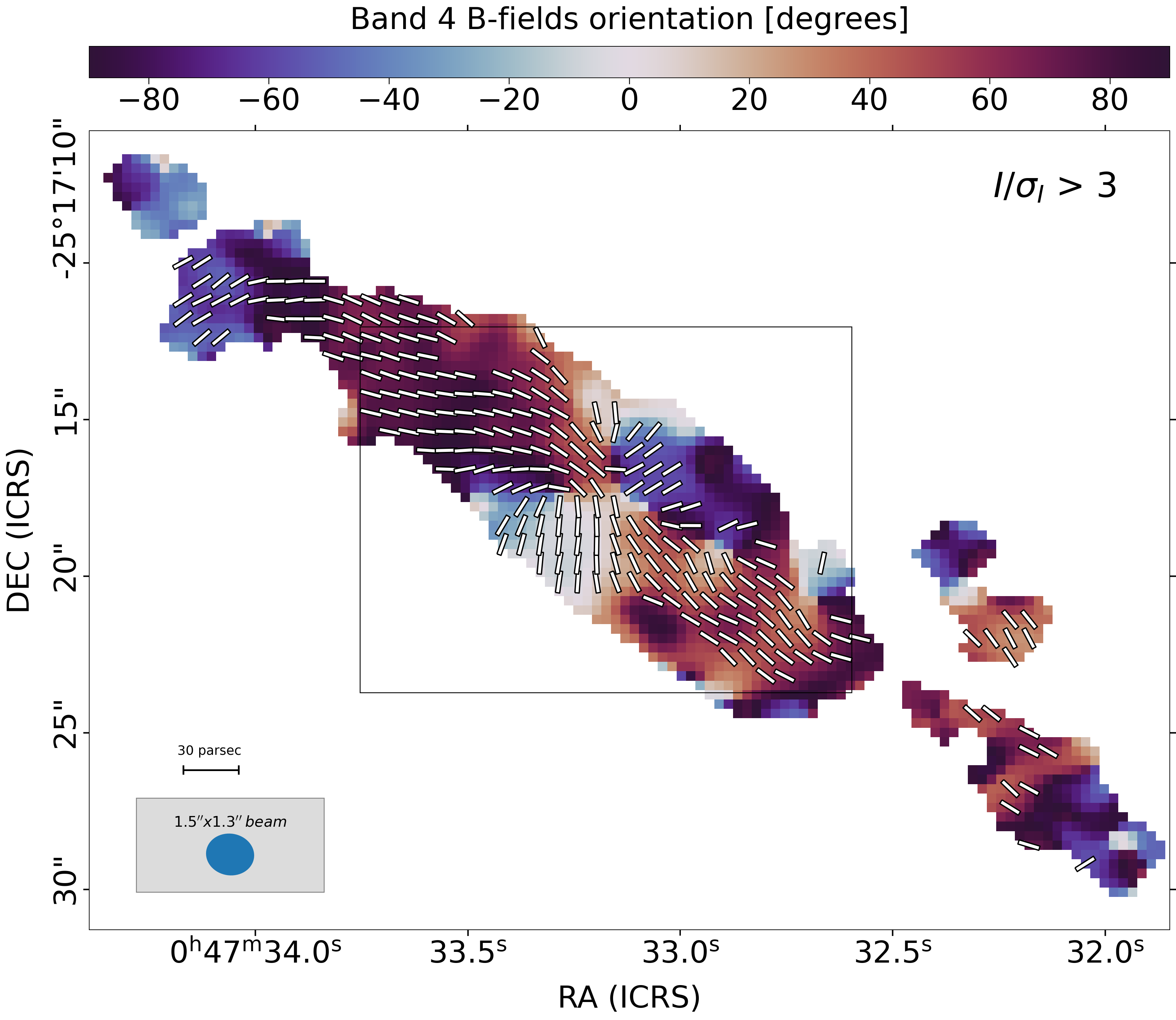} 
    \end{subfigure}\hfil
    \begin{subfigure}{0.48\linewidth} 
        \centering
        \includegraphics[width=\linewidth, height=0.87\linewidth]{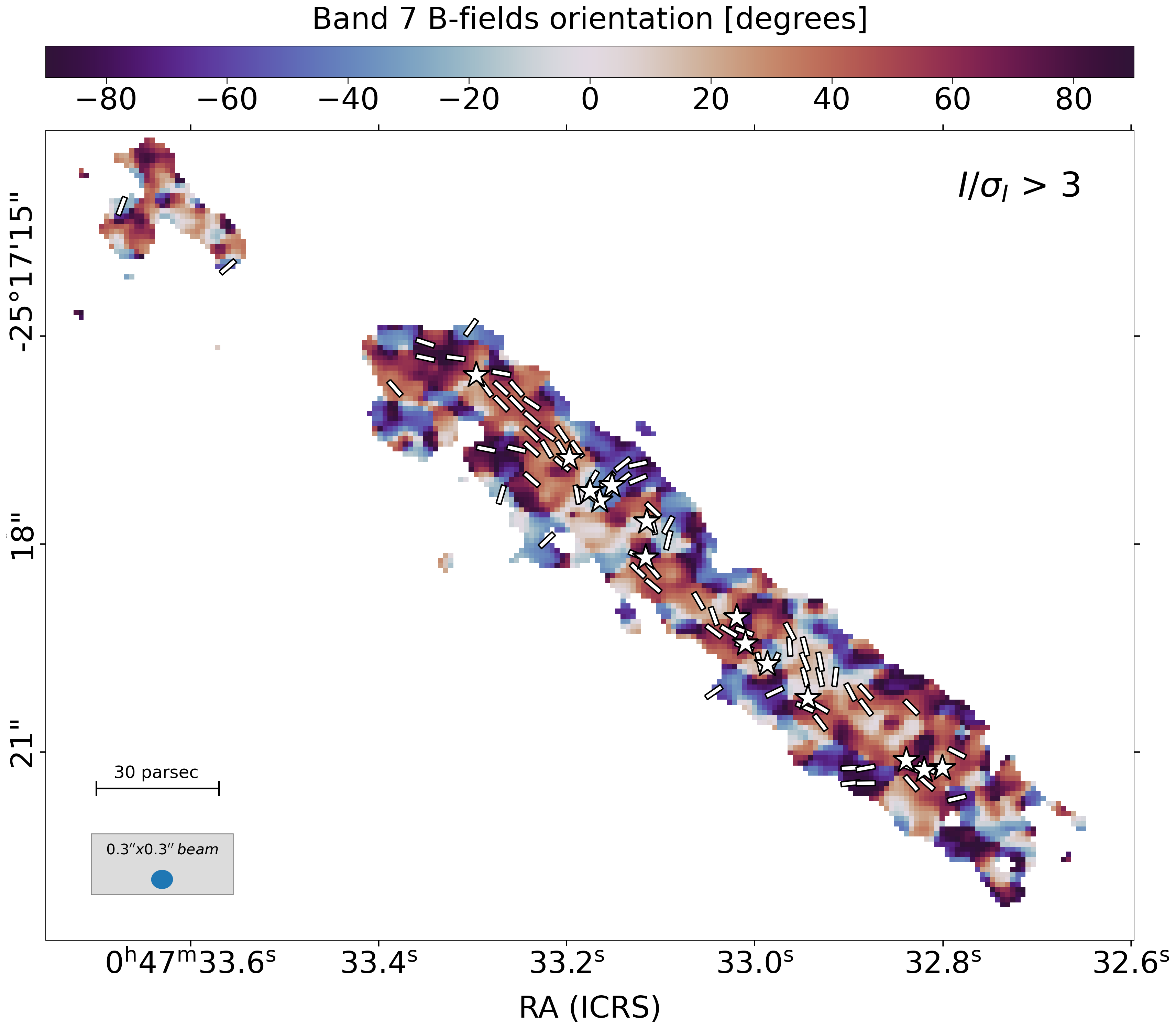} 
    \end{subfigure}

\medskip

    \begin{subfigure}{0.48\linewidth} 
        \centering
        \includegraphics[width=\linewidth, height=0.87\linewidth]{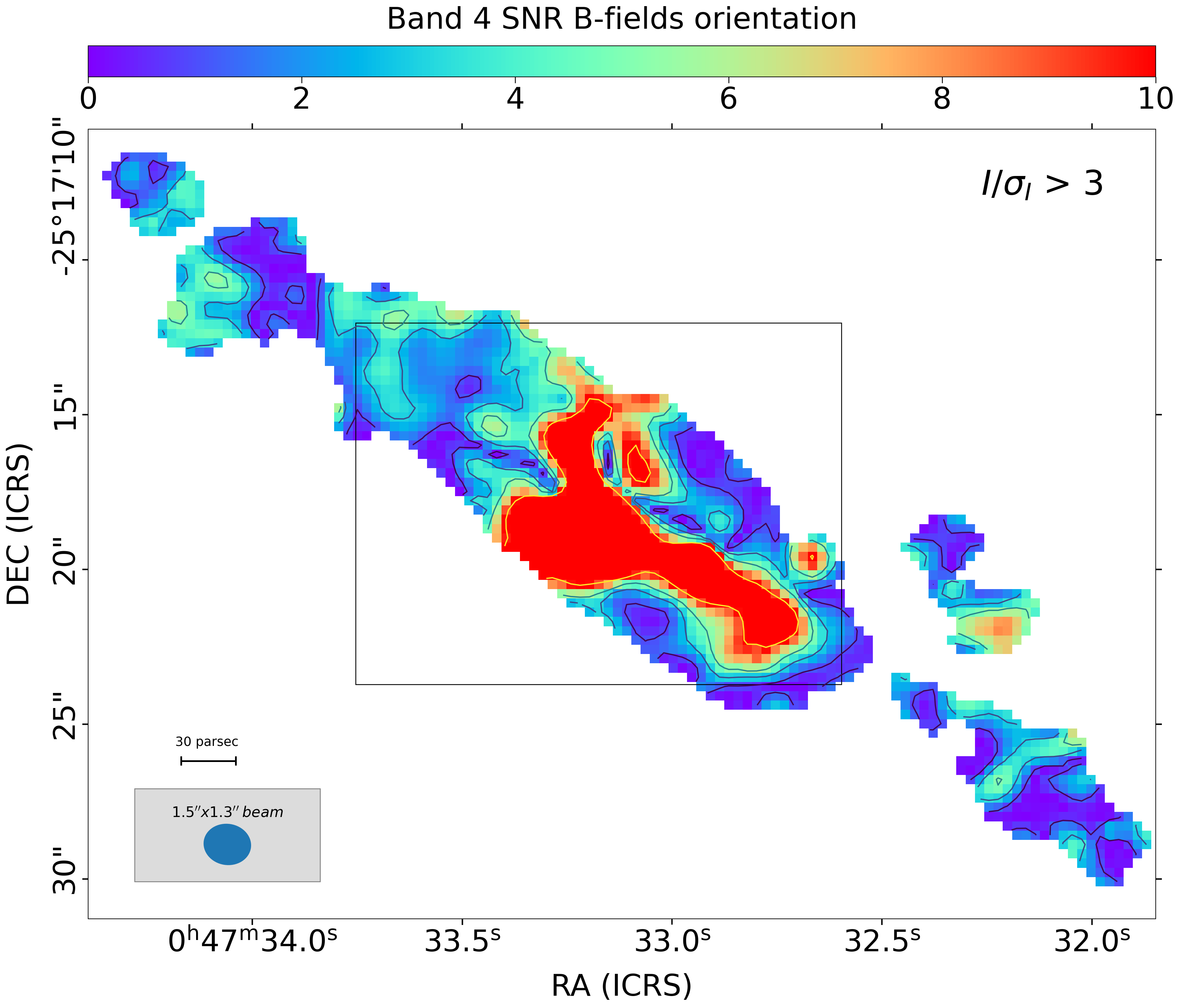} 
    \end{subfigure}\hfil
    \begin{subfigure}{0.48\linewidth} 
        \centering
        \includegraphics[width=\linewidth, height=0.87\linewidth]{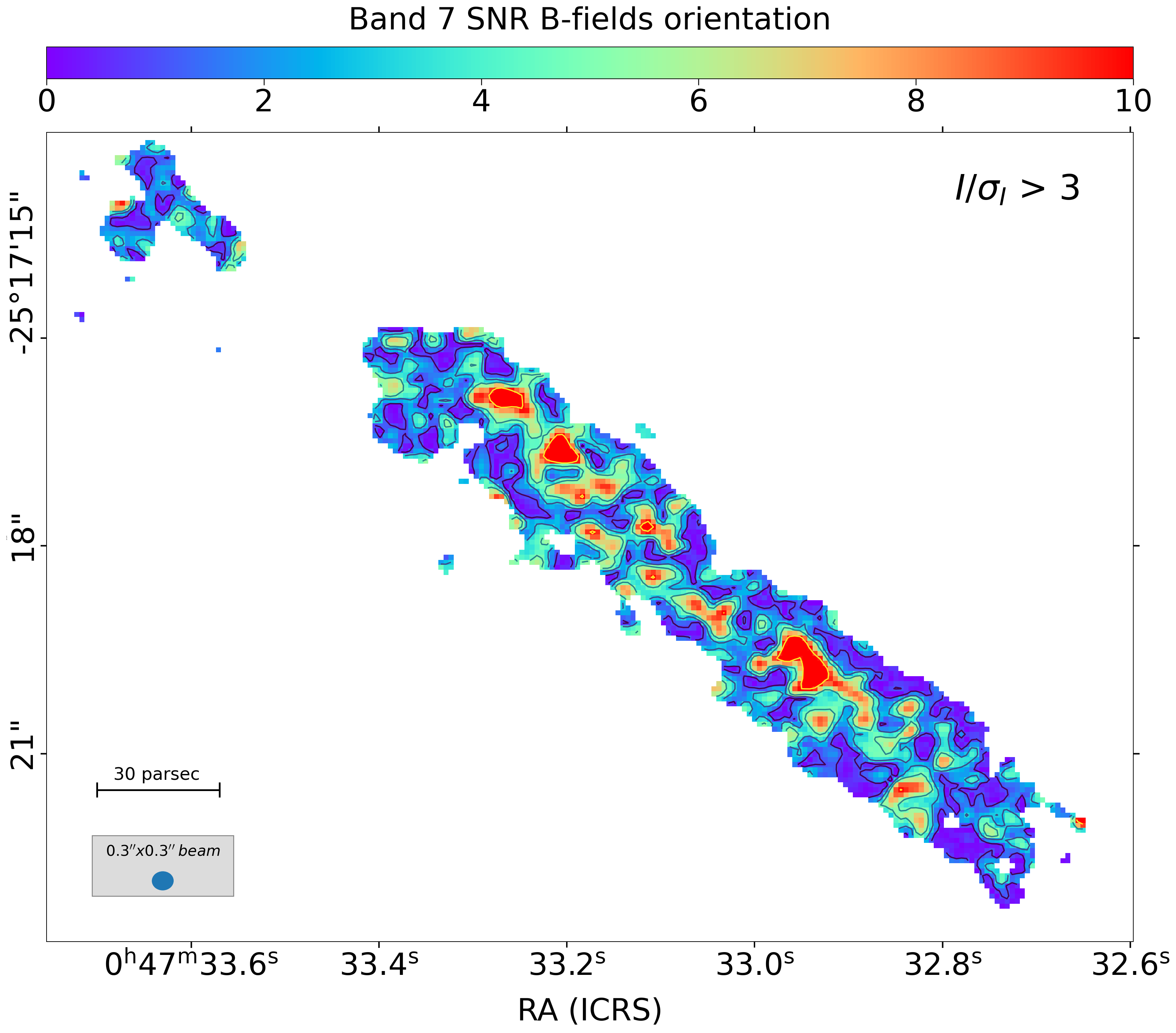} 
    \end{subfigure}
    
\medskip

    \begin{subfigure}{0.37\linewidth} 
        \centering
        \includegraphics[width=\linewidth, height=0.75\linewidth]{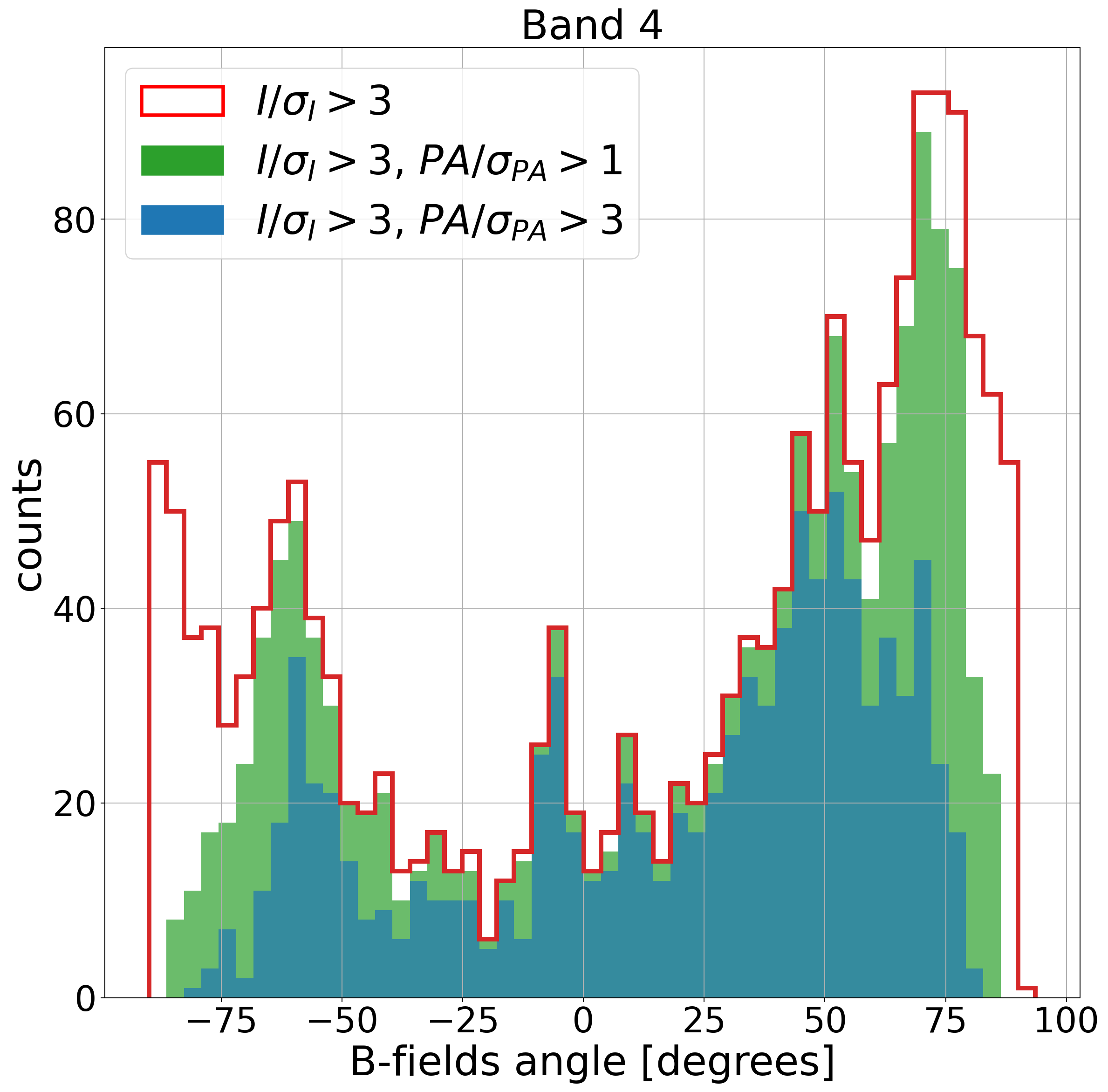} 
    \end{subfigure}\hfil
    \begin{subfigure}{0.37\linewidth} 
        \centering
        \includegraphics[width=\linewidth, height=0.75\linewidth]{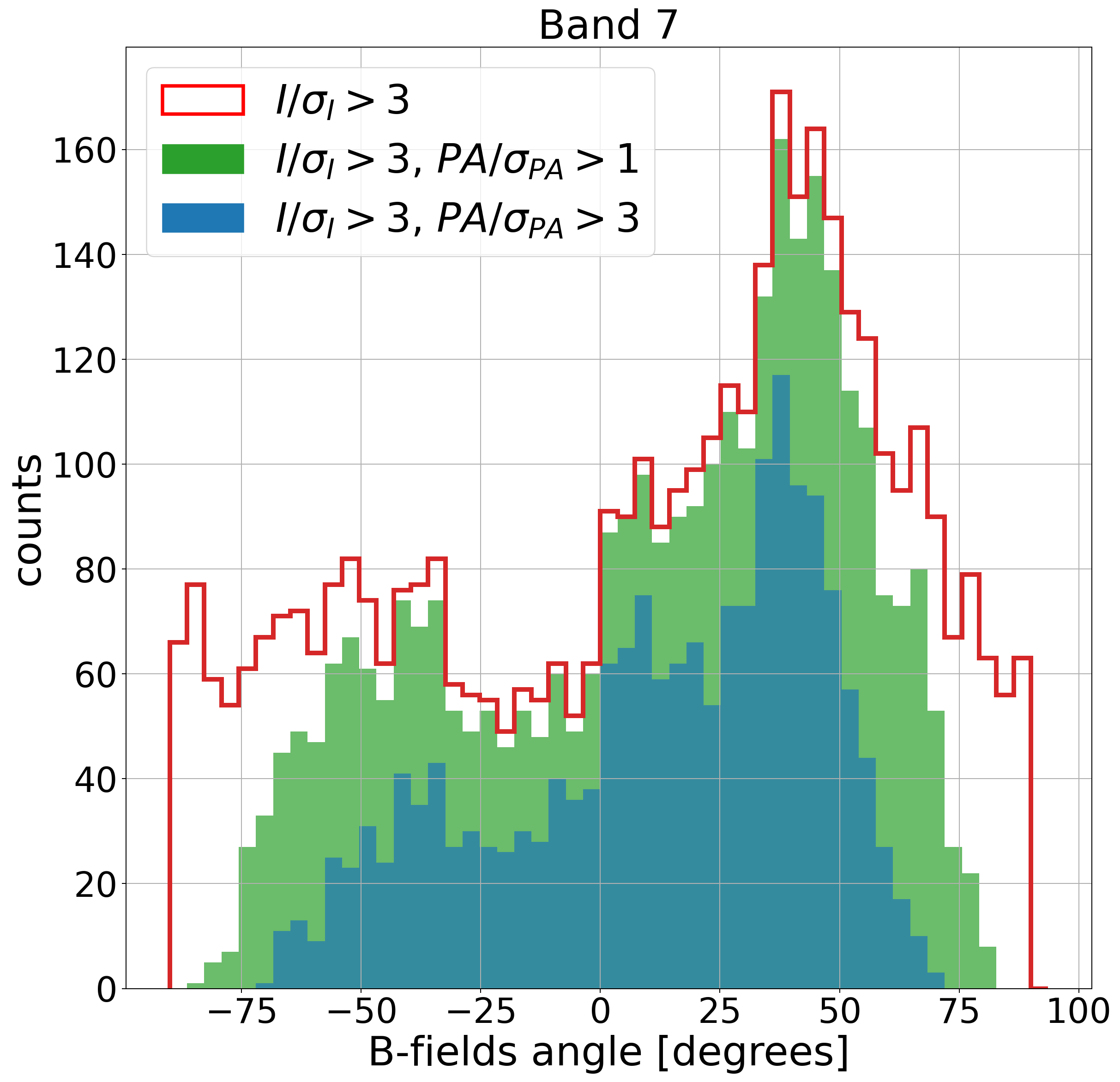} 
    \end{subfigure}

\caption{{\bf{Magnetic field orientation in the centre of NGC253.}} {\it{Upper panels:}} Maps and corresponding vectors of the magnetic field orientation, obtained rotating the observed polarisation angle by 90~degrees. \rev{The vectors indicate the magnetic field orientation and are displayed only at pixels where $\polint/\sigma_{\polint} > 3$}. {\it{Centre panels:}} SNR maps of the B-field orientation with contours corresponding to $[1, 3, 5, 10] \times PA/\sigma_{PA}$. White stars in the lower left panel indicate the positions of the SSCs catalogued by \citealp{Leroy2018}. 
Both maps have been obtained selecting pixels with total intensity above 3 sigma.
{\it{Bottom panels:}} Histograms of the B-field orientation pixel values in each map. 
Results from Band 4 and Band 7 observations are shown in the top row and the bottom row, respectively. The plot annotations and pixel selection criteria are the same as in Figure~\ref{fig:largeP_analysis}.} 
\label{fig:B-fields_orientation_analysis}
\end{figure*}

\begin{figure*}[tbp]
    \centering
    \begin{subfigure}{0.48\linewidth} 
        \centering
        \includegraphics[width=\linewidth, height=0.88\linewidth]{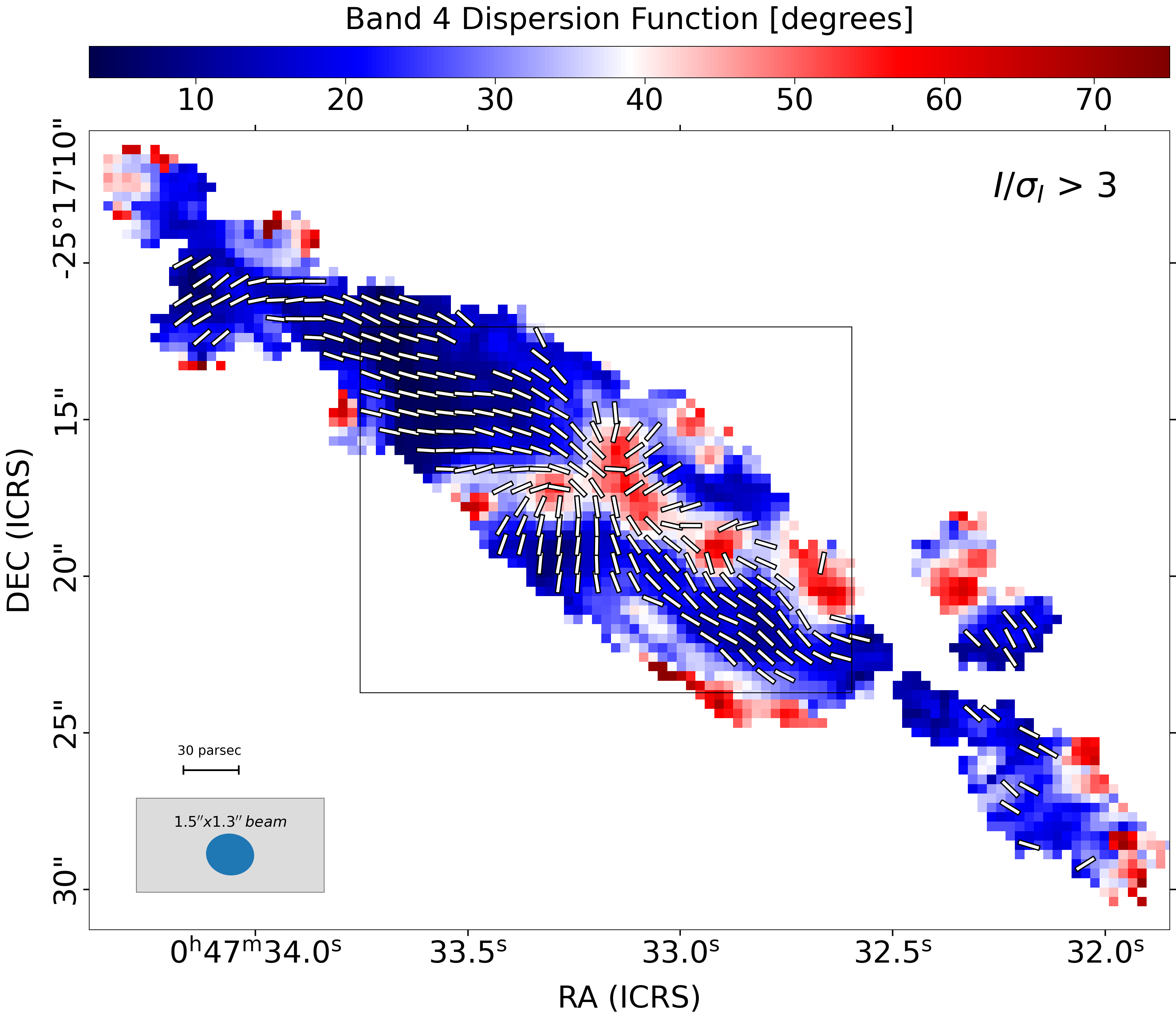} 
    \end{subfigure}\hfil
    \begin{subfigure}{0.48\linewidth} 
        \centering
        \includegraphics[width=\linewidth, height=0.88\linewidth]{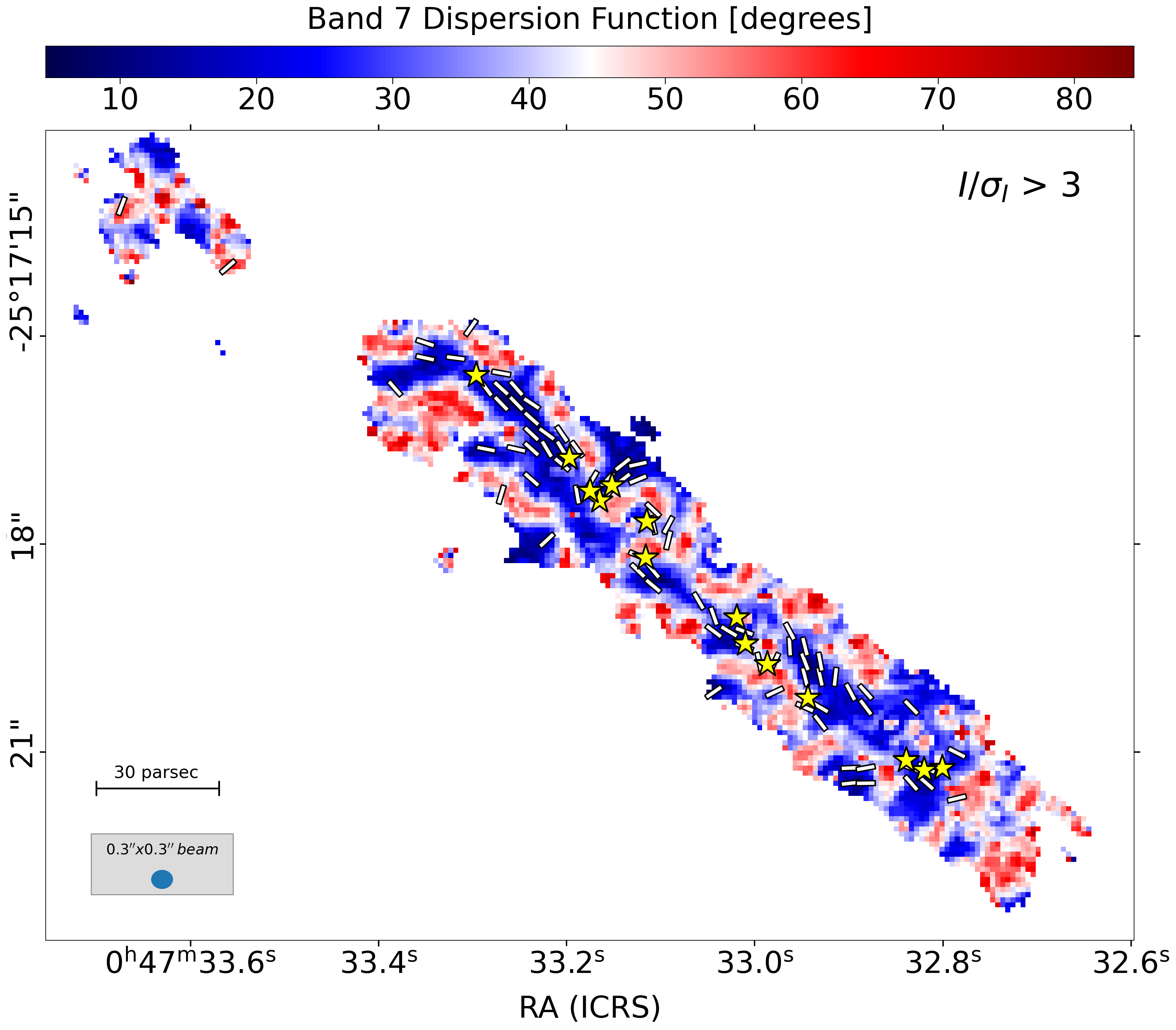} 
    \end{subfigure}
  \caption{{\bf{Polarisation angle dispersion function in the centre of NGC253.}} Maps of the polarisation angle dispersion function obtained using the Band~4 (left) and Band~7 (right) data. Only pixels with $I/\sigma_I > 3$ in total intensity are shown. Vectors indicating the orientation of the B-field vectors are overlaid \rev{only in correspondence of pixels where $\polint/\sigma_{\polint} > 3$}. Yellow stars indicate the positions of the SSCs catalogued by (\citealp{Leroy2018}}
  \label{fig:dispersion_function}
\end{figure*}

\subsection{Ancillary data} \label{ancillary_data}

\noindent For the SED analysis presented in \newrev{Section~\ref{analysis}}, we assembled archival data covering the frequency range from 1.4~GHz ($\lambda \sim $20~cm) to 690~GHz ($\lambda \sim$ 0.4~mm) taken with ALMA and the Very Large Array (VLA). The high frequency data are taken from the ALMA  archive, and have resolutions ranging from $\sim$1 to 1.7$\arcsec$. The low frequency data are taken from VLA archive, and the image at the lowest frequency (1.4~GHz) has a resolution of 2.2$\arcsec$. The data at all frequencies was convolved to this resolution for the SED analysis. The original resolution of each image is reported in Table~\ref{tab:ancillary}, as well as the largest angular scale recovered, and the corresponding project code.

The images at 22 and 33~GHz used in the SED analysis in Section~\ref{analysis} were presented by \citet{Gorski2017, Gorski2019}. 
The resolutions of these datasets are $0.204\arcsec \times 0.121\arcsec$ and $0.096 \arcsec \times 0.045 \arcsec$, with maximum recoverable scales of $2.4\arcsec$ and $1.6\arcsec$ at 22 GHz and 33~GHz, respectively.

Finally, for the comparison of the magnetic field orientation with molecular gas distribution, we used archival ALMA observations of the CO(3-2) emission in NGC253. Of the available data, presented in \cite{Leroy2018}, we selected those obtained with a compact configuration providing a resolution of $0.3 \times 0.4$~arcsec, very similar to that of our full polarisation data in Band~7.

\section{ALMA polarisation images} \label{polimages}

Our final images of the polarised intensity and polarisation fraction at both ALMA bands at their native resolution are shown in the upper panels of Figures~\ref{fig:largeP_analysis} and \ref{fig:smallp_analysis}, respectively. The upper panels of Figure~\ref{fig:B-fields_orientation_analysis} show the orientation of the B-field on the plane of the sky, computed by rotating the polarisation angle (PA) by $90^{\circ}$. For each polarisation quantity, we also show the corresponding SNR map, computed using the uncertainty $\sigma$ defined in Section~\ref{Noise}. 
In the lower panel of each Figure, red empty histograms indicate the distribution of pixels with SNR in total intensity above 3. The green and blue solid histograms represent the distribution of pixels after applying an additional threshold on the corresponding polarisation quantity of 1-sigma and 3-sigma, respectively.

\subsection{Polarised intensity and polarisation fraction}    

\noindent Figure~\ref{fig:largeP_analysis} shows the polarised intensity of the galaxy at Band~4 and Band~7, for all map pixels with a $I/\sigma_I > 3$. The white vectors represent the orientation of the magnetic field on the plane of the sky. \rev{For clarity, one vector is shown every two pixels (for Band 4 maps) and three pixels (for Band 7 maps).}

In Band~4, the median value of the polarised intensity is $\approxeq 0.96 \: MJy/sr$ with a standard deviation of  $\approxeq 0.51 \: MJy/sr$, measured over a region where $I/\sigma_I > 3$ and $\polint/\sigma_{\polint} > 3$. The $PI$ map shows local maxima of $PI\sim 2.5 \: \rm MJy/sr$ from four elongated bright spots situated to the North, South, East and West of the galaxy centre (from here after referred to as $PI_N$, $PI_S$, $PI_E$ and $PI_W$). These maxima (\rev{identified} in Figure \ref{fig:regions}) do not have counterparts in total intensity.

The Band~7 observations access smaller physical scales within the starburst region. At this band, the polarised intensity distribution is more uniform, with a median of $\approxeq 40.28 \: \rm MJy/sr$ and standard deviation of $\approxeq 13.22 \: \rm MJy/sr$ (again considering only pixels with $I/\sigma_I > 3$ and $\polint/\sigma_{\polint} > 3$). The North and South elongated bright spots visible in Band~4 are resolved into several smaller local maxima with $PI \sim 50 \: \rm MJy/sr$. Particularly noteworthy are three of these maxima, two to the north and one to the south, which achieve values surpassing $80 \: \rm MJy/sr$. As for the Band~4 data, these local maxima do not clearly correspond to structures observed in total intensity.

\noindent Figure~\ref{fig:smallp_analysis} shows the PF maps measured in Band~4 and Band~7, superimposed with vectors that represent the orientation of the magnetic field. As in Figure~\ref{fig:largeP_analysis}, signal-to-noise maps and histograms of the PF pixel values are also shown. The PF distribution measured in Band~4 with $I/\sigma_I > 3$ and $\polfrac/\sigma_{\polfrac} > 3$ has a median value of $\overline{PF} \approxeq 1.69$\% and standard deviation of $\approxeq 2.47$\%. Low PF values are concentrated in the innermost region (where the elongated bright spots in PI are located). This inner low PF area (with a median value of around 0.5\%), coincides roughly with the region characterized by $I/\sigma_I>3$ in Band~7.

The higher resolution of the Band~7 data allows us to study the distribution of PF in this region at spatial scales of $\sim5$\,pc. The median computed with $I/\sigma_I > 3$ and $\polfrac/\sigma_{\polfrac} > 3$ is  1.06\% with standard deviation of 0.90\%. The map is not uniform, with several local minima apparently aligned along the galaxy midplane in the NE-SW direction.

\rev{In regions where $I/\sigma_I > 3$ and $\polfrac/\sigma_{\polfrac} > 3$ the minimum values of $\polfrac$ are 0.15\% in Band 7 and 0.09\% in Band 4. These values are comparable to the nominal ALMA minimum detectable degree of linear polarisation of 0.1\% \footnote{\url{https://almascience.eso.org/documents-and-tools/cycle11/alma-technical-handbook}}}.

\subsection{Magnetic field orientation}


In Figure~\ref{fig:B-fields_orientation_analysis}, we show maps of the B-field orientation as projected in the plane of the sky inferred from the Stokes Q and U of Band~4 and Band~7 images. The B-field orientation is also shown as white vectors overlaid over the PI, and PF 
maps in Figures~\ref{fig:largeP_analysis}, and  \ref{fig:smallp_analysis} \rev{in correspondence of pixels where $\polint/\sigma_{\polint} > 3$}.

The B-field structure probed by Band~4 observations is characterized by two main ordered components, one parallel to the mid-plane and one perpendicular to it. 
The two components meet in the central region of the galaxy, where the total intensity distribution observed in Band~4 peaks.
A similar configuration has been seen in other edge-on galaxies at larger scales \citep[$\sim$ 0.1-1 kpc, ][]{Krause2020, Krause2006}. The B-field component parallel to the disk may be driven by the large-scale galactic B-field, while the perpendicular component is likely driven by the galactic outflow pushed away by the starbursts activity.

The North and South maxima in the PI map align with areas where the magnetic field is perpendicular to the midplane. We discuss these regions in more detail in Section~\ref{analysis}.

The orientation of the B-field inferred from the Band~7 observations appears more chaotic than at Band~4. 
The histogram of inferred B-field angles nonetheless indicates two main components that are parallel and perpendicular to the galactic midplane. This is evident in the red histogram ($I/\sigma_I>3$) in the bottom right panel of Figure~\ref{fig:B-fields_orientation_analysis}, where a prominent B-field component with angle $\sim 50^{\circ}$ can be seen, as well as a broader peak with angles ranging from $-80^{\circ}$ to $-30^{\circ}$. This perpendicular component becomes a more obvious peak $\sim -30^{\circ}$ when we restrict our comparison to pixels with $I/\sigma_I > 3$ and $PA/\sigma_{PA} > 3$ (blue histogram in the right panel of Figure~\ref{fig:B-fields_orientation_analysis}).

We note that the B-field orientation that we infer in NGC253 differs from the Central Molecular Zone in our Galaxy. Both the Planck data at 850~GHz and the PILOT balloon-borne data at $240 \: \mu m$ \citep[][]{Mangilli2019} indicate a homogeneous magnetic field orientation that is tilted with respect to the Galactic plane by $\simeq 22^\circ$. This difference could be explained by the different viewing angle or by the vastly different star formation efficiency which characterize the central regions of NGC253 and the Milky Way.
Signs of a parallel and perpendicular B-field component has been observed at $214 \mu m$ by the FIREPLACE survey \rev{\citep{Butterfield2024, pare2024}}.

\subsection{Polarisation Angle Dispersion Function} \label{dispersion}

From our polarisation data, we also constructed a map of the polarisation angle dispersion function $\DeltaAng$, as defined in \cite{Kobulnicky1994} and  \cite{Hildebrand2009}.
\rev{The angle dispersion function is computed as the median angular difference between a central pixel and the surrounding ones. The median is derived over an annulus of radius $\mathbf{l}$ (called \newrev{\textquotesingle lag\textquotesingle} ) and width $\mathbf{\delta l}$ around the central pixel.}


Regions where the magnetic field is well-ordered are characterized by $\DeltaAng \approxeq 0^{\circ}$, while regions with irregular and disordered B-fields will show progressively greater $\DeltaAng$ values. 
A fully random polarisation vector field will have $\DeltaAng \approxeq 52^{\circ}$ \citep{Alina2016}.

Maps of the polarisation angle dispersion function $\DeltaAng$ for our Band~4 and Band~7 ALMA maps are shown in Figure\,\ref{fig:dispersion_function}. We chose to use a lag \rev{$\mathbf{l}$} and a width value corresponding to half the major axis of the clean beam of the ALMA maps (i.e. $\mathbf{l}=0.15 \arcsec$ and $\mathbf{\delta l}=0.15 \arcsec$ for Band~7 and $\mathbf{l}=0.70 \arcsec$ and $\mathbf{\delta l}=0.70 \arcsec$ for Band~4).

\rev{We also computed the uncertainties of $\DeltaAng$ using Equation 9 from \citep{planckcollaborationXIX}}.

At Band~4, we identify three quasi-filamentary features with high values of $\DeltaAng$ ($\sim 50^{\circ}$, red regions in Figure~\ref{fig:dispersion_function}), radiating out from the centre of the starburst region (referred to as SC in Section~\ref{analysis}) where the perpendicular and parallel components of the polarisation vectors meet. These three regions of \rev{high values of $\DeltaAng$ (which we will refer to $\DeltaAng$-filaments)} separate more extended areas where the polarisation angle directions are much more uniform (blue regions in Figure~\ref{fig:dispersion_function}). 

At Band~7, the structure of $\DeltaAng$ is more complex, with an intricate web of \rev{$\DeltaAng$-filaments} across the region with robust polarisation detections. Similarly to the $\DeltaAng$ distribution observed at Band~4, the \rev{$\DeltaAng$-filaments} separate areas with distinct but homogeneous directions.

The complex network of \rev{$\DeltaAng$-filaments} seen at Band~7 is only partially recovered at Band~4, primarily in the form of a long high $\DeltaAng$ region roughly parallel to the major axis. Due to the lower resolution, most of the structure observed in Band~7 is not discernible at Band~4.

\begin{figure*}[tbp]
    \centering
    \begin{subfigure}{0.50\linewidth} 
        \centering
        \includegraphics[width=\linewidth, height=0.78\linewidth]{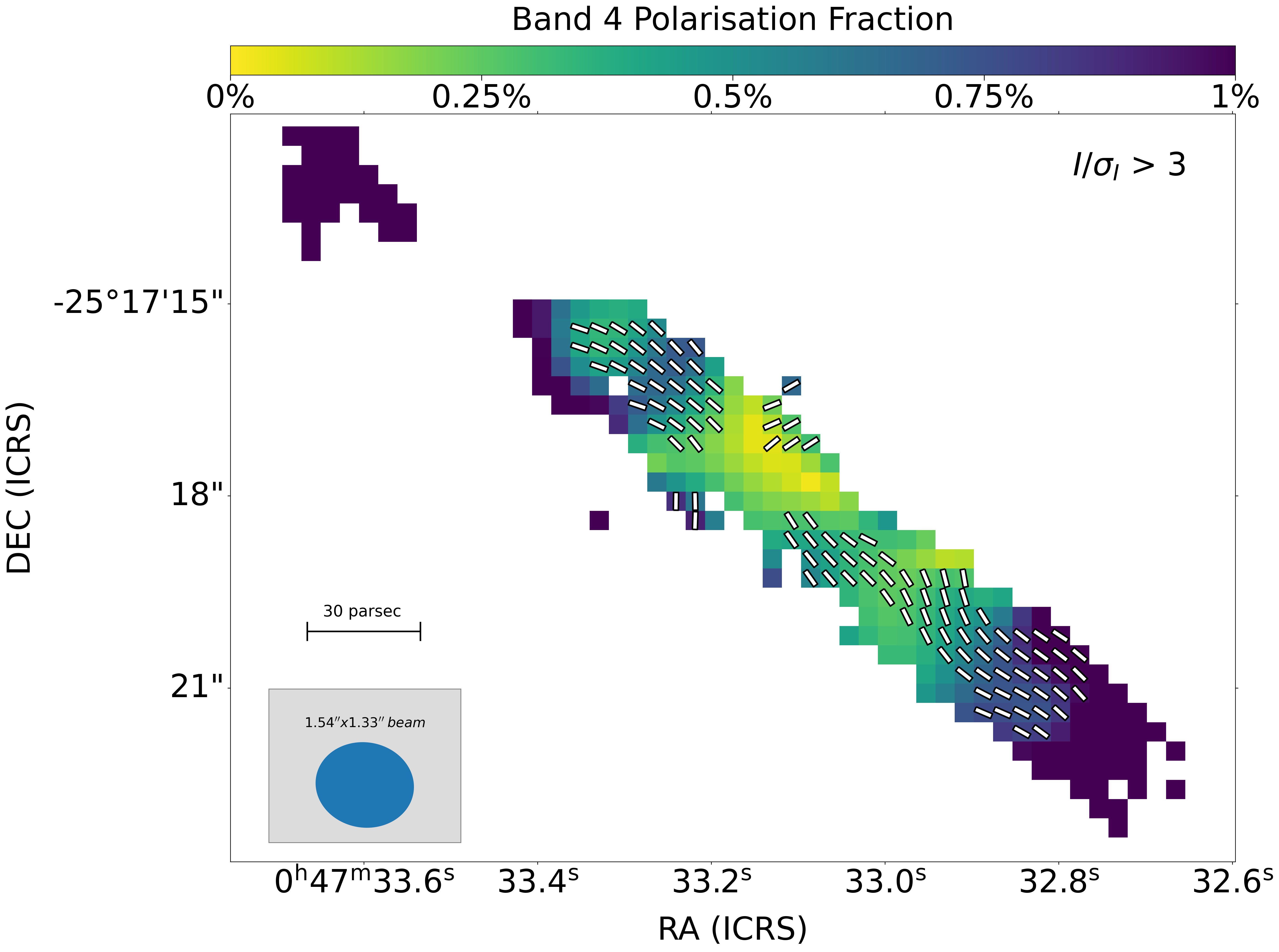} 
    \end{subfigure}\hfil
    \begin{subfigure}{0.50\linewidth} 
        \centering
        \includegraphics[width=\linewidth, height=0.78\linewidth]{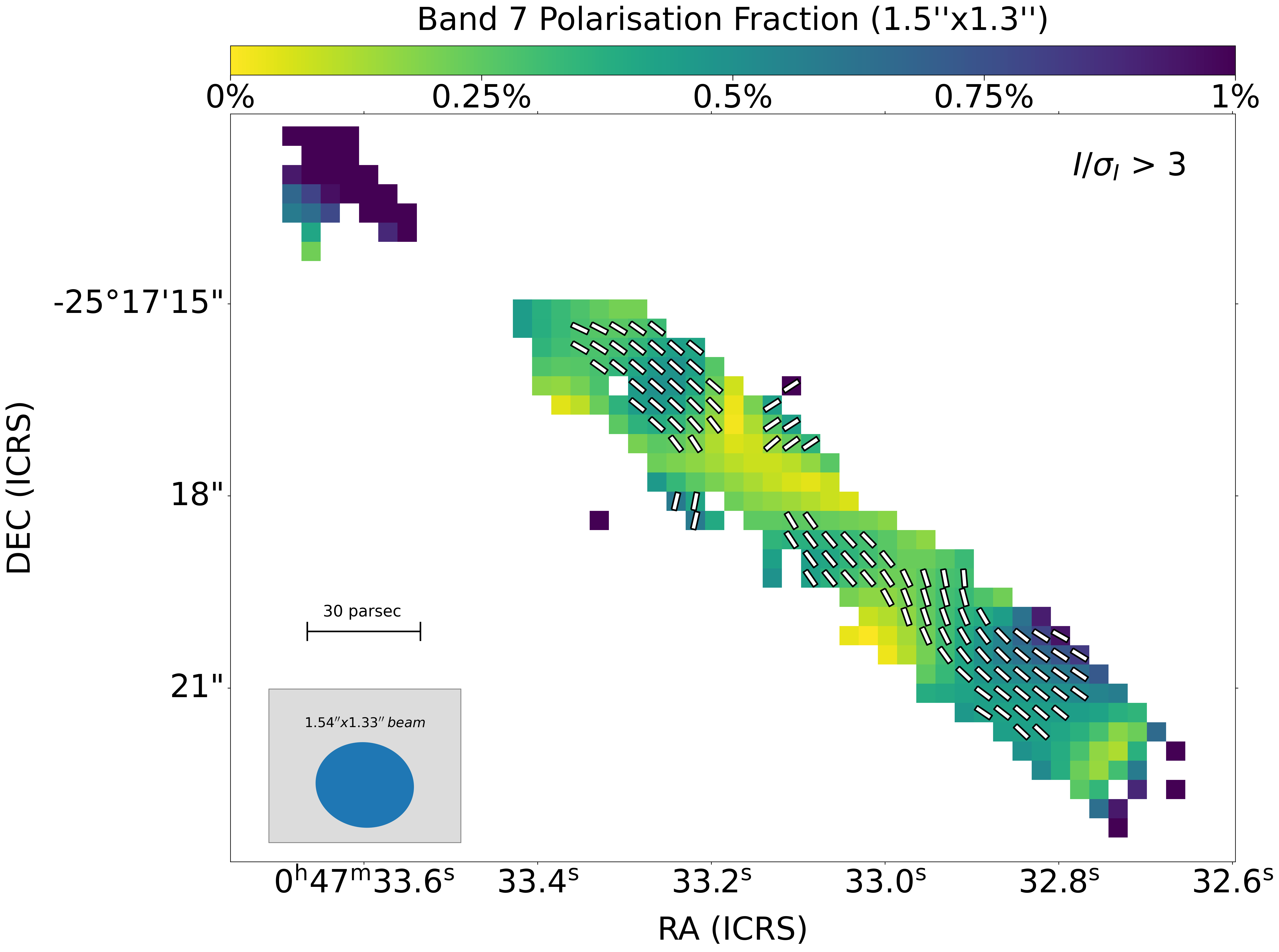} 
    \end{subfigure}

\medskip

    \begin{subfigure}{0.50\linewidth} 
        \centering
        \includegraphics[width=\linewidth, height=0.78\linewidth]{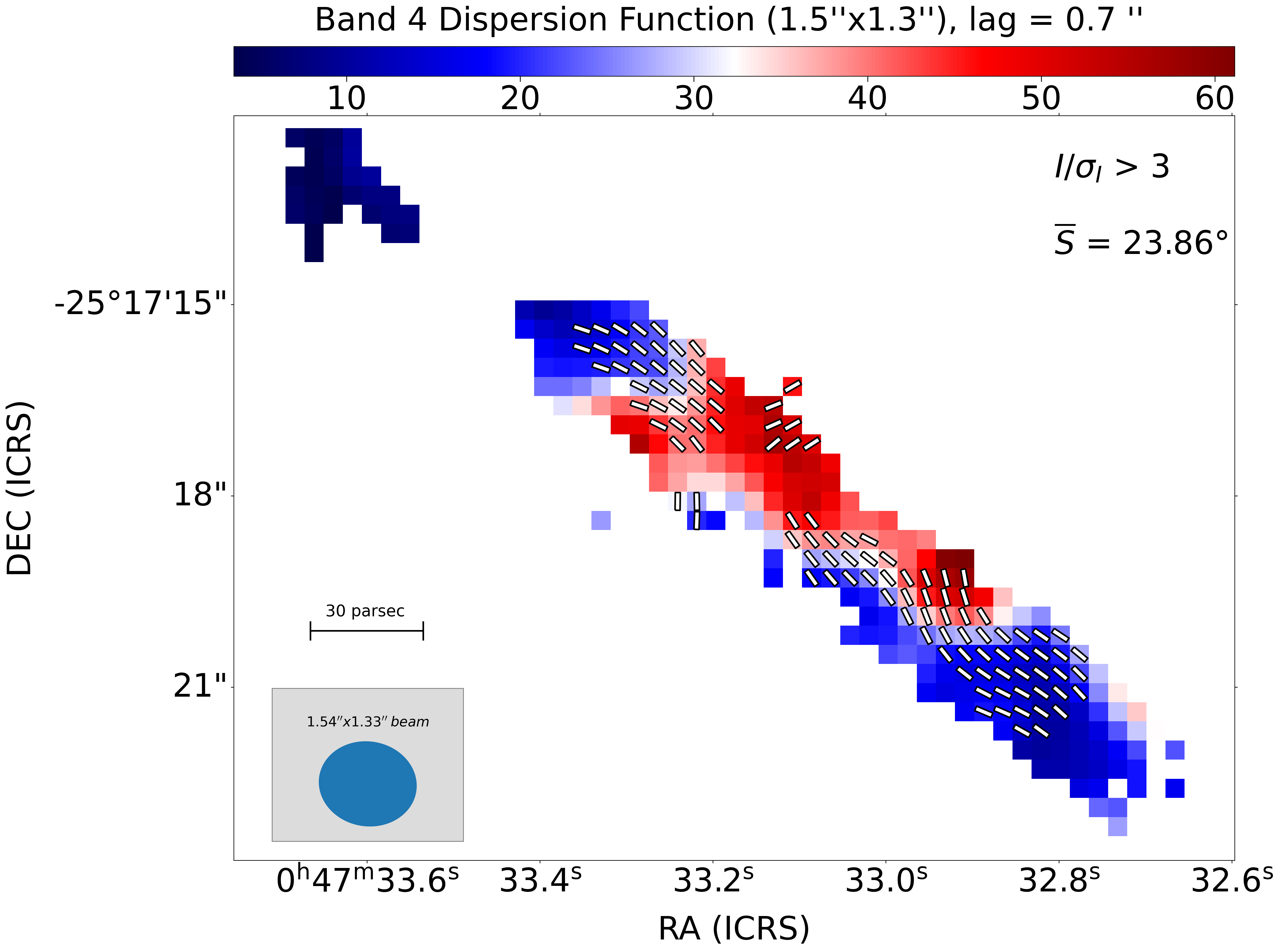} 
    \end{subfigure}\hfil
    \begin{subfigure}{0.50\linewidth} 
        \centering
        \includegraphics[width=\linewidth, height=0.78\linewidth]{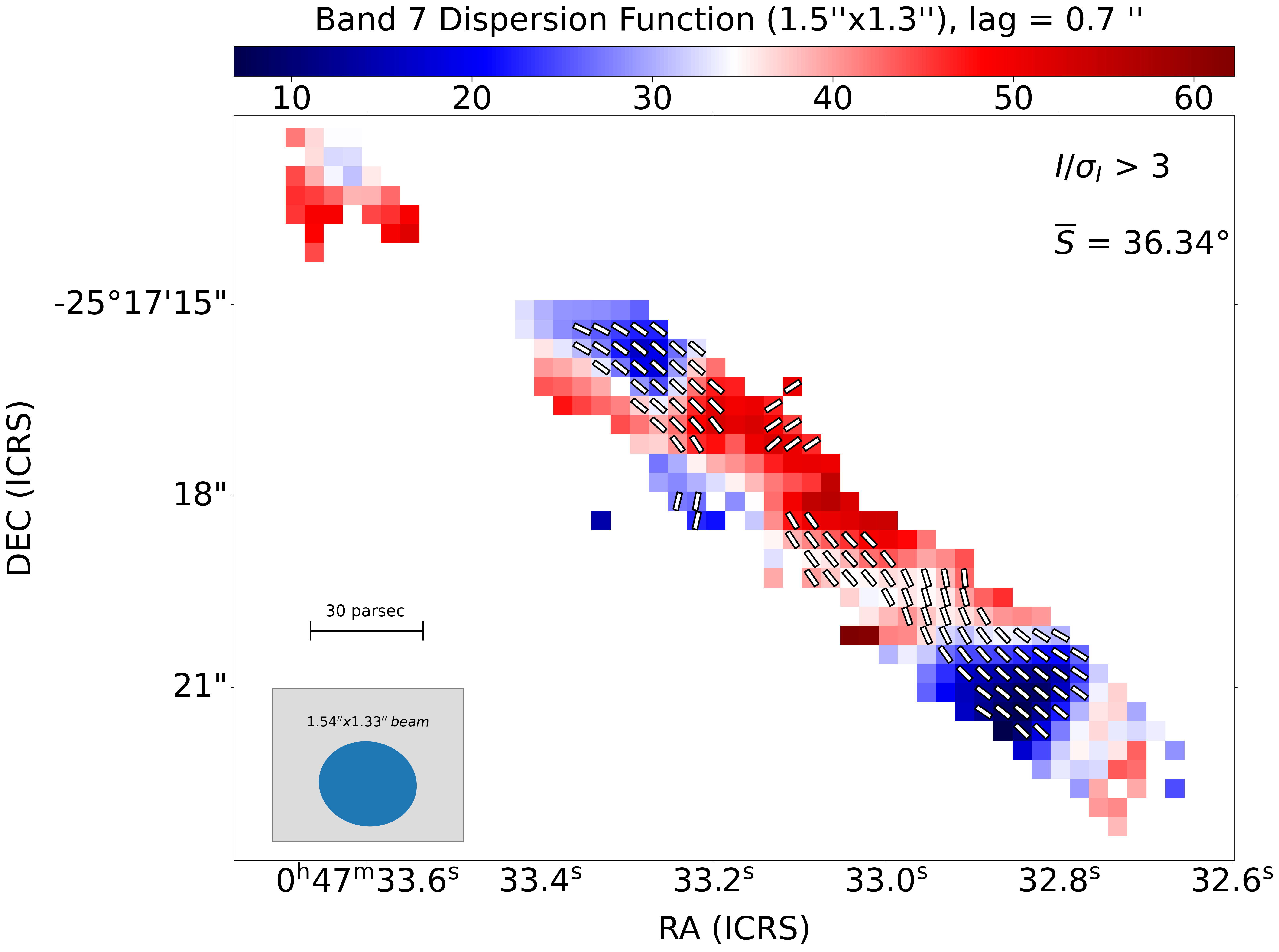} 
    \end{subfigure}
    
\medskip

    \begin{subfigure}{0.50\linewidth} 
        \centering
        \includegraphics[width=\linewidth, height=0.80\linewidth]{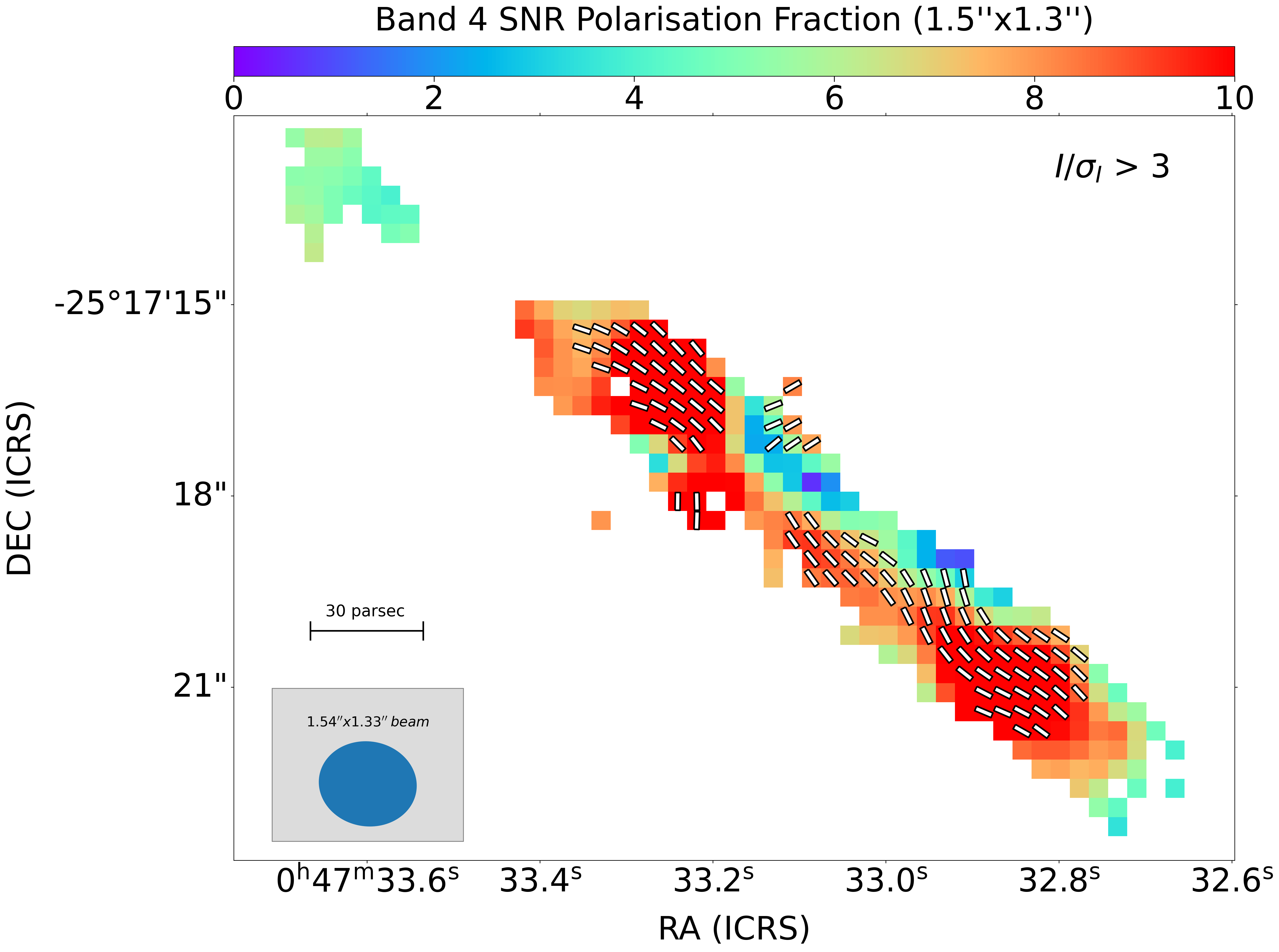} 
    \end{subfigure}\hfil
    \begin{subfigure}{0.50\linewidth} 
        \centering
        \includegraphics[width=\linewidth, height=0.80\linewidth]{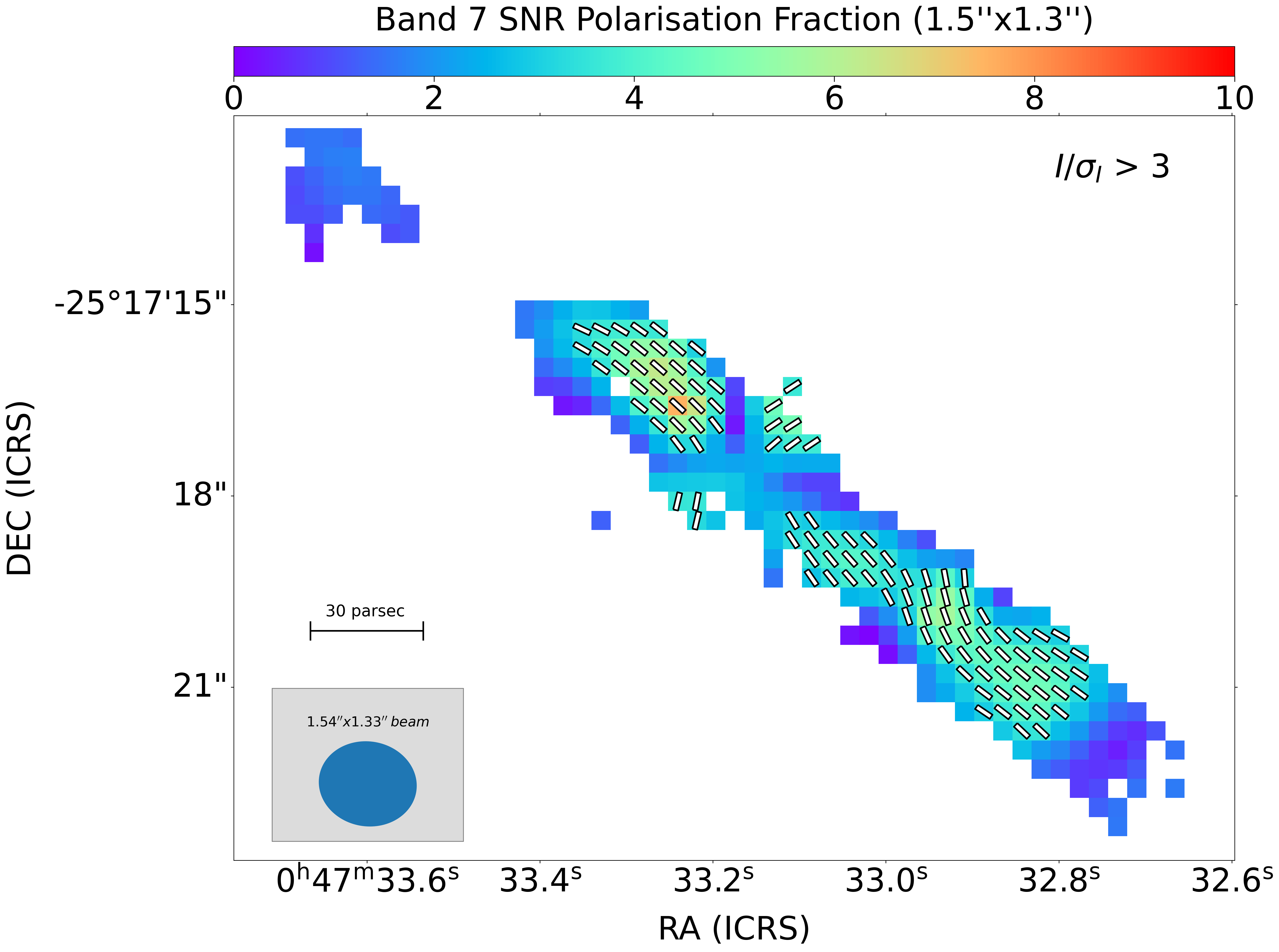} 
    \end{subfigure}
    \caption{{\bf{Polarisation properties in the centre of NGC253 measured at the same resolution.}} Band~4 (left) and Band~7 (right) polarisation fraction (up), $\DeltaAng$ (centre) and SNR of $\polfrac$ (bottom) maps at the common resolution of 1.5 $\times$ 1.3$\arcsec$.
    In all panels, the inferred B-field orientations are indicated with white vectors in correspondence of pixels where $\polint/\sigma_{\polint} > 3$ in Band 7. The images  only show the central starburst region, inside the mask constructed from Band~7 data at its native resolution that includes pixels with $I/\sigma_I > 3$.}
    \label{fig:images_sameres}
\end{figure*}

\begin{figure*}[tbp]
    \centering 
\begin{subfigure}{0.35\linewidth}
  \includegraphics[width=\linewidth, height=0.80\linewidth]{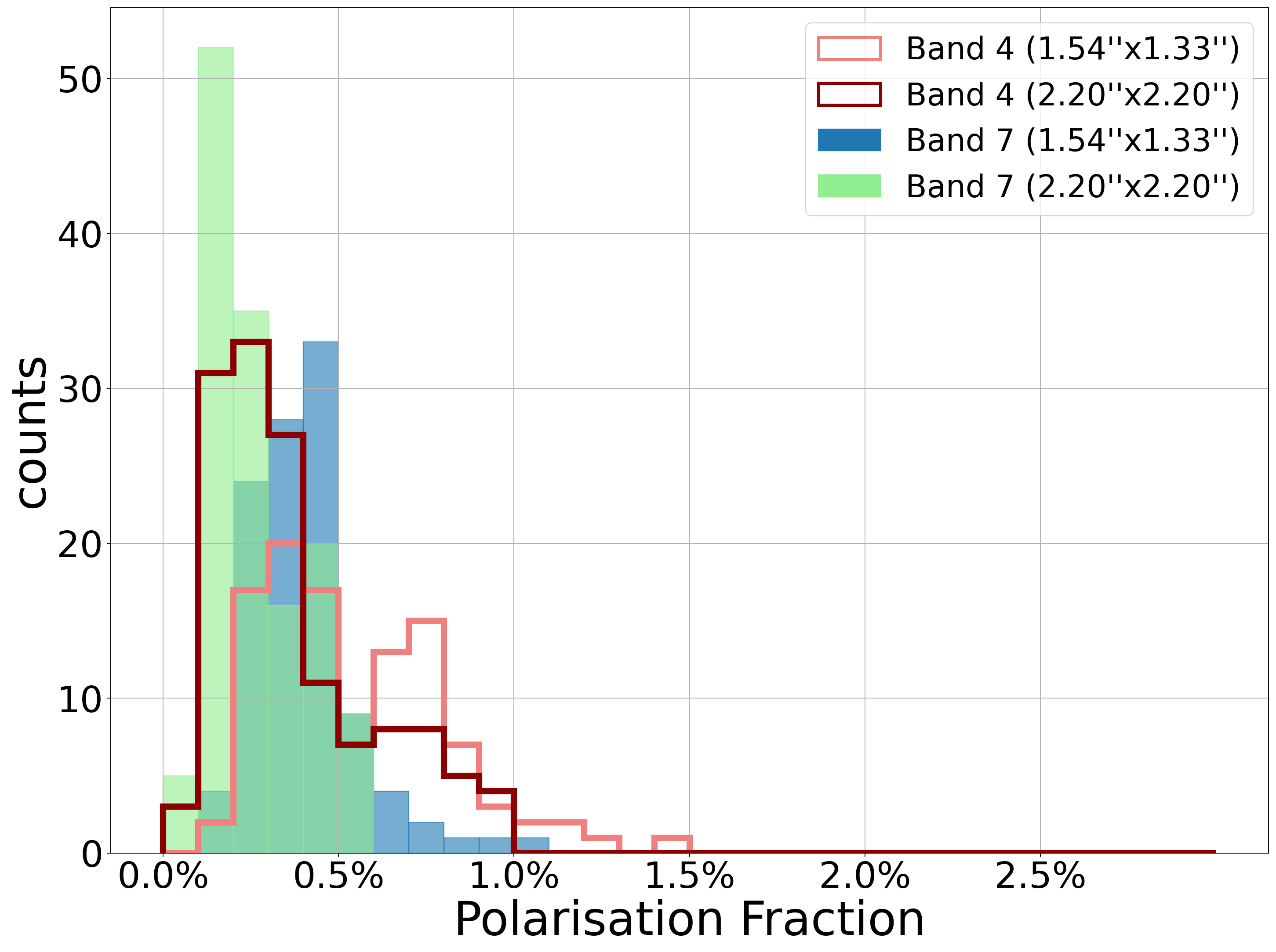}
\end{subfigure}\hfil 
\begin{subfigure}{0.35\linewidth}
  \includegraphics[width=\linewidth, height=0.80\linewidth]{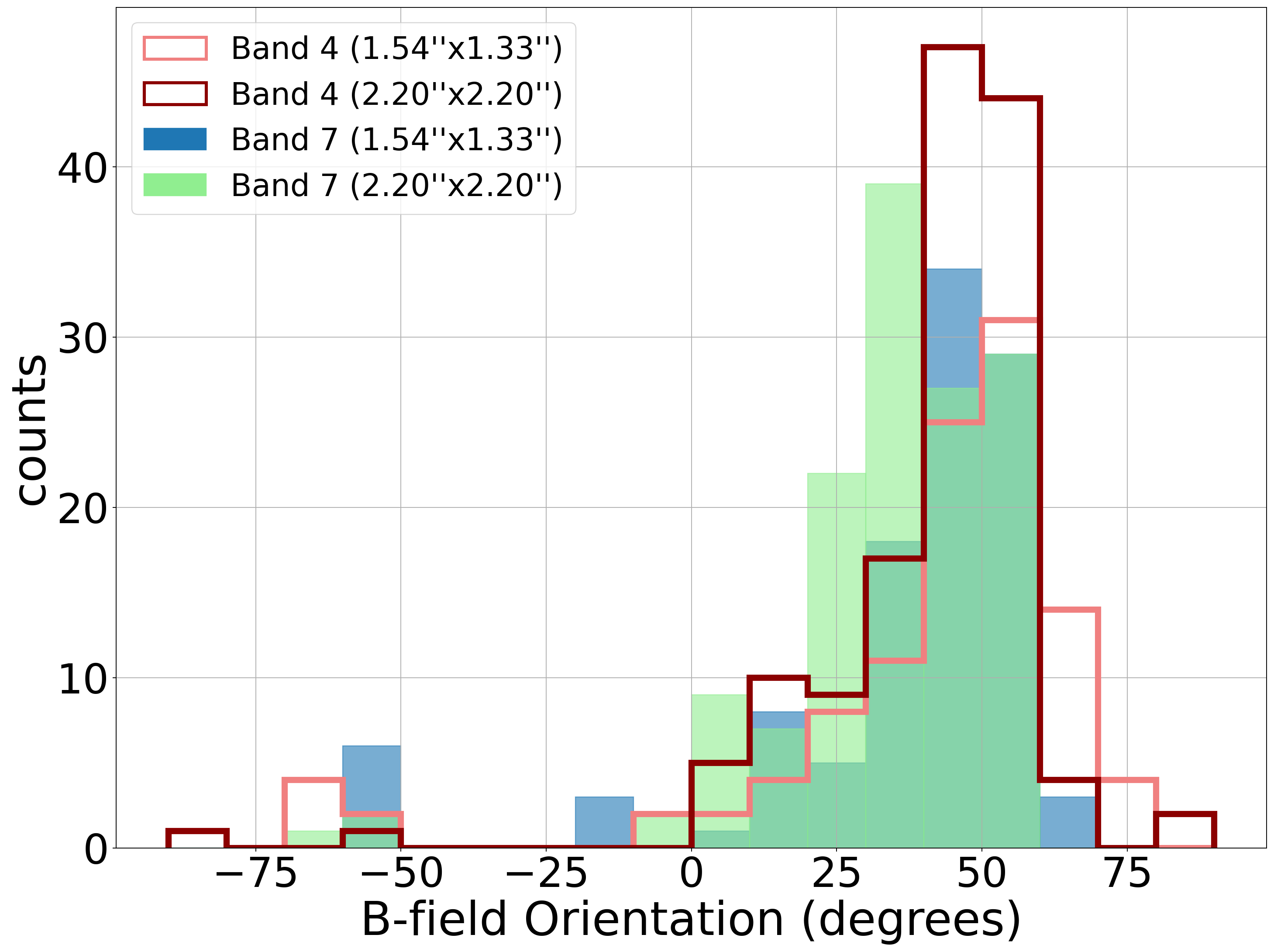}
\end{subfigure}\hfil 
\caption{{\bf{Polarisation properties in the centre of NGC253 measured at the same resolution.}} Distributions of pixel values of the polarisation fraction (left panel) and B-field orientation (right panel). Empty histograms represent the distributions of Band~4 pixels in the images at 1.5 $\times$ 1.3$\arcsec$ (pink) and at 2.2 $\times$ 2.2$\arcsec$ resolutions (red). The solid blue and green histograms represent the distributions at Band~7. Only pixels with SNR $\polfrac/\sigma_{\polfrac} > 3$ have been used for the computation of the histograms.}
\label{fig:images_sameres_hist}
\end{figure*}

\subsection{Polarisation images at matched resolution} \label{common}

\begin{table}[tbp]
\centering
\scalebox{0.80}{
\begin{tabular}{|c|c|c|c|c|c|} 
\toprule

      \multicolumn{2}{|c|}{Quantity} &
      \multicolumn{2}{c|} {1.5$\times$1.3 arcsec} & 
      \multicolumn{2}{c|} {2$\times$2 arcsec}    \\

\multicolumn{2}{|c|}{} & \textbf{Band 4} & \textbf{Band 7} & \textbf{Band 4} & \textbf{Band 7} \\
\toprule
\toprule
\multirow{2}{*}{I [MJy/sr]} & median & 318.65 & 1936.09 & 361.00 & 1637.37  \\ 
                            & std    & 268.82 & 1028.43 & 222.36 & 670.13 \\ 
\midrule
\multirow{2}{*}{PI [MJy/sr]} & median & 1.92 & 8.86 & 1.46 & 4.01  \\ 
                             & std    & 0.51 & 2.40 & 0.39 & 1.85 \\
\midrule
\multirow{2}{*}{PF [\%]} & median & 0.49 & 0.39 & 0.31 & 0.22  \\ 
                        & std & 0.90 & 0.16 & 0.81 & 0.13  \\ 
\midrule
\multirow{2}{*}{B-field [degrees]} & median & 47.93 & 46.44 & 44.74 & 35.04 \\
                                   & std    & 31.49 & 27.11 & 20.05 & 20.11 \\
\midrule
\multirow{2}{*}{S [degrees]} & median & 27.15 & 32.87 & 13.43 & 26.12  \\ 
                             & std    & 12.55 & 11.51 & 15.38 & 15.95 \\

\bottomrule
\end{tabular}}

\caption{ {\bf Median and standard deviation values of key polarisation quantities}. These measurements have been computed using our total intensity and polarisation images of NGC253 at common resolution. The measurements are made using the area defined in Section~\ref{common}. Additionally, for $\polint$, $\polfrac$, $\DeltaAng$ and B-field orientation only pixels with $\polfrac/\sigma_{\polfrac} > 3$ have been used for the median and standard deviation computation.
}
\label{tab:median_observed}
\end{table}

Measurements of polarisation properties strongly depend on resolution. To compare the results at different frequencies, we brought both images to the native resolution of the Band~4 data: 1.5$\arcsec$ $\times$ 1.3$\arcsec$. Figure~\ref{fig:images_sameres} shows the resulting $\polfrac$, $\DeltaAng$ and SNR of $\polfrac$ maps at both bands, overlaid with the B-field orientation.
The images have been masked to show only pixels characterized by $I/\sigma_I > 3$ in Band~7 at native resolution, \rev{while only B-filed vectors corresponding to regions where $\polint/\sigma_{\polint} > 3$ in both Band 4 and Band 7 are shown}. The PF values of the smoothed Band~7 data have been debiased using the covariance matrix computed from the smoothed Stokes I, Q and U maps, following the same process described in Section~\ref{Noise}. The resulting median and standard deviation values of I, $\polint$, $\polfrac$, B-field orientation and $\DeltaAng$ maps are summarised in Table~\ref{tab:median_observed}.

Considering only pixels with $I/\sigma_I > 3$ and $\polfrac/\sigma_{\polfrac} > 3$, the PF distribution of the two images at common resolution is characterized by a median PF of 0.49 \% for Band~4 and 0.39 \% for Band~7. The overall PF configuration is similar in both bands when examined at the same resolution: the observed region appears more polarised at the edges of the mask, showing very low PF to the left and right of the galaxy's centre.

The median values of $\DeltaAng$ measured at the same resolution are 27.15~degrees in Band~4 and 32.87~degrees in Band~7, indicating that there is still a slightly greater dispersion of the polarisation angles in Band~7. Overall, however, the $\DeltaAng$ maps at matched resolution are very similar, with local maxima situated in roughly the same regions of our observed field.

For the SED analysis, we smoothed the Band~4 and Band~7 images to 2.2 $\times$ 2.2$\arcsec$, corresponding to the resolution of the ancillary data described in Section~\ref{ancillary_data}. At this lower resolution, the median PF is 0.31\% for Band~4 and 0.22\% for Band~7. The decline in median PF with the increasing size of the observational beam is likely the result of beam depolarisation. This occurs when the polarisation and B-field structures are smaller than the observational beam, such that the combination of their heterogeneous polarisation angles yields a net polarisation lower than the intrinsic value at smaller scales. 

The distributions of polarisation fraction pixel values in the Band~4 and Band~7 maps at the two matched resolutions (1.5 $\times$ 1.3$\arcsec$ and 2.2 $\times$ 2.2$\arcsec$) are shown in 
Figure~\ref{fig:images_sameres_hist}.
When smoothed to the same resolution, the $\polfrac$ distribution peaks at $\sim 0.3\% - 0.5\%$ at both frequencies. The Band~4 distributions, however, show an extended tail of pixels with higher $\polfrac$ values. The pixels with high $\polfrac$ are mostly located in the SW edge of the regions where we have robust polarisation measurements (see Figure~\ref{fig:images_sameres}). We discuss this feature further in Section~\ref{analysis}.

The B-field orientation inferred at both common resolutions is also very similar in the two bands. The histograms of the orientation of the B-field (bottom panels of Figure~\ref{fig:images_sameres_hist}) peak strongly near $\sim 50^{\circ}$, indicating a component parallel to the galaxy midplane \rev{(characterized by a position angle of \newrev{$52\degree \pm 1\degree$}, \citet{Lucero2015})}, with a weaker peak in both bands at $\sim -60^{\circ}$, i.e. roughly orthogonal to the plane. From Figure~\ref{fig:images_sameres}, it is evident that this perpendicular component is located in the same region at the two bands, corresponding to the local maximum of Band~4 total intensity.


The $\polfrac$ and B-field distributions at the two bands at the same resolution do not show significant differences over most of the region that we analyse. The differences in the $\polfrac$ and B-field distributions that are evident in Figures \ref{fig:smallp_analysis} and \ref{fig:B-fields_orientation_analysis}, are lost when the images are convolved to a larger beam, meaning that Band~4 and Band~7 observations probe similar B-field and polarisation structures at 1.3$\arcsec$ scales ($\sim 25 \: pc$).

\section{SED-fitting analysis}
\label{analysis}

The differences observed in the native resolution polarisation measurement at Band~4 and Band~7 are at least partially the result of the difference in resolution between the two bands. Nevertheless, the two bands may also trace different components of the ISM. The total intensity ratio between Band~4 and Band~7 in our spatially resolved data is higher than what we would expect from thermal dust emission. This indicates an additional emission component in Band~4. Previous analysis of the galaxy emission with multi frequency data also suggested a complex situation in terms of contribution of different emitting mechanisms.
\cite{Peel2011} used data over the whole galaxy to determine a global spectral energy distribution (SED), which suggested the presence of free-free, dust and synchrotron emission across the $10 - 100$ GHz range in the starburst region of the galaxy. 
Previous ALMA observations of the 99~GHz continuum and $H40\alpha$ emission with a spatial resolution of $\sim 30$ pc in the central region indicated that free-free emission with an electron temperature $T_e = 3700-4500$ K contributes $\sim 70\%$ to the total flux at $\nu = 99$ GHz \citep{Bendo2015}. 
\rev{In \citealp{Humire2025} a detailed SED-fitting analysis spanning from near-UV to cm wavelengths was performed targeting at a resolution of $\sim 50$ pc, revealing contributions of synchrotron, free-free and dust emission in the central molecular zone.}
Higher resolution observations at spatial scales of $\sim 5$ pc showed a complex situation in which the relative contributions of dust, synchrotron and free-free varies spatially across the starburst region \citep{Mills2021}.

\begin{sidewaysfigure*}
    \centering
    
    \begin{minipage}{\linewidth}
        \centering
        \scalebox{0.81}{
        \begin{tabular}{|c||c|c||c|c|c||c|c|c||c|c|c|c|c|} 
        \toprule
        \multicolumn{3}{|c||}{Measured Values} & \multicolumn{11}{|c|}{Fit results with $\alpha_{ff}$ = 0.118, $T_{d}$ = 30 K and $\alpha_{syn} = [-0.45, -0.75]$} \\  
        \toprule
        \toprule
        \multicolumn{3}{|c||}{} &
        \multicolumn{3}{c||}{Band 7} &
        \multicolumn{3}{c||}{Band 4} &
        \multicolumn{5}{|c|}{} \\
        \toprule

                        Region 
        & \rev{\textbf{PF$_{\text{meas.},\text{Band~7}}$}[\%]}& \rev{\textbf{PF$_{\text{meas.},\text{Band~4}}$}[\%]} & I$_{dust}$/I$_{tot}$ & I$_{syn}$/I$_{tot}$ & I$_{ff}$/I$_{tot}$ & I$_{dust}$/I$_{tot}$ & I$_{syn}$/I$_{tot}$ & I$_{ff}$/I$_{tot}$ &  $\beta_{d}$ & $\chi^2_{red}$ & PF$_{dust}[\%]$ & PF$_{syn}[\%]$ & $\frac{PF_{syn}}{PF_{dust}}$ \\
        \toprule
        SA & 0.31 & 0.55 & [0.89, 0.91] & [0.03, 0.01] & [0.08, 0.08] & [0.22, 0.32] & [0.26, 0.14] & [0.51, 0.55] & [1.71, 1.36] & [0.10, 0.07] & [0.25, 0.27] & [1.78, 3.44] & [7.2, 12.7] \\ 
        \midrule
        SC & 0.10 & 0.17 & [0.80, 0.82] & [0.04, 0.01] & [0.16, 0.17] & [0.06, 0.13] & [0.21, 0.08] & [0.73, 0.79] & [2.62, 1.84] & [0.03, 0.04] & [0.08, 0.09] & [0.83, 2.37] & [11.0, 26.8] \\ 
        \midrule
        SO & 0.32 & 0.58 & [0.91, 0.93] & [0.03, 0.01]\% & [0.06, 0.06] & [0.31, 0.42] & [0.29, 0.16] & [0.40, 0.42] & [1.45, 1.15] & [0.12, 0.09] & [0.27, 0.29] & [1.78, 3.29] & [6.6, 11.3] \\ 
        \midrule
        $PI_E$ & 0.29 & 0.44 & [0.90, 0.91] & [0.02, 0.01] & [0.08, 0.08] & [0.26, 0.34] & [0.21, 0.10] & [0.54, 0.56] & [1.65, 1.38] & [0.08, 0.05] & [0.29, 0.31] & [1.65, 3.36] & [5.7, 11.0] \\
        \midrule
        $PI_S$ & 0.30 & 0.60 & [0.77, 0.79] & [0.04, 0.01] & [0.19, 0.20] & [0.04, 0.08] & [0.20, 0.11] & [0.76, 0.81] & [2.90, 2.17] & [0.07, 0.05] & [0.27, 0.32] & [2.86, 5.60] & [10.5, 17.7] \\ 
        \midrule
        $PI_W$ & 0.47 & 0.70 & [0.94, 0.95] & [0.01, 0.01] & [0.04, 0.04] & [0.41, 0.49] & [0.19, 0.10] & [0.40, 0.42] & [1.56, 1.39] & [0.11, 0.08] & [0.44, 0.46] & [3.71, 7.37] & [8.5, 16.1] \\
        
        \bottomrule
        \end{tabular}}
        \captionof{table}{{\bf SED fitting results for the different regions analysed in band 7 and band 4.}
        For each band, the actual \rev{PF$_{meas.}$} values measured from the data are reported in bold in the columns 2-3, while the columns 4-9 report the range of values of the relative contribution of dust, synchrotron, and free-free to the total intensity. These values are obtained from the SED fitting, assuming as synchrotron spectral $\alpha_{syn}$ the values $\alpha_{syn}=0.45$ and $\alpha_{syn}=0.75$. Column 10 reports the resulting fitted value of the emissivity index $\beta_d$, while column 11 reports  the range of values of the reduced $\chi^2$ of the fits. The dust polarisation fraction PF$_{dust}$
        and the synchrotron polarisation fraction PF$_{syn}$ derived (see Section \ref{analysis} for more details) are reported in the columns 12 and 13. The last column report the ratio of the derived polarisation fractions of the synchrotron and dust components.
        The regions in the first column are described in details in \newrev{Section \ref{analysis}} and indicated in Figure \ref{fig:regions}
        }
        \label{tab:SED_PF}
    \end{minipage}
    
    \vspace{1cm} 
    
    \begin{minipage}{\linewidth}
        \centering
        \begin{subfigure}{0.29\textwidth}
            \includegraphics[width=\linewidth]{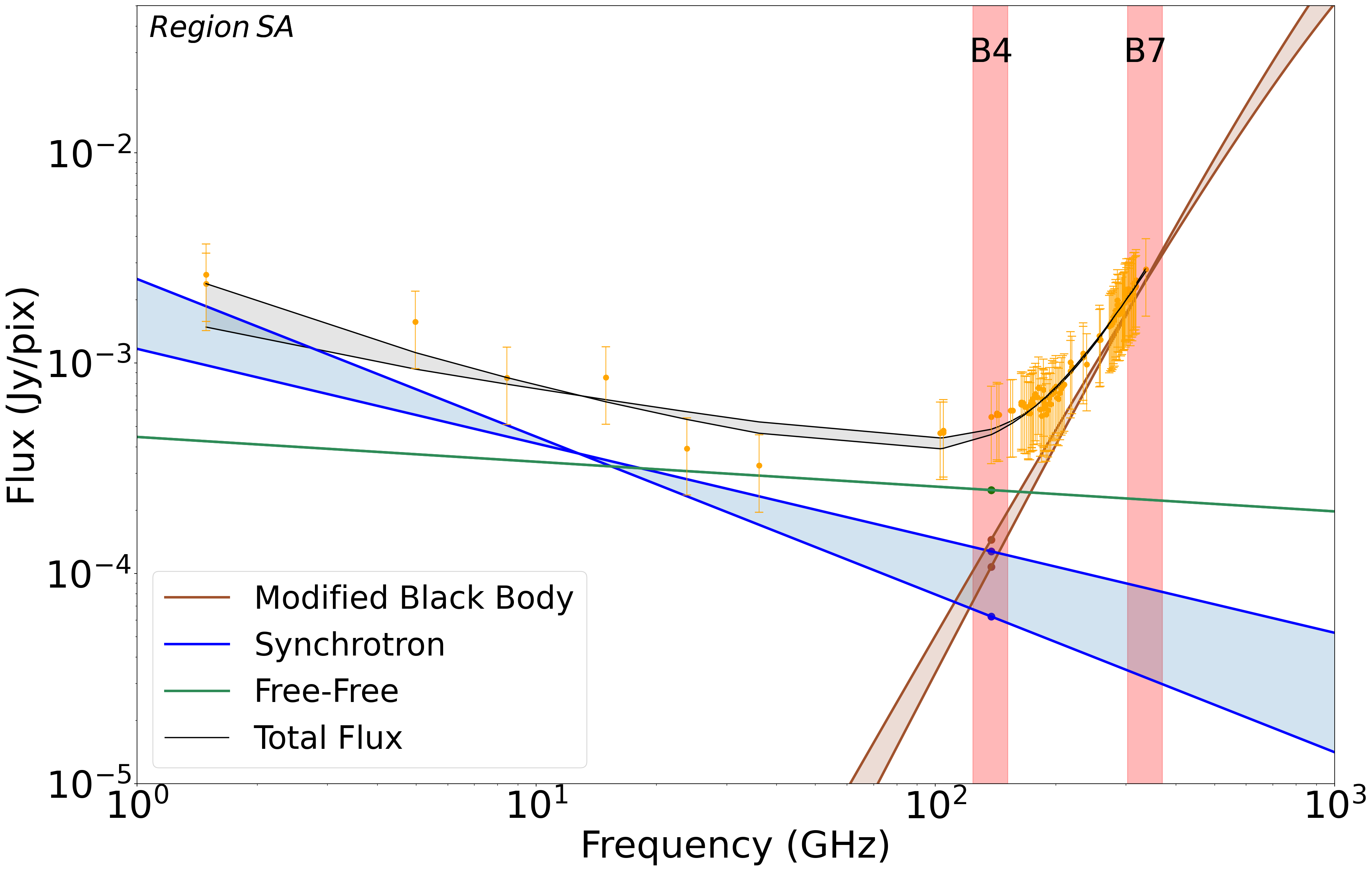}
        \end{subfigure}\hfill
        \begin{subfigure}{0.29\textwidth}
            \includegraphics[width=\linewidth]{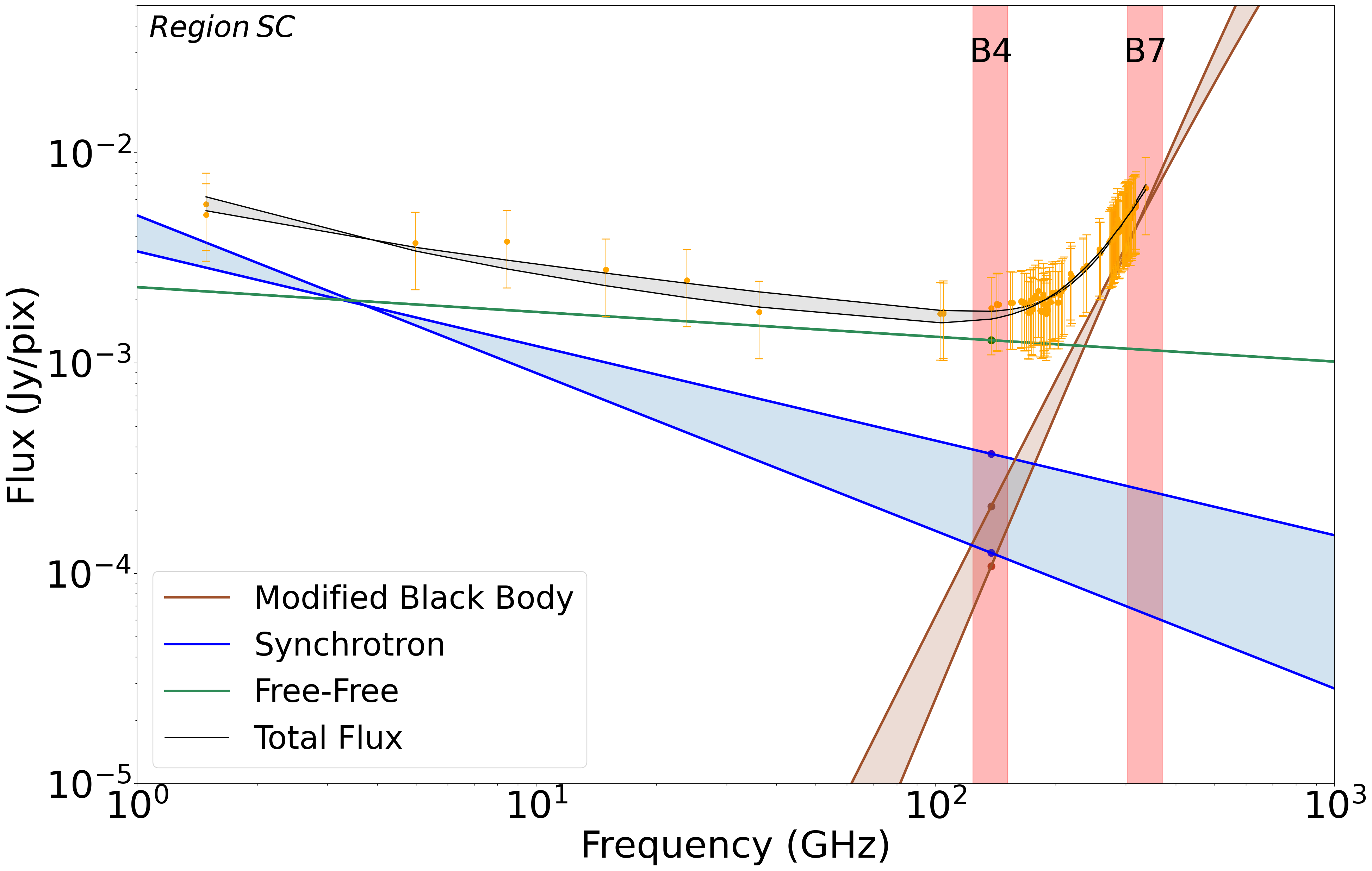}
        \end{subfigure}\hfill
        \begin{subfigure}{0.29\textwidth}
            \includegraphics[width=\linewidth]{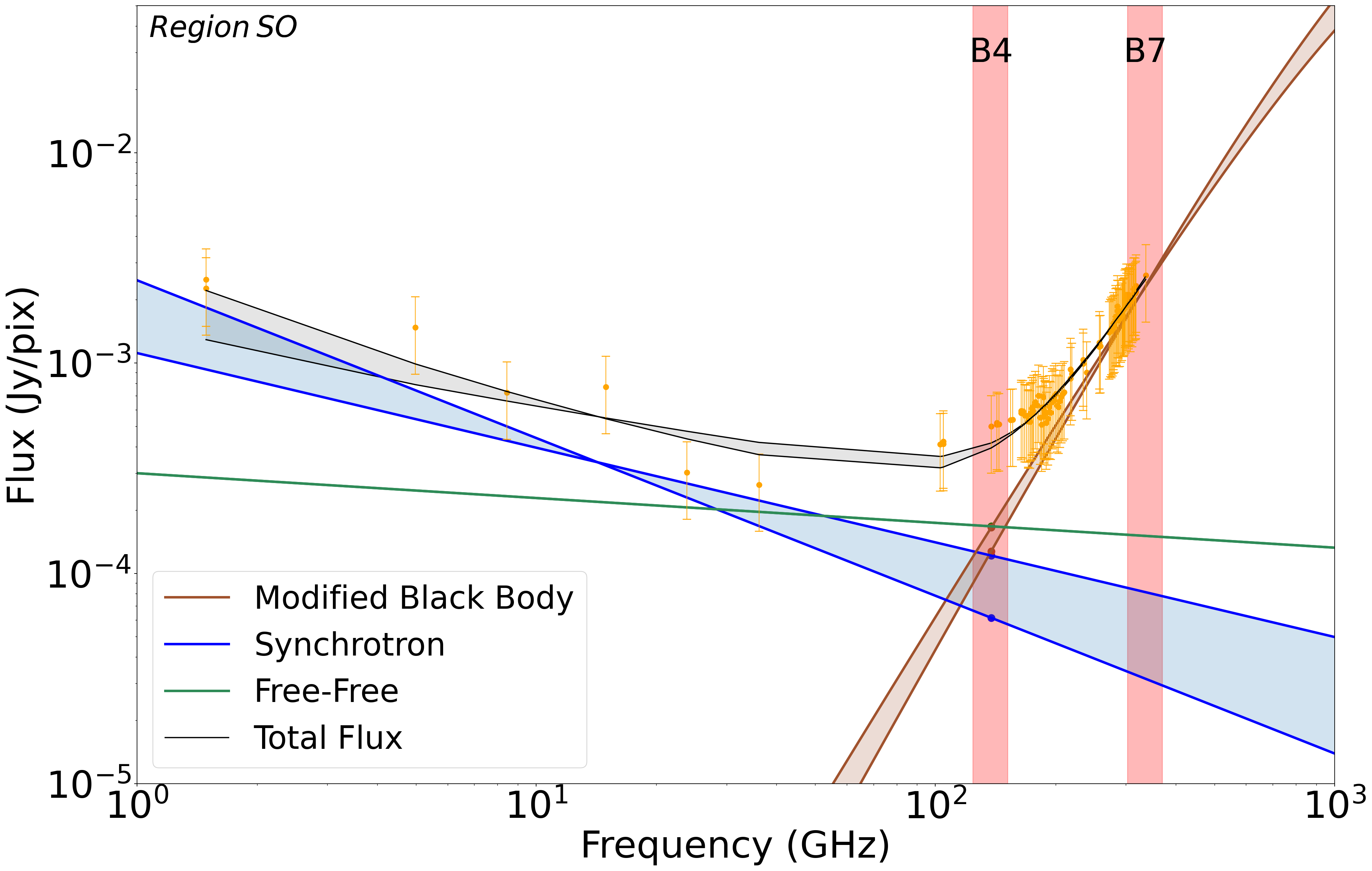}
        \end{subfigure}
        \medskip
        \begin{subfigure}{0.29\textwidth}
            \includegraphics[width=\linewidth]{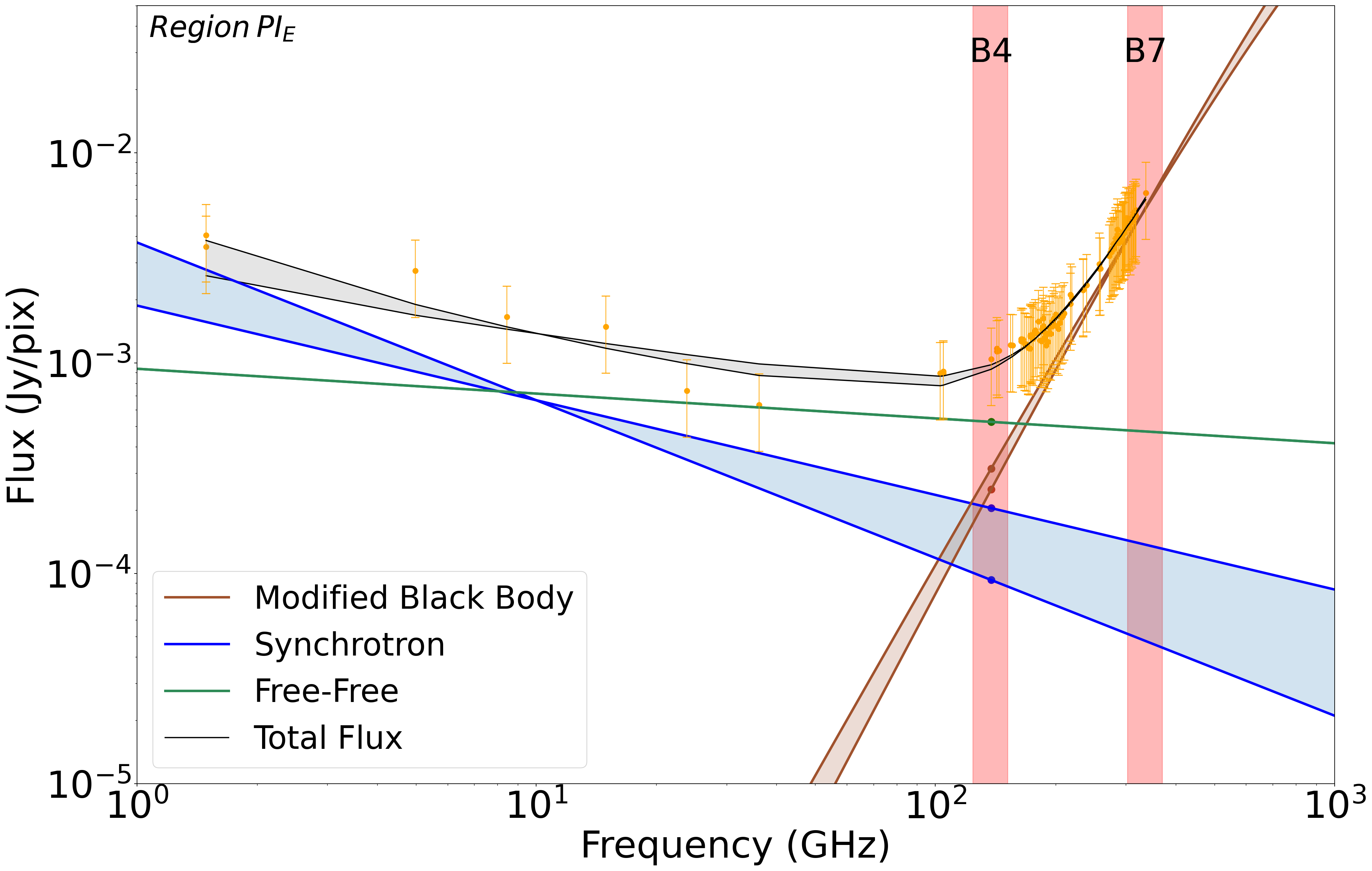}
        \end{subfigure}\hfill
        \begin{subfigure}{0.29\textwidth}
            \includegraphics[width=\linewidth]{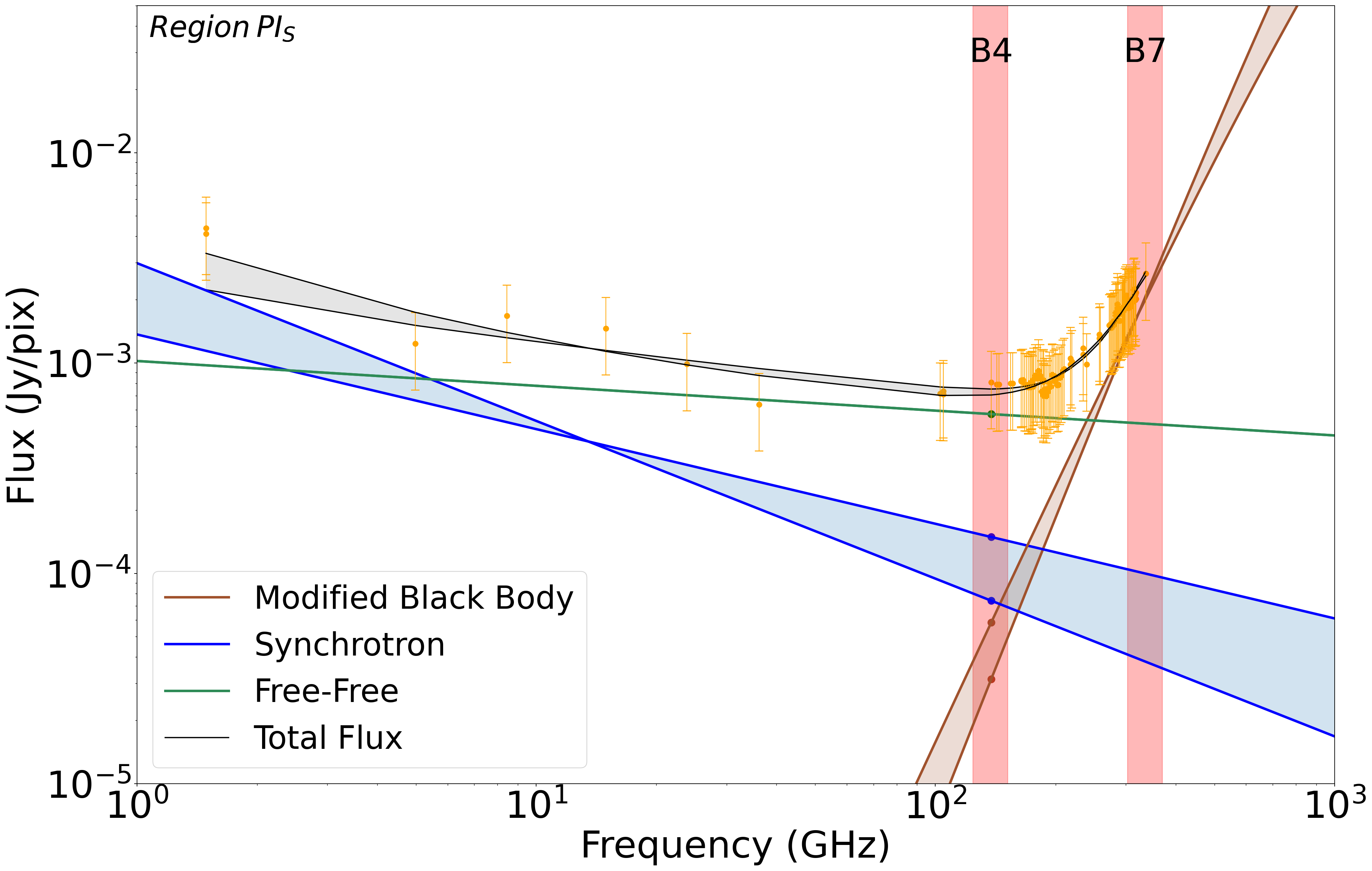}
        \end{subfigure}\hfill
        \begin{subfigure}{0.29\textwidth}
            \includegraphics[width=\linewidth]{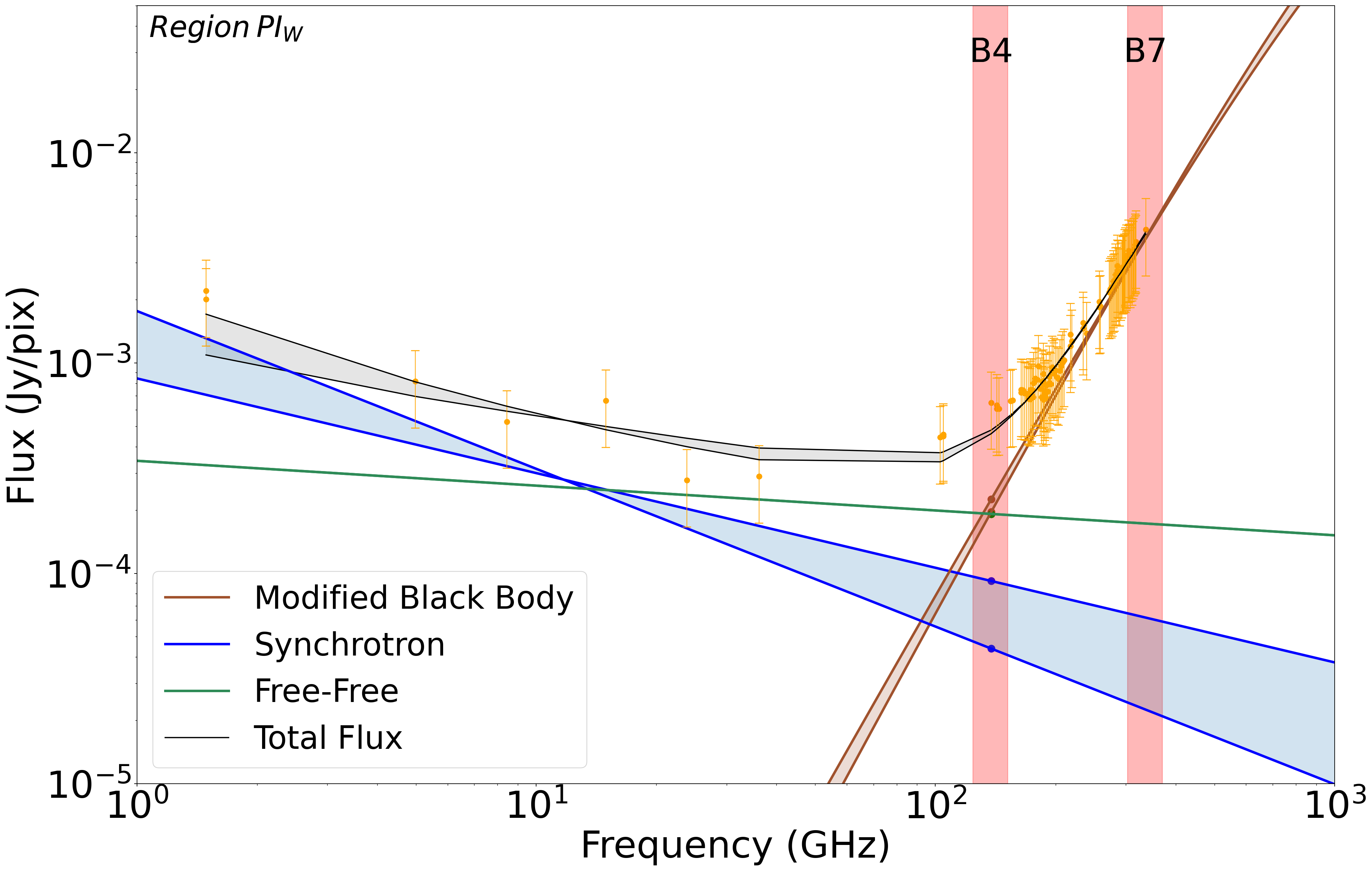}
        \end{subfigure}
        \captionof{figure}{{\bf{SED fitting analysis.}} 
        \label{fig:sed_fitting}
        The panels show the SED decomposition for the different regions (indicated in the top-left corner) discussed in \newrev{Section \ref{analysis}}. The observed measurements, shown by orange points with their uncertainties ($\pm 3 \sigma$) have been modelled using DustemWrap. The three fitted components: dust (in orange), synchrotron (in blue), and  free-free (in green) are shown with shaded area indicating their range, assuming two different spectral index values.
        }
    \end{minipage}

\end{sidewaysfigure*}

\subsection{Estimating the Band~4 emission components}

In order to estimate the relative contribution of the different emission mechanisms to the polarisation fraction observed in Band~4, we analysed the total intensity spectral energy distribution (SED) of the target, collecting archival data in the frequency range $1-300$\,GHz (see Table~\ref{tab:ancillary} for details about the archival data used). All images were convolved to a 2.2$\arcsec$ round beam, corresponding to the largest beam of the available archival data, and regridded to have common pixelization. 

The SED-fitting was performed using the public code DustEMWrap \citep{Compiegne2011} \footnote{http://dustemwrap.irap.omp.eu/}. We constructed input SEDs using our full observed region, as well as a few regions with exceptional polarisation properties or intensities (see below for precise definitions). Our model fits describe the thermal dust emission using a Modified Black Body (MBB) and allow for contributions from synchrotron and free-free emission using the plugins provided in DustEMwrap. \rev{Since there are not noticeable signs of anomalous microwave emission (AME), we exclude any possible contribution from spinning dust. The absence of significant AME emission in NGC253 is also suggested in \citep{Mills2021} and \citep{Peel2011}.}

The mask used for the SED-fitting of the whole emission has been defined as the largest mask that include pixels with $I/\sigma_I > 3$ in all archival images. This region is labelled SA in Figure~\ref{fig:regions} and shown in pink contours.
The variances used for the SED-fitting process are all assumed to be $0.4 \times I^2$, where I is the total intensity of the corresponding data point.

Due to the degenerate contributions of the three processes in the frequency range $1-300$~GHz, we fixed or restricted the range of several key physical parameters. The spectral index of the free-free emission was set to a fixed value of 0.118, a commonly observed value at sub-millimeter wavelengths \citep{Condon1992, Draine2011}. The temperature of dust emission was assumed to be $T_d=30\,K$, a reasonable value for the centre of NGC253 \citep{Melo2002, Weiss2008}.
For the synchrotron spectral index $\alpha_{syn}$ we tested two extreme values: 0.45 and 0.75. These values cover the typical range observed in high-density starburst galaxies \citep{Heesen2022}.
The intensities of the dust, synchrotron, and free-free emission and the emissivity index $\beta_d$ of the MBB were allowed to vary independently during the fit. \newrev{Table~\ref{tab:SED_PF}} reports the results of the SED fitting, quoted as a range of values obtained when adopting the two extreme values of $\alpha_{syn}$.

The decomposition obtained for the SED of the whole starburst region is shown in the upper left panel of Figure~\ref{fig:sed_fitting}. Both fits with $\alpha_{syn}=0.45$ and $\alpha_{syn}=0.75$ are shown.
The model was able to fit the data with a reduced $\chi^2_{red} =$ 0.10 and 0.07, respectively.
The $\beta_d$ values derived from SED-fitting across various regions align with those observed in other starburst galaxies \citep{McKay2023}.


The SED fit results for the SA region indicate that the thermal dust emission contributes $\approxeq 90\%$ of the Band~7 total intensity, while free-free and synchrotron contribute $\approxeq 8\%$ and $\approxeq 1\%$ respectively. In the spectral range $10-200$ GHz, none of the three emissions components are negligible. At Band~4, the contribution of dust, synchrotron and free-free emission to the total intensity is $22\% - 32\%$, $26\% - 14\%$ and $51\% - 55\%$, respectively.

In order to estimate how the different contributions vary across the central starburst region, we repeated the analysis using a subset of pixels within different areas that are indicated in Figure \ref{fig:regions}. These areas were selected by identifying pixels with elevated values of both total intensity and $\polint$.
Specifically, the SC area is defined by selecting the brightest 25\% of pixels in the 2.2$\arcsec$ map of Band~4 total intensity. This area corresponds to the region where the parallel and perpendicular components of the magnetic field structure observed in Band~4 converge. The region SO is the complementary region of SC in SA.
The areas labeled $PI_E$, $PI_S$, and $PI_W$ are identified by selecting the pixels with the highest 25\% observed values of polarised intensity at Band~4. The subscripts E, S, and W denote the positions relative to the midplane of the galaxy.
The SED decompositions for these regions are shown in the panels of Figure \ref{fig:sed_fitting}. The names of the regions are indicated in the upper left corner of each panel.
The SED-fitting results for these subregions are also listed in \newrev{Table {\ref{tab:SED_PF}}}.
The overall trend observed in the SA area is roughly consistent across all other subregions, where \rev{Band 4} maps a similar contribution of free-free, synchrotron, and dust emissions, while \rev{Band 7} primarily traces dust. 

\subsection{Contributions to Band-4 polarisation}

The above SED decomposition indicates that free-free emission, which is unpolarised \citep{Rybicki1986}, dominates the total Band~4 intensity. Our polarisation fraction measurements at Band~4 are thus due to a combination of synchrotron and dust emission, both polarised, diluted by the unpolarised free-free emission.
However, we can use our SED decomposition results to estimate the polarisation fractions of the dust and synchrotron emission independently. We assume that only synchrotron and thermal dust contribute to the polarised emission in both {\alma} bands.
The polarisation fraction of the synchrotron emission is not expected to strongly vary across the frequency range of our data \citep[see][]{Sokoloff1998}, and we assume that it is constant. Not much is known about possible frequency variations of the thermal dust polarisation fraction. The analysis of the all sky Planck data detected no systematic trend over the range 100-353~GHz, albeit with large uncertainties. Combining this result with very limited observations of specific sky regions in the FIR has led models  \citep[e.g.,][]{Hensley2021} to assume no spectral variations of the dust polarisation fraction as long as dust particles are large enough to align with magnetic field. Here we also follow this assumption and assume no frequency variations of the thermal dust polarisation fraction over the range observed with ALMA.
Under these assumptions, the observed polarised intensities in Bands~4 and~7 can be written simply as

\begin{equation}
\label{eq:polacontrib}
\begin{pmatrix} \displaystyle\polint_{B4} \\ \displaystyle\polint_{B7} \end{pmatrix}
=
\begin{pmatrix} \displaystyle\rev{I_{syn,B4}} & \displaystyle\rev{I_{dust,B4}} \\ \displaystyle\rev{I_{syn,B7}} & \displaystyle\rev{I_{dust,B7}} \end{pmatrix}
\times
\begin{pmatrix} \displaystyle\polfracsynch \\ \displaystyle\polfracdust \end{pmatrix}
\end{equation}

\noindent where $I_{syn}$  and $I_{dust}$ are the total intensity of the synchrotron and thermal dust emission, $\polfracsynch$ and $\polfracdust$ are the polarisation fraction of the synchrotron and dust emission \rev{(constant in Band 4 and 7 due to the assumptions we made)}, and the subscripts B4 and B7 refer to {\alma} measurements in Bands~4 and 7, \newrev{respectively}.

The polarisation fractions can be deduced by inverting Equation\,\ref{eq:polacontrib}, leading to
\begin{eqnarray}
\label{eq:polafracs}
\polfracsynch &= \displaystyle\frac{I_{dust,B7} \times \polint_{B4} - I_{dust,B4} \times \polint_{B7}}{I_{syn,B4} \times I_{dust,B7} - I_{syn,B7} \times I_{dust,B4}} \\
\polfracdust &= \displaystyle\frac{I_{syn,B4} \times \polint_{B7} - I_{syn,B7} \times \polint_{B4}}{I_{syn,B4} \times I_{dust,B7} - I_{syn,B7} \times I_{dust,B4}}.
\end{eqnarray}

\begin{figure}[tbp]
\centering
\includegraphics[scale=0.15]{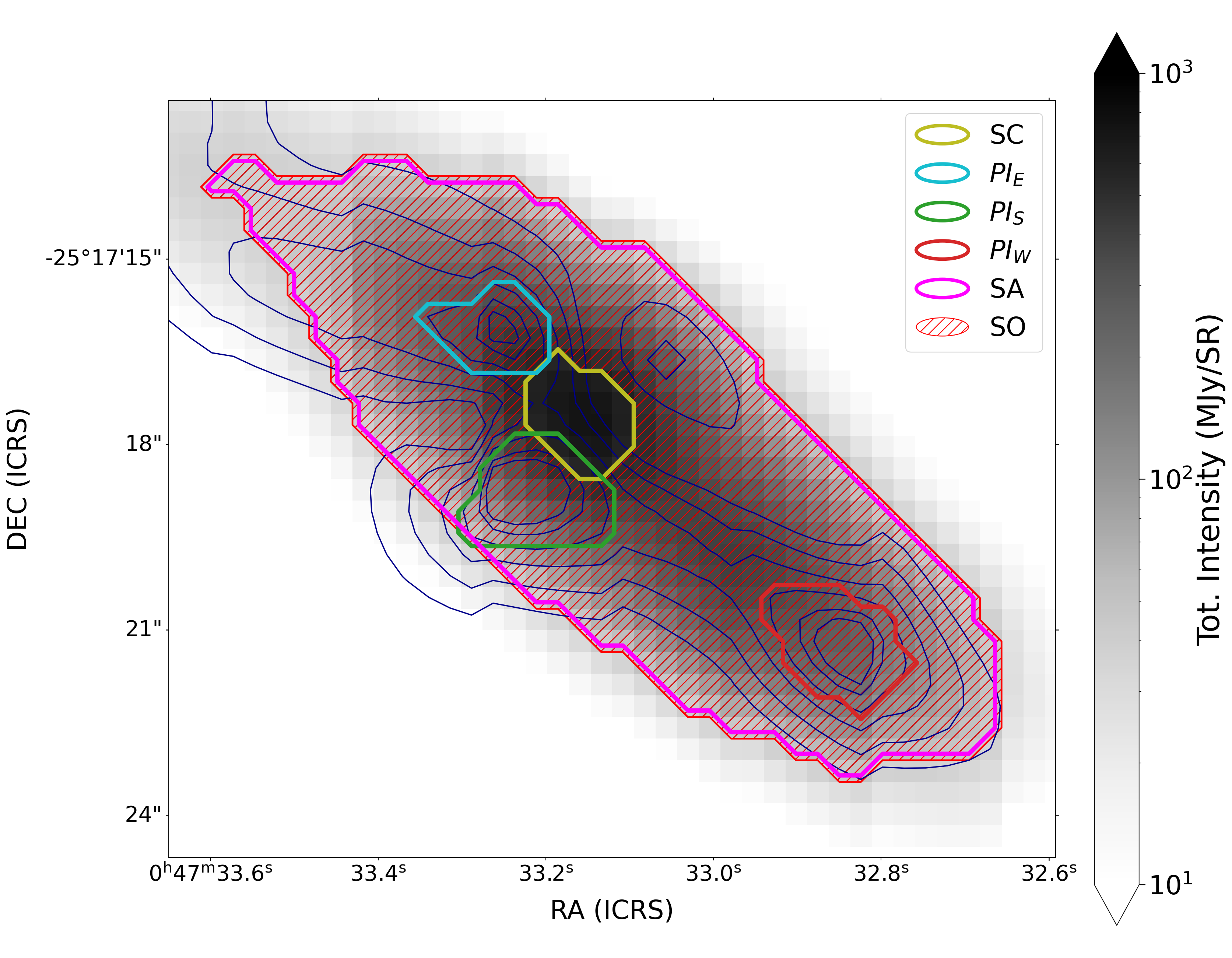}
\caption{{\bf Regions used for the SED fitting analysis.} Band~4 images of total (colour scale) and polarised (contours) intensity. The regions analysed in detail in Section~\ref{analysis} are indicated. }
\label{fig:regions}
\end{figure}

We measured $\polint_{B4}$ and $\polint_{B7}$ from the PI maps in Band~4 and Band~7, and derived the $PF_{dust}$ and $PF_{syn}$ using the SED-fitting results for $I_{syn}$, $I_{ff}$ and $I_{dust}$ for the two bands. 
Inside the SA mask (pink contour in Figure~\ref{fig:regions}), the estimated polarisation fraction for the synchrotron component ranges from $\polfrac_{syn} = 1.78 - 3.44\%$ (depending on the adopted synchrotron spectral index) and the dust polarisation ranges from $\polfrac_{dust}=0.25 - 0.27\%$. The measured polarisation fraction, \rev{which includes contributions from both dust and synchrotron components}, is \rev{$\polfrac_{meas.,B4}=0.55\%$} at Band~4 and \rev{$\polfrac_{meas.,B7}=0.31\%$} at Band~7.

For all regions where we conducted the SED-fitting, the synchroton and dust polarisation fractions obtained using the above approach are listed in the last two columns of \newrev{Table~\ref{tab:SED_PF}}. The reduced $\chi^2$ of the SED fits are also reported.

\section{Discussion} \label{discussion}

\subsection{Variations in the polarisation fraction} 

The measured polarisation fraction in Band~4 is $\sim 0.50\%-0.70\%$ across the starburst area, dropping to $\sim 0.17\%$ in the SC region. From our SED-fitting analysis, this corresponds to $PF_{dust}$ of $0.25\%$ for SA and $0.08\%$ for SC. A decrease in the measured $\polfrac$ in sub-mm data is usually attributed to a gradual loss of alignment of dust grains, fluctuations in the orientation of the B-field along the line-of-sight (LOS), \rev{or tangling of the B-fields on scales smaller than the observational resolution}, causing depolarisation \citep{planckcollaborationXIX}. 
In addition to these effects, the observed decrease of $\polfrac$ in Band~4 can be also attributed to a contamination of non-polarised free-free emission, as our SED-fitting suggests. 
Interestingly, the decrease in polarisation fraction in the SC region is more pronounced for the dust component compared to the synchrotron component. The ratio between the polarisation fractions of the synchrotron and dust components, $\frac{\polfrac_{sync}}{\polfrac_{dust}}$, is 7-11 in the SO region, but it increases to 11-27 in the SC region.
The regions $PI_E$, $PI_S$ and $PI_W$ are all characterized by a relative strong synchrotron component of PF. The region $PI_S$ and $PI_W$ are also characterized by a large $\frac{\polfrac_{sync}}{\polfrac_{dust}}$ ratio (around 10-18). This may be linked to the presence of the extended multi-phase outflow (see \newrev{Section} \ref{environment}).

We can summarize the SED fitting results as follows:

\begin{itemize}

    \item The different polarisation fractions observed at Band~4 and~7, after being smoothed to the same resolution, can be attributed to the distinct contributions of synchrotron, dust and free-free 
    to the total emission, combined with the different characteristic polarisation fraction of the dust and synchrotron components at these frequencies. At Band~7, the total intensity is dominated by dust, while at Band~4 the emission is mostly due to free-free, with non-negligible, roughly equivalent contributions from dust and synchrotron.
    The high polarisation fraction of the synchrotron component is likely the main factor behind the higher observed polarisation fraction in Band 4.
    
    \vspace{0.1cm}

    \item In all the regions, the polarisation fraction of the synchrotron component is higher than that of the dust component. \rev{In particular, The ratio between the polarisation fractions of the synchrotron and dust components, $\frac{\polfrac_{sync}}{\polfrac_{dust}}$, is 7.2-12.7 in the region SA, 11.0-26.8 in the SC, 6.6-11.3 in SO, 5.7-11.0 in $PI_E$, 10.5-17.7 in $PI_S$ and 8.5-16.1 in $PI_W$.}

    \vspace{0.1cm}

    \item The estimated $\polfrac_{dust}$ and measured $\polfrac$ in Band~7 exhibit comparable values. 
    Given the assumption of constant polarisation fraction of both the synchrotron and dust components over the frequency range $100-353$ GHz, we can deduce that in Band~7 the polarisation fraction is due almost entirely to dust with negligible effect due to synchrotron. 

    \vspace{0.1cm}
    
    \item 
    The inferred magnetic field structure (shown in Figure \ref{fig:images_sameres}) is common to both bands, and thus appears relatively insensitive to the different emission processes that dominate the polarisation in the two bands.
    
    \vspace{0.1cm}
    
    \item The depolarisation within the central region (SC) appears to be more pronounced for the dust component than for the synchrotron component. This suggests that the degradation of the alignment of dust grains, contributing to $\polfrac_{dust}$, may play a role in the observed depolarisation.  Multiple phenomena could explain the dust depolarisation in this region, including a differential tangling of the B-field between dense and more diffuse part of the LOS. 
      
\end{itemize}

\subsection{The B-field morphology and outflow in the centre of NGC253} \label{environment}

The nuclear starburst of the galaxy NGC253 is responsible for launching a large multiphase outflow which extends to several kpc. This structure is detected across the electromagnetic spectrum, including X-ray \citep{Strickland2000, Lopez2023}, ionized gas \citep{Westmoquette2011}, molecular gas \citep{Bolatto2013, Krieger2019} and radio \citep{Heesen2011}.

The morphological structure of the outflow varies significantly depending on the observational wavelength. Optical observations (particularly $H\alpha$) and X-ray observations reveal a prominent biconical morphology that extends to several kpc in the SE and NW directions. From $H\alpha$ kinematic modelling, the opening angle of the outflow is $\sim60\degree$ \citep{Westmoquette2011}. Due to the orientation of the galaxy, the SE cone is more prominent and extended than the NW cone. 

The molecular phase of the outflow, traced by CO(1-0) and CO(3-2) emission lines, is characterized by a more concentrated structure situated along the edges of the biconical wind. One of the most prominent structures is a SW streamer placed along the direction of the minor axis of the starburst region (see e.g. \citealp{Bao2024}). 
Enhancements in the CO emission, such as the SW streamer and other distinct features, might be due to individual clouds that are trapped, disrupted and ejected in the outflow \citep{Zschaechner2018}.

Radio continuum emission indicates the presence of four prominent filaments (scale height of 150 $\pm$ 20\,pc) surrounding the SE and NW cone \citep{Heesen2011}. In that study, the large scale B-field structure was probed by radio polarisation observations at \rev{wavelengths of $6 \: \text{cm}$ and $3 \: \text{cm}$}, with resolutions of approximately 500 pc and 120 pc, respectively. The large scale B-field lines inferred from the radio data slowly open away from the midplane of the galaxy, forming a characteristic X-shaped field centered on the nucleus. The total B-field strength measured in the filaments is $46 \pm 10 \: \mu$G, with an ordered field strength of $21 \pm 5 \: \mu$G, which is high enough to counteract the pressure in the nuclear outflow \citep{Heesen2011}, suggesting a role of the B-field in collimating the outflow.

\begin{figure*}[tbp]
    \centering
    \begin{subfigure}{0.95\linewidth} 
        \centering
        \includegraphics[width=\linewidth]{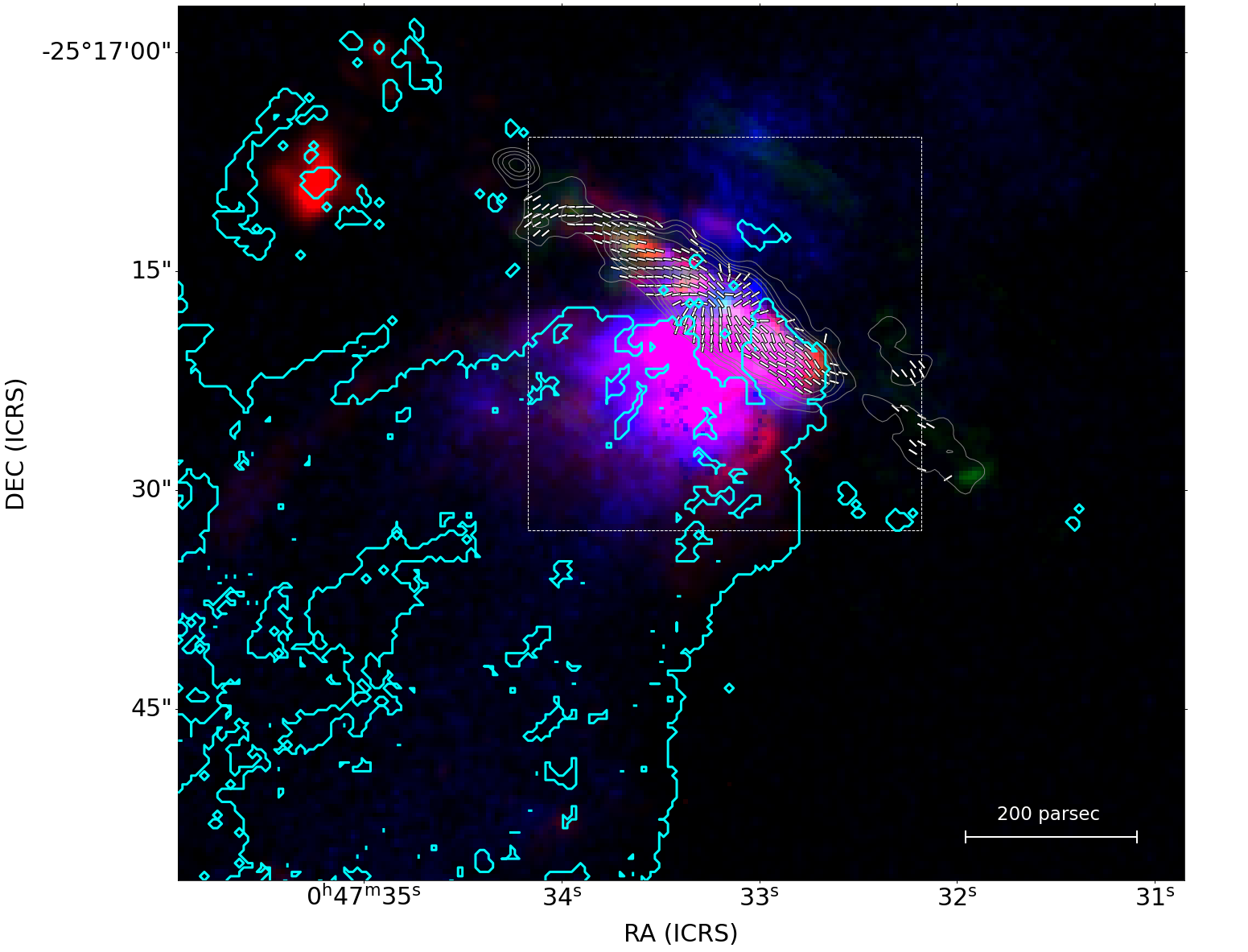}
        \caption{}
    \end{subfigure}

    \caption{{\bf Three-colour composite, multi-wavelength view of the central region in NGC253.} Blue, red, and green show the X-ray emission from \citealp{Lopez2023}, optical $[NII]$ data from Cronin et al. in prep., and CO(3-2) emission from \citealp{Leroy2018}, respectively. The B-field lines and the emission of Stokes I flux of our ALMA Band~4 data are also shown as white vectors and yellow contours, respectively. \rev{The vectors are displayed every two pixels and only where $\polint/\sigma_{\polint} > 3$}. The grey contours correspond to $[5, 7, 10, 30, 100] \times \sigma_I$. The red arrow marks the position of the molecular southwest streamer. The white rectangle marks the region shown on the upper panel of Figure \ref{fig:env_2}.}
    \label{fig:env}
\end{figure*}

\begin{figure*}[tbp]
    \centering
    \begin{subfigure}{0.67\linewidth} 
        \includegraphics[width=\linewidth]{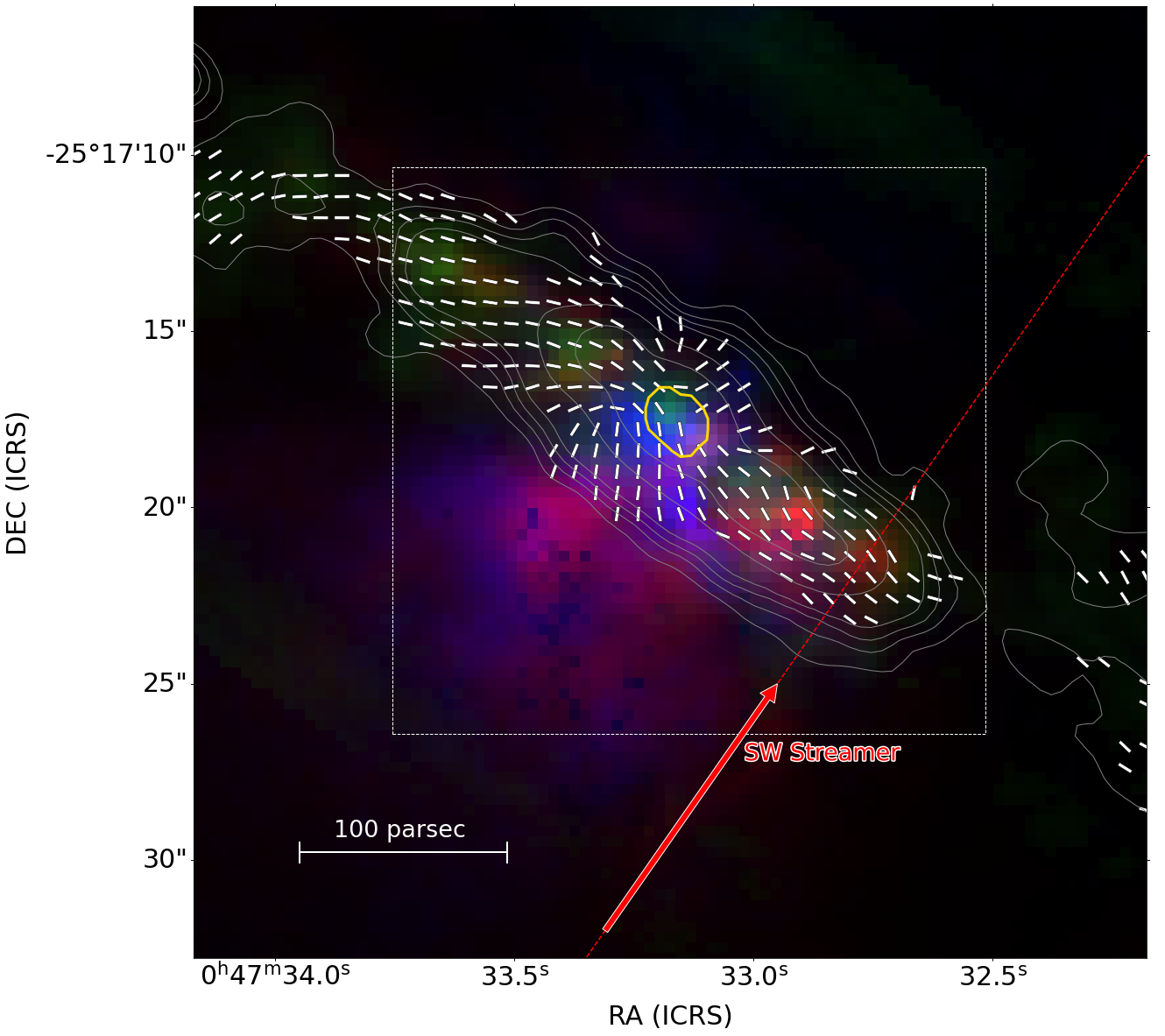}
    \end{subfigure}
    \hfill 
    \begin{subfigure}{0.67\linewidth} 
        \includegraphics[width=\linewidth]{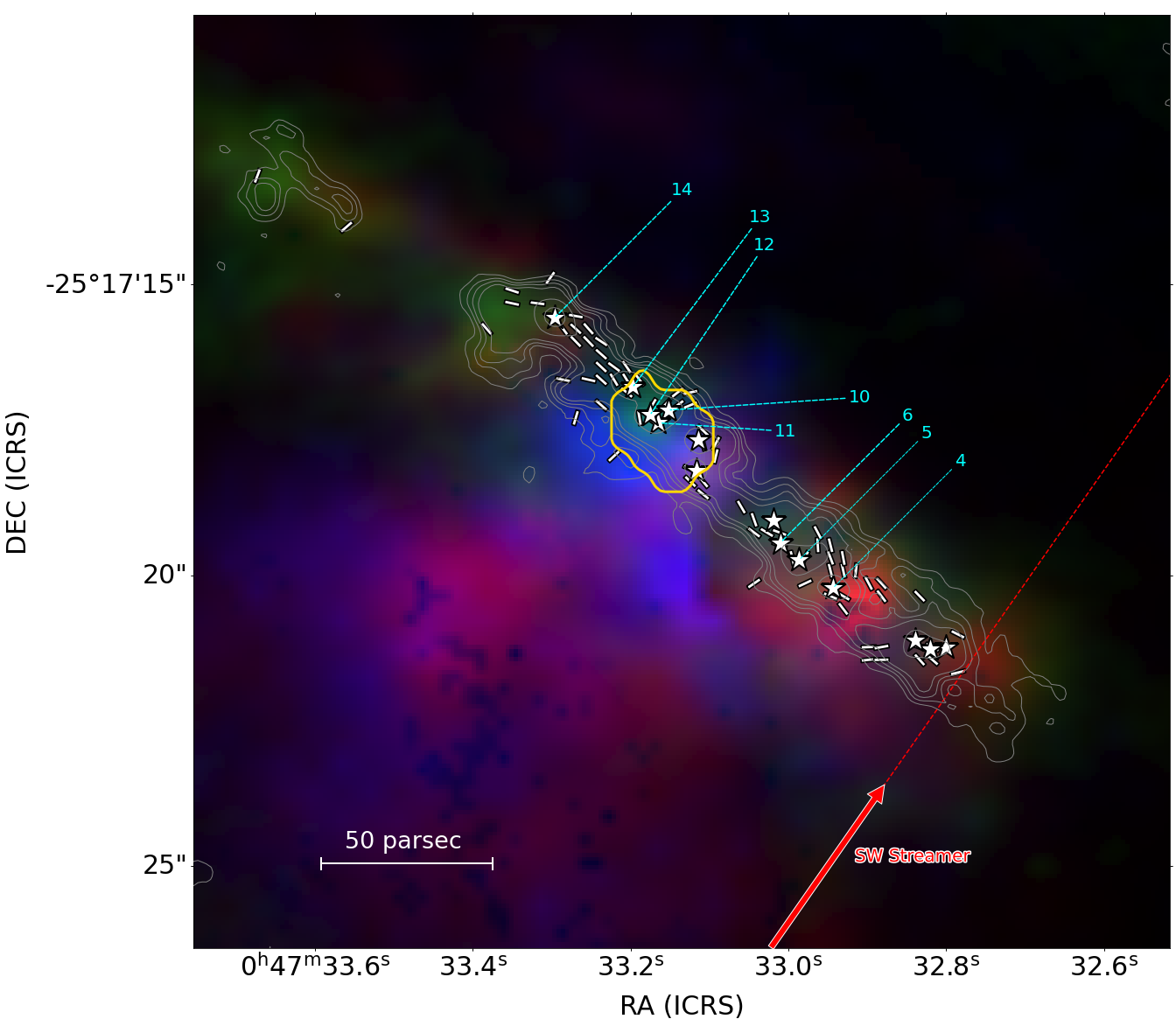}
    \end{subfigure}
    \caption{{\bf Three-colour composite, multi-wavelength view of the central region in NGC253.} Blue, red, and green show the X-ray emission from \citealp{Lopez2023}, optical $[NII]$ data from Cronin et al. in prep., and CO(3-2) emission from \citealp{Leroy2018}, respectively. The B-field lines and the emission of Stokes I emission of our ALMA Band~4 data (upper panel) and Band~7 data (lower panel) are also shown as white vectors and yellow contours, respectively. \rev{The vectors are displayed every two pixels and only where $\polint/\sigma_{\polint} > 3$}. The grey contours correspond to $[5, 7, 10, 30, 100] \times \sigma_I$. The red arrow marks the position of the molecular southwest streamer. The white rectangle on the upper panel marks the region shown on Figure 
    \ref{fig:env}. The SSCs are marked as white stars and labelled following the notation of \citealp{Leroy2018}. \rev{The yellow contour mark the position of the SC region defined in Section \ref{analysis}}.}
    \label{fig:env_2}
\end{figure*}

Figure~\ref{fig:env} shows a three colour, multi wavelength view of the central region of NGC253. The X-ray (blue colour), $[NII]$ emission line (red) and CO(3-2) (green) emissions are shown in relation to the B-field structure revealed by our Band~4 ALMA data. The white contours indicate the Stokes~I emission in Band~4. The B-field lines inferred from our Band~4 data are shown as white vectors. \rev{The vectors are displayed every two pixels and only where $\polint/\sigma_{\polint} > 3$}.
The cyan contour highlights the regions that show outflow components, identified using 
$[NII]$ and $H \alpha$ line integrated intensity from VLT/MUSE data (Cronin et al. in prep.).
The SW streamer location, taken from \citealp{Zschaechner2018}, is indicated as a red arrow and labelled as SW.
\rev{In Figure~\ref{fig:env_2}, the upper panel shows the same three-colour image as Figure~\ref{fig:env} but zoomed in to provide a closer view of the central region, while the lower panel presents the RGB image superimposed with the Band~7 B-field vectors.}
The white stars mark the position of the SSCs identified by \citealp{Leroy2018} (see also \citealp{Levy2021, Levy2022, Mills2021}), which account for a large fraction of the star formation activity in the central starburst region. Three of these SSCs (labelled 4, 5 and 14 in \citealp{Leroy2018}) have been observed to show P-Cygni proﬁles in multiple lines, which suggest the presence of small outflow activities. 
Furthermore, some of these SSCs (4, 6, 10, 10NE, 11, 12, and 13) host $H40\alpha$ emission with exceptionally broad linewidths ($\delta v_{FWHM} \sim 100-200 \: km s^{-1}$), suggesting that they contribute to driving the ionized component of the multi-phase outflow \citep{Mills2021}.

The base of the ionized component of the biconical outflow is a strong source of X-ray and $[NII]$ emission. The B-field inferred from our Band~4 data (Figure \ref{fig:env} and upper panel of Figure  \ref{fig:env_2}) seem to closely track the profile of the base of the outflow: the B-lines start to open away from the midplane at the peak position of the X-ray emission, assuming the same conical structure of the outflow. Away from the base of the outflow towards the SE, the B-field lines seem to assume an ordered horizontal position relative to the midplane, until they are disrupted at the position the SW molecular streamer. The same behaviour is observed on the northern side of the midplane, with the horizontal B-lines being disrupted towards a more vertical orientation along the direction of the SW streamer. 
The signal-to-noise of our observations decreases away from the midplane. Consequently, it is not possible with our data to determine if the B-field structure continues to follow the biconical outflow and align with the SW streamer's direction.

Both the B-field lines and the outflow cone are centred on the SC region  defined in Section~\ref{analysis}. This area is characterized by high molecular gas column density (as traced by the CO(3-2) map), and hosts four of the seven SSCs characterized by broad linewidth $H40\alpha$ emission. In the SC region, the SED-fitting analysis 
(see Section~\ref{analysis}) also suggests a high synchrotron polarisation fraction 
indicating a coherent and ordered B-field. A possible interpretation is that the high star formation rate drives the large scale winds which compress and order the B-field, aligning it to the outflow cone. This interpretation is coherent with the picture of the SSCs characterized by broad linewidth $H40\alpha$ emission being the main driver of the large scale winds forming the biconical outflow.  

\rev{The launch location of the wind seems to corresponds roughly to the SC region defined in Section \ref{analysis}, which is highlighted as yellow contour in both panels of Figure \ref{fig:env_2}}. The highly tangled B-field lines and the limited area for which we have robust Band~7 polarisation measurements complicate a more detailed comparison between the B-field structure and the outflow morphology. From our data, it is not clear if the outflows due to SSCs 4,5 and 14 in \citep{Levy2021} have a counterpart in the small scale B-field morphology. 

All the SSCs show a spatial relation with the measured PF in Band~7. This can be seen in the upper-right panel of Figure \ref{fig:smallp_analysis}, where the positions of the SSCs are indicated by white stars. Locations with at least one SSC are characterized by local minima in PF measured in Band~7. 
Measuring the PF within one beam located on the SSCs, we get an average value below 0.4\%, compared to a median PF value of 1.06\% across the entire mask with $I/\sigma_{I} > 3$ and $\polfrac/\sigma_{\polfrac} > 3$.
The correlation between local minima in PF and the position of SSCs is part of a more general relation that links the gas column density and polarisation fraction, which we discuss in the next subsection.

\subsection{The $\polfrac$ vs $\NH$ relation} \label{sect:pfvsNH}

A trend for the thermal dust polarisation fraction to decrease with increasing gas column density was identified from the all-sky Planck data at 353~GHz \citep{planckcollaborationXIX}, over a wide range of column densities from $10^{20}$ to $10^{23}\,H/cm^{2}$. This confirmed results from earlier ground-based observations of very bright regions that typically detected low polarisation fractions. Analysis of the Planck data revealed that the $\polfrac$-$\NH$ anti-correlation extends to dense regions with no star formation. It is important to stress that since polarisation fraction measurements are positively biased upward by noise and photon noise is always higher in low brightness regions, an anti-correlation can arise due to noise bias only. Therefore, only good quality de-biased polarisation data should be used to address this issue.

The physical origin for the anti-correlation between $\polfrac$ and $\NH$ is not completely clear. As dust alignment with magnetic field could result from radiative alignment torques (RATs, e.g. \citealt{Dolginov1976, Andersson2015}), the lack of photons in dark dense clouds could explain the observations. However, since polarisation sums as a spin-2 vector within the observational beam and along the LOS, the observed depolarisation could also result from the tangled structure of the magnetic field in dense regions. Additionally, since all models of dust alignment with magnetic field predict a better alignment for large dust grains, the observed trend could also result from variations of the grain size distribution with column density. Explanations involving destruction of large dust grains \citep[e.g.][]{Lee2020, Hoang2021, Lopez-Rodriguez2024} have been proposed for highly active star-forming regions.

Analysis of the all-sky Planck maps at $350\,GHz$ did not reveal a systematic change in the relationships between the polarisation dispersion angle (see \newrev{Section~\ref{dispersion}}) and polarisation fraction with column density, indicating that tangling of the B-field is the best explanation for the observed trend over the $\NH$ range sampled by the Planck data \citep{planckcollaborationXIX}. 

We used our debiased polarimetric observations and the archival ALMA observations of the CO(3-2) line \citep{Leroy2018} to investigate the correlation between the polarisation fraction and the column density in NGC253. We calculated the column density $\NH$ from the CO(3-2) line fluxes using the relation $N_{H2} = X_{CO} I_{CO(1-0)}$, where $X_{CO}$ is the $CO$-to-$H_2$ conversion factor for the CO(1-0) line integrated intensity $I_{CO(1-0)}$ \citep{Bolatto2013b}.
We converted $I_{CO(3-2)}$ to $I_{CO(1-0)}$ using the empirical line ratio $r_{31}$ = $\frac{I_{CO(3-2)}}{I_{CO(1-0)}}$. 
We adopted $X_{CO} = 0.5 \cdot 10^{20} [K \: km \: s^{-1} \: cm^{-2}]$, a value consistent with other estimates for starburst galaxies \citep{Bolatto2013b}. Following \citealp{Schinnerer2024} we assumed $r_{31} = 0.31$.
We then estimated $\NH$ assuming $\NH \approxeq 2N(H_2)$. 

\begin{figure}[tbp]
  \begin{center}
  \begin{subfigure}{0.8\linewidth}
    \includegraphics[width=\linewidth, height=0.80\linewidth]{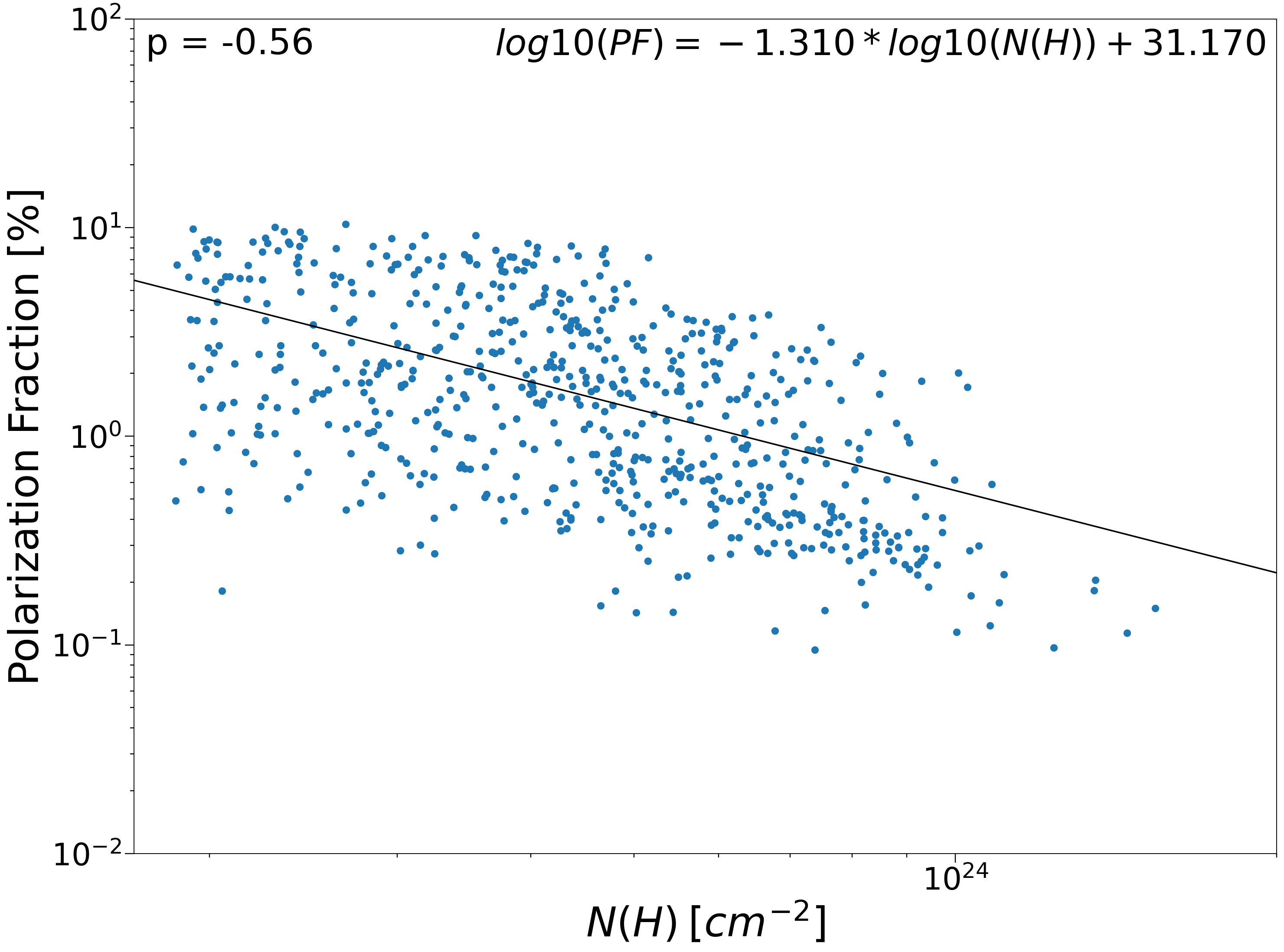}
  \end{subfigure}
  \vspace{1em} 
  \begin{subfigure}{0.8\linewidth}
    \includegraphics[width=\linewidth, height=0.80\linewidth]{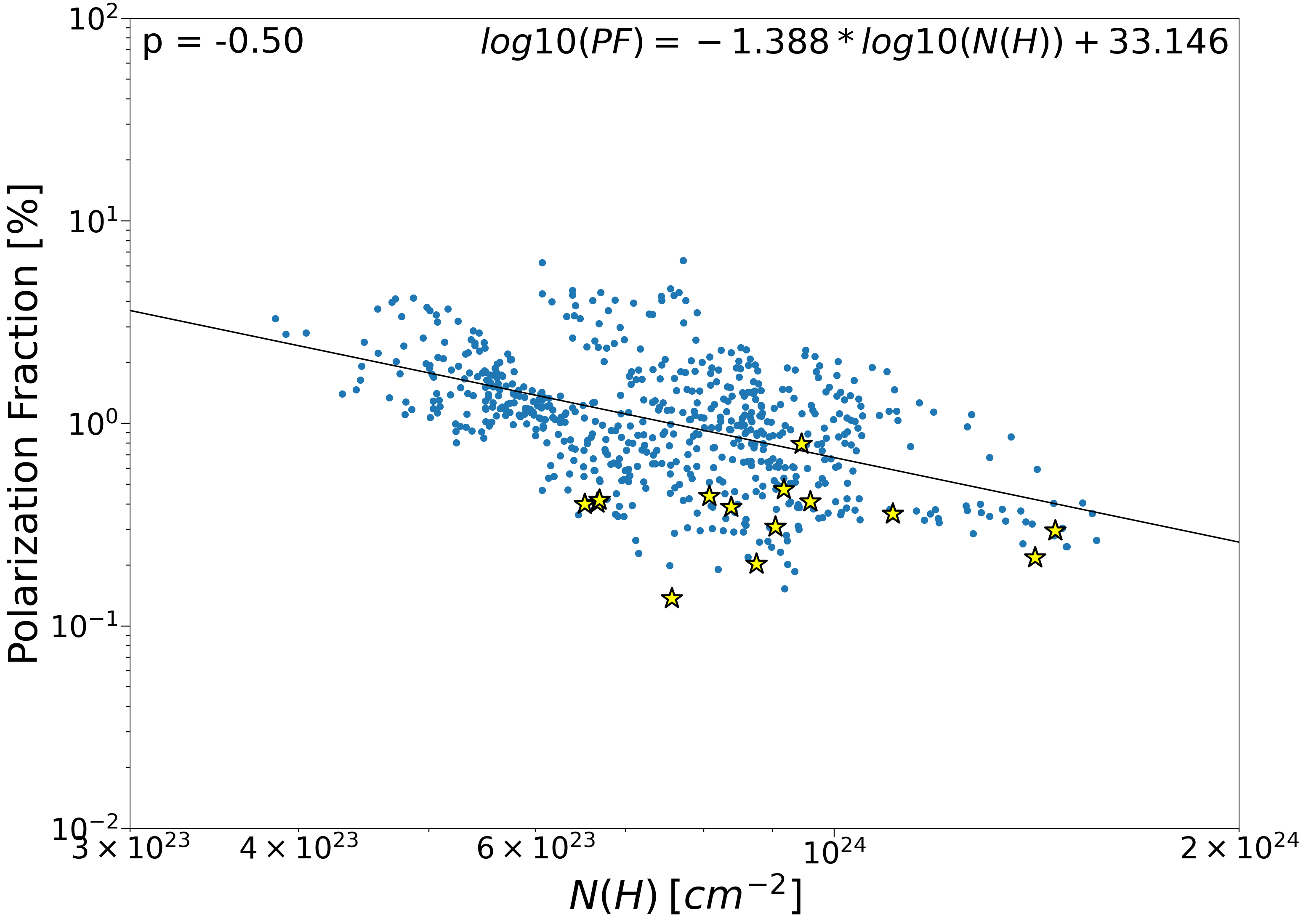}
  \end{subfigure}
  \end{center}  
\caption{
 {\bf Scatter plot of the observed polarisation fraction $\polfrac$ as a function of gas column density in NGC253.} Band~4 data are shown on top panel, Band~7 in the bottom panel.   
The black line shows the trend derived from a fit to our data (see Equation\,\ref{eqn:psvsNH}. The yellow stars in the bottom panel indicate values computed at the location of the 14 SSCs (see Figure~\ref{fig:images_sameres}). 
} 
\label{fig:smallp_SSC}
\end{figure}

Figure~\ref{fig:smallp_SSC} shows the observed relation between $\polfrac$ and $\NH$ in our data computed at Band~4 (upper panel) and Band~7 (lower panel). The CO(3-2) data were convolved to the same resolution as the ALMA data, and only pixels with good signal-to-noise in CO(3-2) intensity ($S/N>3$), Stokes~I ($S/N>3$) and $\polfrac$ ($S/N>3$) are included. 

Our data show an anti-correlation between $\polfrac$ and $\NH$ 
given by:
\begin{equation}
log_{10}(\polfrac)=-1.310 \times log_{10}(\NH)+31.17
\label{eqn:psvsNH}
\end{equation}

\noindent with a Pearson correlation coefficient $\pearson = -0.56$ 
for Band~4, and by:  
\begin{equation}
log_{10}(\polfrac)=-1.388 \times log_{10}(\NH)+33.146.
\label{eqn:psvsNH}
\end{equation}

\noindent with $\pearson = -0.50$ for Band~7.

\rev{The linear fit was done taking into account the uncertainties in $\polfrac$ using the orthogonal distance regression method.}
The trends in Figure~\ref{fig:smallp_SSC} indicate that the observed decrease of $\polfrac$ with $\NH$ in the Planck data continues at even higher column densities. A similar trend has been observed in Galactic cold cores and star-forming regions using JCMT and ALMA data \citep[e.g.,][]{Koch2022,Lin2024}, although it is often reported as a trend between polarisation fraction and total intensity rather than $\NH$.
We note that the values of $\polfrac$ observed in NGC253 across the column density range \rev{$10^{23}<\NH< 4\times10^{23}$ cm$^{-2}$} tend to sit slightly higher than the trend line identified by Planck \cite{planckcollaborationXIX}, although both our results and the Planck results show a large scatter. This discrepancy might be due to a more favourable orientation of the magnetic field direction in our observed field-of-view in NGC253.

\begin{figure*}[tbp]
    \centering
    \begin{subfigure}{0.45\textwidth}
        \centering
        \includegraphics[width=\linewidth, height=0.80\linewidth]{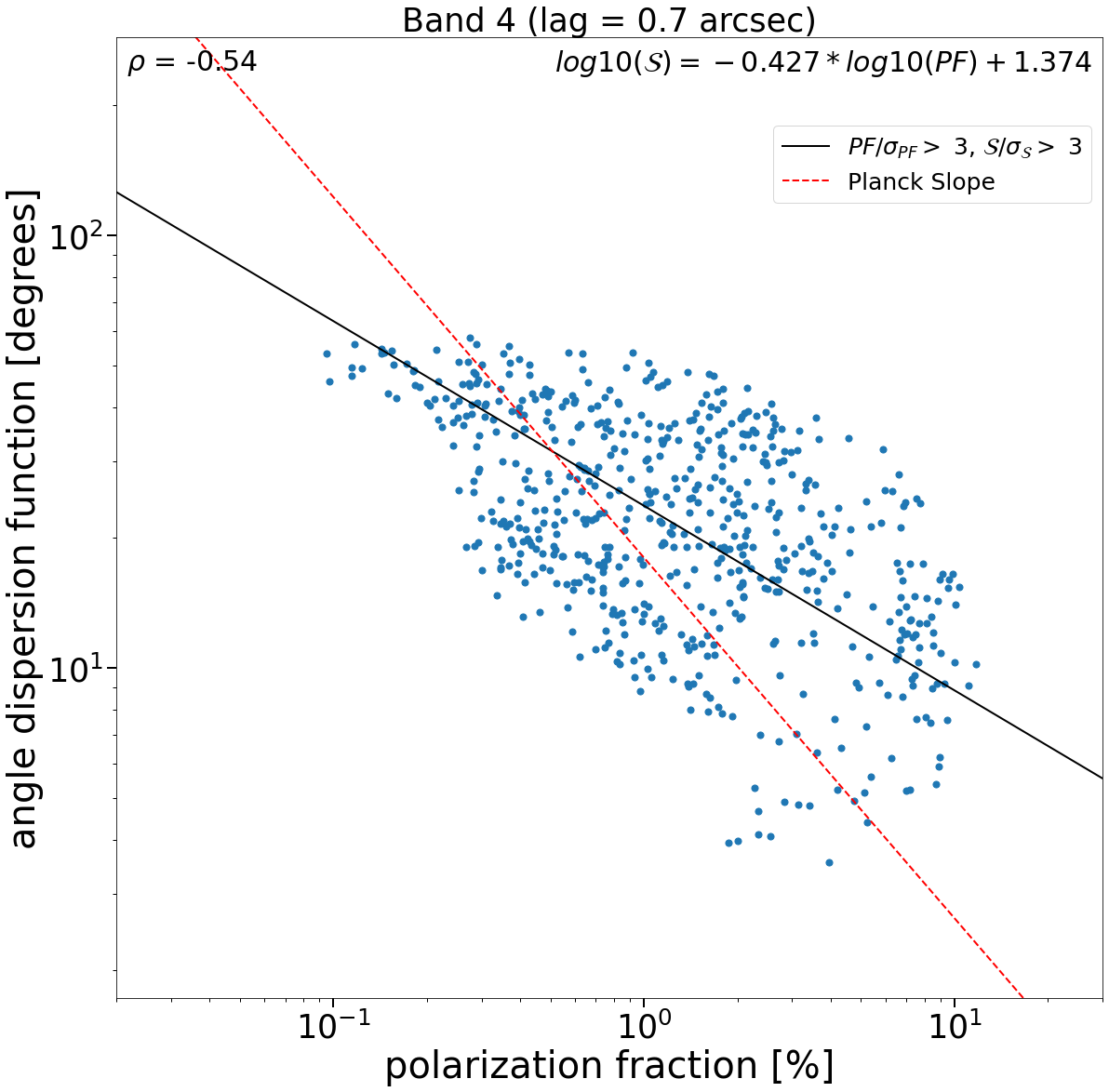}
    \end{subfigure}
    \hfill
    \begin{subfigure}{0.45\textwidth}
        \centering
        \includegraphics[width=\linewidth, height=0.80\linewidth]{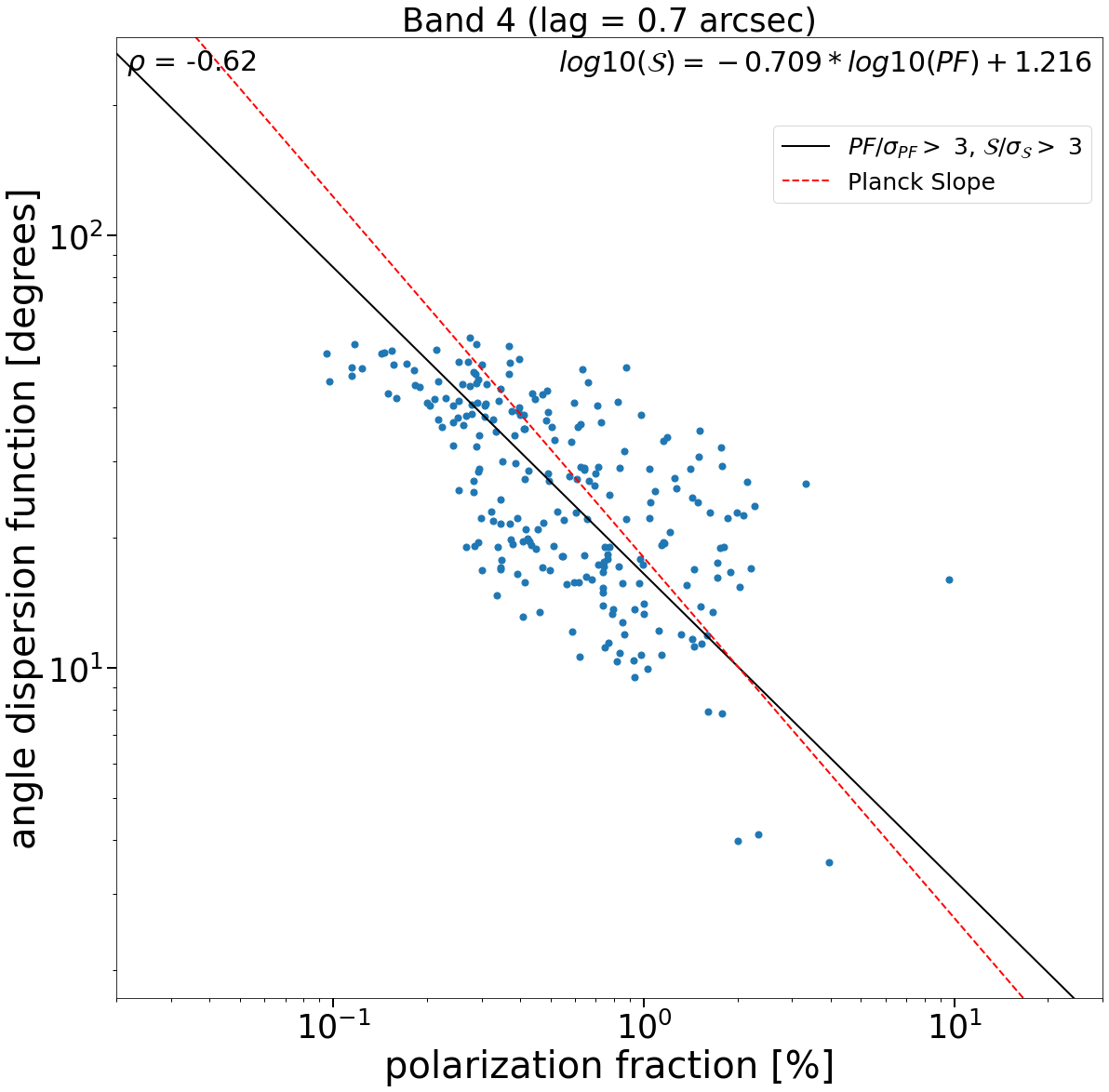}
    \end{subfigure}
    \vskip\baselineskip
    \begin{subfigure}{0.45\textwidth}
        \centering
        \includegraphics[width=\linewidth, height=0.80\linewidth]{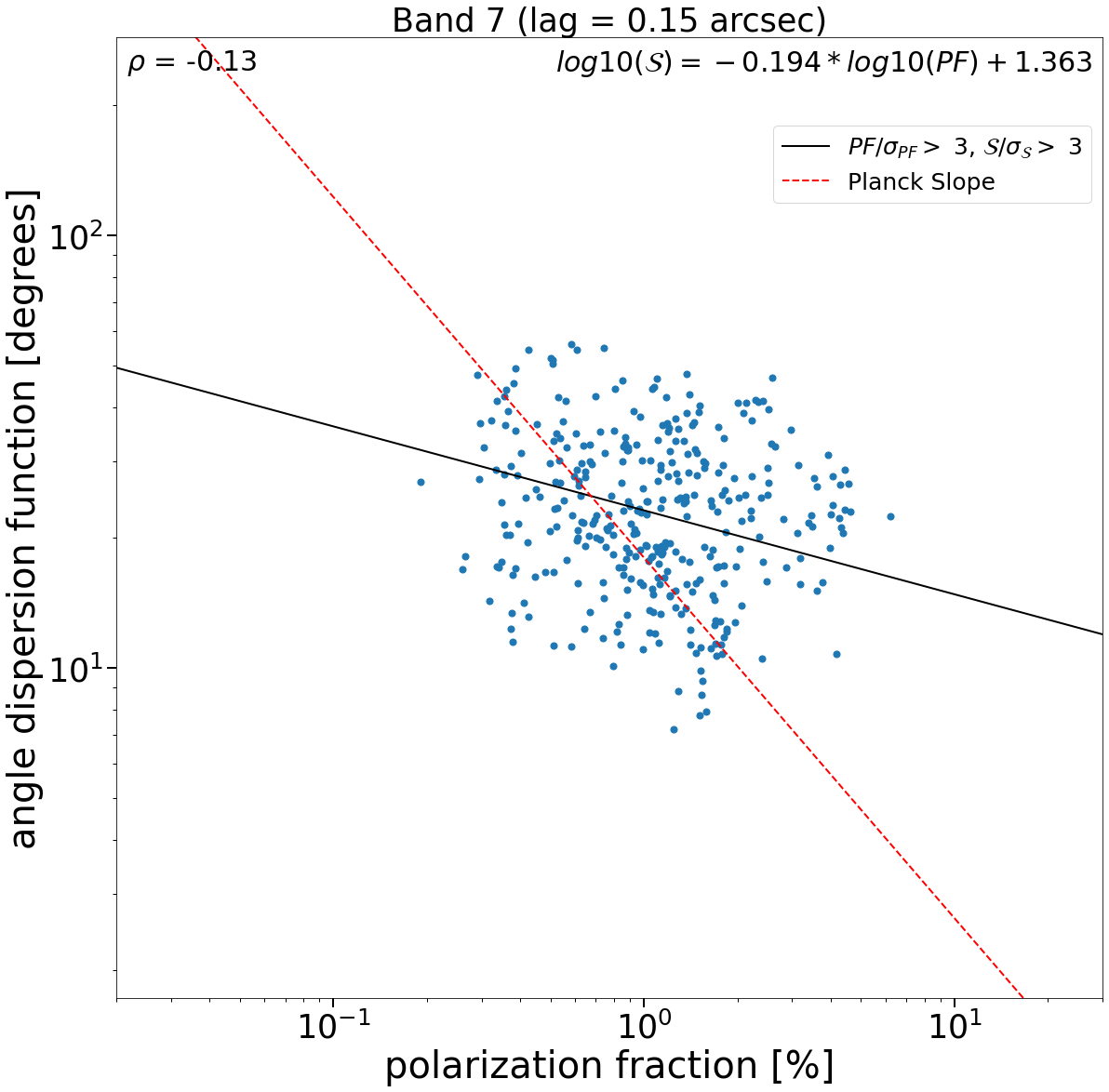}
    \end{subfigure}
    \hfill
    \begin{subfigure}{0.45\textwidth}
        \centering
        \includegraphics[width=\linewidth, height=0.80\linewidth]{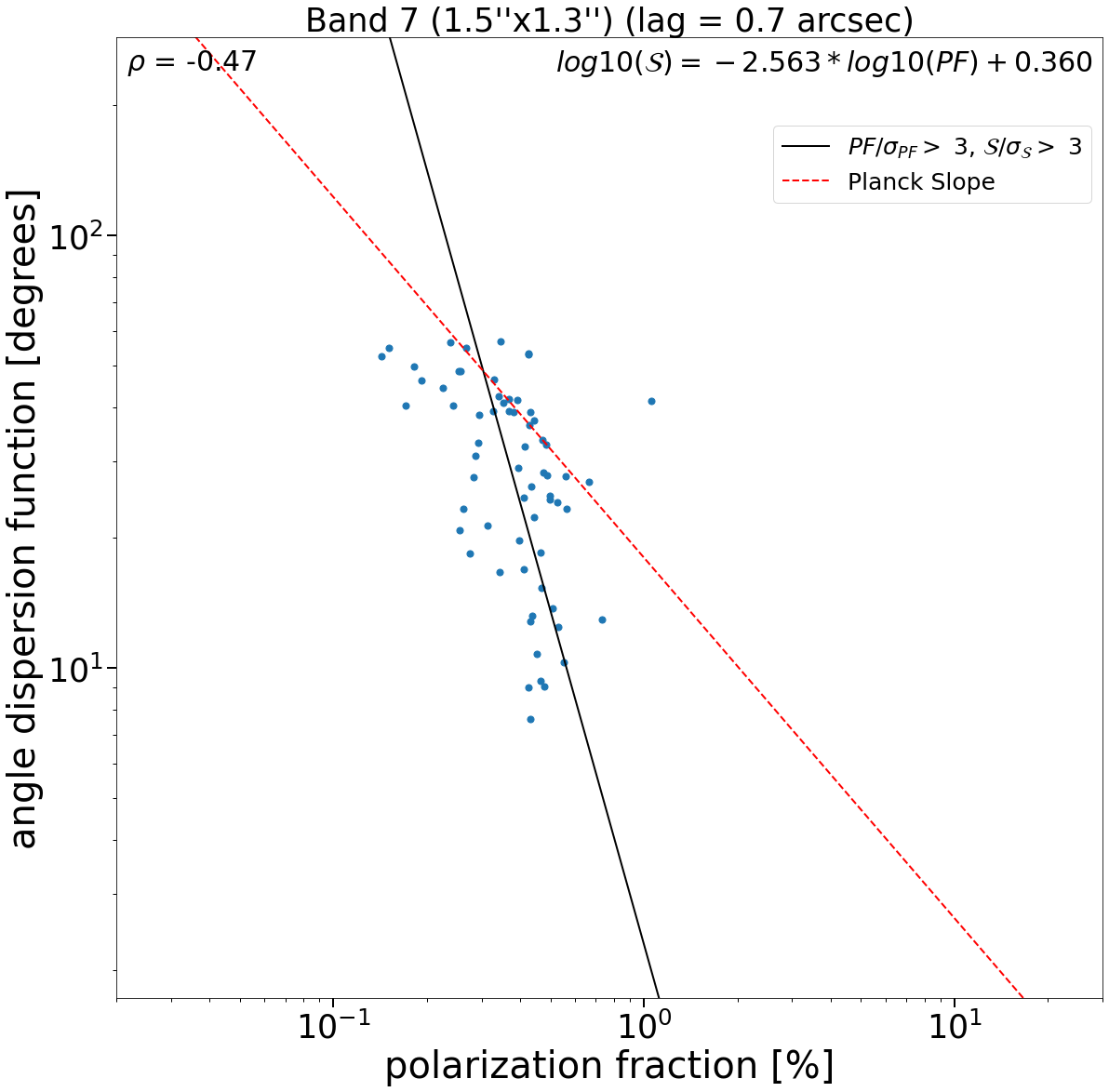}
    \end{subfigure}
    \caption{ { {\bf Observed polarisation angle dispersion $\DeltaAng$ as a function of the polarisation fraction $\polfrac$.} Band 4 data are shown on the top left panel, Band 7 on the bottom left. On each plot, the black dashed line shows the trend derived from best fitting the data (see Equation\,\ref{eqn:psvsS}) excluding all pixels characterized by  $I/\sigma_{I} < 3$, $\polfrac/\sigma_{\polfrac} < 3$ and $\DeltaAng/\sigma_{\DeltaAng}$ < 3, }
    while the red line shows the best correlation obtained for the solar neighborhood in the MW from the Planck data for reference. The upper right panel and the lower right panel shows the same plots computed in Band~4 and Band~7 smoothed at the common resolution of $1.5 \arcsec \times 1.3 \arcsec$ using pixels belonging exclusively in the central starburst region, defined by the mask identified in Band 7 (at its native resolution) including pixels with SNR in total intensity larger than 3. \rev{On top of each panel the lag used to computed the angle dispersion function $\DeltaAng$ is indicated}.
    }
    \label{fig:smallpvsS}
\end{figure*}

\subsection{$\polfrac$ vs $\DeltaAng$ relation}


Recent polarisation studies also analyse the relation between $\polfrac$ and polarisation angle dispersion function $\DeltaAng$ (see \rev{Section~\ref{dispersion}}), which quantifies the local order of $\polang$, the magnetic field direction as projected on the sky \rev{\citep{Alina2023, planckcollaborationXIX}}. The all-sky maps of $\DeltaAng$ constructed from the Planck polarisation data at $350\,GHz$ show intricate filamentary structures of high $\DeltaAng$ values that share no common structures with column density filaments, except possibly along the Galactic plane. These ridges appear to separate regions with rapidly varying polarisation angles from regions with more coherent $\polang$. The Planck analysis showed a clear anticorrelation between $\polfrac$ and $\DeltaAng$, indicating that most of the variation in $\polfrac$ in the Planck data is depolarisation due to the tangling of the magnetic field within the beam and/or along the LOS. 

In \cite{PlanckCollaborationXX}, predictions based on MHD simulations
were compared to the data, showing a similar anti-correlation between $\polfrac$ and $\DeltaAng$, again suggesting that tangling of the field was responsible for the observed variations of $\polfrac$.
The simulations, however, were limited to column density values less than \rev{$\NH= 3\times10^{22}\,cm^{-2}$}, and the Planck data at the resolution used for polarisation analysis did not exceed $\NH= 10^{23}\,cm^{-2}$ where other physical effects could become important. 
The central starburst region of NGC253 allows us to explore a higher $\NH$ regime.

The upper and lower left panels of Figure\,\ref{fig:smallpvsS} show the correlation between $\polfrac$ and $\DeltaAng$ from our observations in Band~4 and Band~7. Only pixels with $I/\sigma_{I} > 3$, $\polfrac/\sigma_{\polfrac} > 3$ and \rev{$\DeltaAng/\sigma_{\DeltaAng} > 3$} are included. 

Following \cite{planckcollaborationXIX}, we fit a straight line to the data in log-log given by

\begin{equation}
log(\DeltaAng) = \alpha \times log(\polfrac) + \beta,
\label{eqn:psvsS}
\end{equation}

\noindent which we overlay on Figure\,\ref{fig:smallpvsS} in black. 
\rev{The linear fit was done taking into account the uncertainties in $\polfrac$ and $\DeltaAng$ using the orthogonal distance regression method.}
The correlation derived in 
the Planck \rev{Milky Way (MW)} data is also shown on the figure as a red dotted line for reference. 
\rev{We find $\alpha = -0.427$ and $\beta = 1.374$ for Band~4 data and $\alpha = -0.194$ and $\beta = 1.363$ for Band~7 data. The Pearson correlation coefficients are $\pearson = -0.54$ and $\pearson = -0.13$ for Band~4 and Band~7, respectively, denoting an overall moderate anti-correlation for Band~4 and a little to no anti-correlation for Band~7.}


The upper and lower right panels in Figure \ref{fig:smallpvsS} show the correlation between $\polfrac$ and $\DeltaAng$ for the two Band~4 and Band~7 images convolved to the Band~4 beam ($1.5 \arcsec \times 1.3 \arcsec $) and using a lag of 0.70$\arcsec$ for the $\DeltaAng$ computation. 
In this case, we applied the same mask used in Section \ref{common}, which only includes pixels with $I/\sigma_{I} > 3$ in Band~7 at native resolution. Additionally, we excluded all pixels with $\polfrac/\sigma_{\polfrac} < 3$ and $\DeltaAng/\sigma_{\DeltaAng}$ < 3.
In this case, the anti-correlation is tight in both bands, with a Pearson index of -0.62 and -0.47 for Band~4 and Band~7, respectively. 
The anti-correlation observed at these scales at both bands is consistent with the similarities observed in maps of $\DeltaAng$ and $\polfrac$, when smoothed to the same angular resolution (see \newrev{Section\,\ref{fig:images_sameres}}). This indicates that the same magnetic field structure is traced by both dust and synchrotron emission in the central regions of NGC253 at $\sim$ 25 pc scales.
Overall, the dependence of $\polfrac$ with $\DeltaAng$ and $\NH$ is quite similar to what was observed in \cite{planckcollaborationXIX} in the MW.

In Band~7 at native resolution (lower left panel of Figure\,\ref{fig:smallpvsS}), we see a different behaviour compared to the MW Planck results. At this scale ($\sim5$\,pc), the anticorrelation between $\NH$ and $\polfrac$ is still present, while we see no anticorrelation between $\DeltaAng$ and $\polfrac$. This could indicate that thermal dust polarisation varies independently of the apparent B-field tangling at column densities larger than \rev{$\NH\simeq 10^{23}$\,cm$^{-2}$} and scales $\lesssim 5$ pc. It might suggest that, at scales of $\sim 5$ pc in NGC253, the physical properties of the ISM, rather than the tangling of the B-fields, control the observed depolarisation. Although we do not have access to $\sim 5$ pc in Band~4 data, this divergent behaviour appears to be primarily caused by the different scales that characterize the Band~4 and Band~7 data, rather than the contributions of synchrotron and dust in these bands, since the correlation between $\polfrac$ and $\DeltaAng$ is recovered in Band~7 when the image is smoothed to the same resolution of Band~4, even though the slope and $\pearson$ index are different.

\section{Conclusions} \label{conclusion}

We have presented ALMA Band~4 (145\,GHz) and Band~7 (350\,GHz) polarisation measurements for the central $\sim$200\,pc of the starburst galaxy NGC253. We have derived maps of the polarisation intensity and fraction, of the B-field orientation and polarisation angle dispersion function, as well as the corresponding pixel-by-pixel signal-to-noise maps. The reported uncertainties on the polarisation results account for the covariance between the Stokes I, Q and U parameters.
The uncertainties allow us to detect PF as low as 0.1\% with a SNR greater than 3. 
The polarisation properties measured at Bands~4 and~7 exhibit distinct characteristics, reflecting the different emission mechanisms that dominate at each band.

To evaluate the contributions of synchrotron, free-free, and dust in the region, we conducted an SED-fitting analysis using archival data from ALMA and VLA across the frequency range $1-350$\,GHz. 
At Band 7, the total intensity is mainly thermal dust emission, while Band 4 comprises significant contributions from free-free, dust, and synchrotron emission. By assuming a constant $\polfrac$ for the synchrotron and dust components in the $150-350$ GHz range, we estimated the $\polfrac$ values for these components for our observed field and various smaller regions. We find that the synchrotron emission is more polarised than the dust emission in all analyzed regions. This higher polarisation fraction of the synchrotron component appears to be the primary reason for the higher than expected polarisation observed in Band~4 maps, despite the significant contribution from (unpolarised) free-free emission at this frequency.

The analysis also revealed a sharp decline in $\polfrac$ estimates for the very centre of the starburst region (labelled as SC), which is characterized by high column density values and the presence of several super star clusters. This decline is likely caused by different factors, including significant non-polarised radiation from free-free emission and magnetic field fluctuations along the line-of-sight. The more pronounced decrease for the dust polarisation fraction in this central region suggests that differential tangling between the dense and more diffuse medium along the line of sight, and/or a dependence of the  grain alignment or destruction of dust particles with environment may also play a role.

Using archival high-resolution ALMA observations of the CO(3-2) line and our debiased polarimetric data, we investigated the correlation between $\polfrac$ and $\NH$ in NGC253. 
We found a significant anti-correlation between $\polfrac$ and $\NH$.
This trend suggests that the decrease in $\polfrac$ with $\NH$ observed in Planck data \citep{planckcollaborationXIX} continues to even higher column densities. 

Furthermore, the $\polfrac$ and the polarisation angle dispersion function $\DeltaAng$ are moderately anti-correlated in both bands when smoothed to spatial scales of $\sim25$\,pc. At its native spatial resolution of $\sim5$\,pc, we find no relationship between  $\polfrac$ and $\DeltaAng$ in the Band~7 data. Our results suggest that the tangling of the B-field is the main cause of the observed depolarisation at scales $\sim 25$ pc, while other factors play a role at scales $\sim 5$ pc. Further studies are required to test the effect of environment and spatial scale on the polarised emission of dust grains.

The similarity of the B-fields orientation maps for Bands 4 and 7 images convolved at common resolution suggests that, on spatial scales $\sim$ 25 pc, the same magnetic field configuration is traced by both bands. This indicates that while the distinct contributions of synchrotron and dust influence the observed polarisation fraction characteristics, the magnetic field structures are largely unaffected at these scales.

From a qualitative analysis, the B-field morphology seems to be coupled with the multi-phase outflow observed in the central region of NGC253. The B-field probed by Band~4 data follows the base of the biconical outflow, mirroring its profile. The conical outflow and the B-fields both seem to originate in the central region SC. This region is characterized by the presence of several super starclusters with exceptionally high broad linewidth $H40\alpha$ emission, which may be the main responsible in driving the ionized component of the multi-phase outflow. The high polarisation fraction of the synchrotron component estimated at the launch site of the outflow could be explained if the B-fields are compressed and ordered by the strong winds produced by the SSCs. A coupling between the B-field and the molecular outflow is seen mostly along the direction of the SW streamer, where the B-field probed by Band~4 data seems to be disrupted on \rev{the northern sides} of the midplane. Overall, our results highlight the interplay between magnetic fields, star formation, and outflows in shaping the central region of NGC253. A more quantitative analysis of the connection between the B-field structure and outflow activity in NGC253 goes beyond the scope of this paper, and will be addressed in a future work.

\begin{acknowledgements}
This paper makes use of the following ALMA data: ADS/JAO.ALMA\#2018.1.01358.S. ALMA is a partnership of ESO (representing its member states), NSF (USA) and NINS (Japan), together with NRC (Canada), NSTC and ASIAA (Taiwan), and KASI (Republic of Korea), in cooperation with the Republic of Chile. The Joint ALMA Observatory is operated by ESO, AUI/NRAO and NAOJ. 
DA acknowledges support from Nazarbayev University Faculty Development Competitive Research Grant No. 201223FD8821.
ADB and SAC acknowledge partial support from grant NSF-AST 2108140.

\end{acknowledgements}

\bibliographystyle{aa} 
\bibliography{bibliography.bib} 

\begin{appendix} 

\section{ALMA and VLA archival images}\label{appA}

For the SED analysis presented in \newrev{Section~\ref{analysis}}, we assembled archival data covering the  frequency range from 1.4 to 36~GHz taken with the Very Large Array (VLA) (\url{http://www.vla.nrao.edu/astro/nvas/}) and from  100 to 690~GHz taken with ALMA  (\url{https://almascience.eso.org/aq/}). The images have different resolutions
reported in Table~\ref{tab:ancillary}, 
and have been convolved to the resolution of 
the lowest frequency (1.4~GHz) image (2.2$\arcsec$).  The Table~\ref{tab:ancillary} reports the details of the archival data.

\begin{table}[tbph!] 
\begin{tabular}{|c|c|c|c|c|}
\toprule
Band & Centre  & Angular & MRS & Project code \\
 &  frequency & resolution &  &   \\
 & [GHz] & $[\arcsec$] & [$\arcsec$] &  \\
\toprule
      \multicolumn{5}{|c|}{VLA} \\ 
\toprule


L &1.499& 1.61 &36&AU30\\
L &1.499& 2.2 &36&AU30\\
C &4.985& 1.0& 8.9& VAH37\\
X& 8.439& 0.63& 17 & TEST$^*$ \\
Ku & 14.94&0.54& 12& AU36\\
K & 23.870 & 0.21 & 2.4 & 16B337 \\
Ka & 36.1693 & 0.096 & 1.6 & 16B337 \\

\toprule
      \multicolumn{5}{|c|}{ALMA} \\ 
\toprule
3& 104.264& 1.672 & 18.547 & 2016.1.00292.S\\
\bf{4} & \bf{145.0} & \bf{1.76} & \bf{17.9}& \bf{This work}\\
5 & 179.849& 0.946 & 10.101 & 2017.1.00028.S\\
5 & 193.003& 1.11 &11.463& 2018.1.00162.S\\
5 & 196.392& 1.175 &11.94 & 2018.1.00162.S\\
5 & 199.702& 1.316 &12.553 & 2018.1.00162.S\\
5 & 203.057& 1.19 &10.96 & 2018.1.00162.S\\
6& 265.669& 0.967& 9.675& 2017.1.00161.L\\
6 & 234.469& 0.907 & 11.457& 2013.1.00099.S\\
7 & 295.572& 0.892& 8.882& 2017.1.00161.L\\
7 & 289.255& 0.868& 8.1& 2017.1.00161.L\\
7 & 285.891& 0.923& 8.965& 2017.1.00161.L\\
7 & 282.344& 0.93& 9.24 & 2017.1.00161.L\\
\bf{7} & \bf{343.5} & \bf{0.3} & \bf{3.8} & \bf{This work}\\ 
8& 484.804 & 1.551 & 12.72 & 2013.1.00368.S\\
9& 690.981& 1.675 & 10.915 & 2018.1.00294.S\\
\hline
7$^{**}$ & 345.141 & 0.43& 9 &2015.1.00274.S\\
\toprule

\end{tabular}

\footnotesize{$^*$: The data have been taken during a TEST run on Jul 3th 1990.

$^{**}$: This dataset contains the CO(3-2) line emission, used in Section \ref{sect:pfvsNH}.   }

\caption{Data used for the SED analysis (see Section \ref{analysis}). The continuum images have been gathered through the VLA  and ALMA public archives. Columns report the band name, the central frequency, the native angular resolution and maximum recoverable scale (MRS) and the project code of the data.}

\label{tab:ancillary}

\end{table}

\end{appendix}


\end{document}